\documentclass[useAMS,usenatbib]{aastex61}

\usepackage{enumitem}
\usepackage{calc}
\usepackage{chngcntr}
\usepackage{epsfig}
\usepackage{amsmath}
\usepackage{relsize}
\usepackage{natbib}
\usepackage{graphicx}
\usepackage[caption=false]{subfig}
\usepackage{wrapfig}
\usepackage[T1]{fontenc}
\usepackage[normalem]{ulem}
\usepackage{textcomp}
\usepackage{longtable}
\usepackage{answers}
\usepackage{array}
\usepackage{rotating}
\usepackage{tablefootnote}
\usepackage{float}
\usepackage{mathtools}
\DeclarePairedDelimiter\abs{\lvert}{\rvert}

\begin{document}

\shorttitle{JCMT Transient Survey: Variability on Several Year Timescales}
\shortauthors{Mairs et al.}
\title{The JCMT Transient Survey: Identifying Submillimetre Continuum Variability over Several Year Timescales Using Archival JCMT Gould Belt Survey Observations }

\correspondingauthor{Steve Mairs}
\email{smairs@uvic.ca}

\author{Steve Mairs}
\affiliation{Department of Physics and Astronomy, University of Victoria, Victoria, BC, V8P 1A1, Canada}
\affiliation{NRC Herzberg Astronomy and Astrophysics, 5071 West Saanich Rd, Victoria, BC, V9E 2E7, Canada}

\author{Doug Johnstone}
\affiliation{Department of Physics and Astronomy, University of Victoria, Victoria, BC, V8P 1A1, Canada}
\affiliation{NRC Herzberg Astronomy and Astrophysics, 5071 West Saanich Rd, Victoria, BC, V9E 2E7, Canada}

\author{Helen Kirk}
\affiliation{NRC Herzberg Astronomy and Astrophysics, 5071 West Saanich Rd, Victoria, BC, V9E 2E7, Canada}

\author{James Lane}
\affiliation{Department of Physics and Astronomy, University of Victoria, Victoria, BC, V8P 1A1, Canada}

\author{Graham S. Bell}
\affiliation{East Asian Observatory, 660 North A`oh\={o}k\={u} Place, University Park, Hilo, Hawaii 96720, USA}

\author{Sarah Graves}
\affiliation{East Asian Observatory, 660 North A`oh\={o}k\={u} Place, University Park, Hilo, Hawaii 96720, USA}

\author{Gregory J. Herczeg}
\affiliation{Kavli Institute for Astronomy and Astrophysics, Peking University, Yiheyuan 5, Haidian Qu, 100871 Beijing, People's Republic of China}

\author{Peter Scicluna}
\affiliation{Academia Sinica Institute of Astronomy and Astrophysics, P. O. Box 23-141, Taipei 10617, Taiwan}

\author{Geoffrey C. Bower}
\affiliation{Academia Sinica Institute of Astronomy and Astrophysics, 645 N. A'ohoku Place, Hilo, HI 96720, USA}

\author{Huei-Ru Vivien Chen}
\affiliation{Department of Physics and Institute of Astronomy, National Tsing Hua University, Taiwan}

\author{Jennifer Hatchell}
\affiliation{Physics and Astronomy, University of Exeter, Stocker Road, Exeter EX4 4QL, UK}

\author{Yuri Aikawa}
\affiliation{Department of Astronomy, University of Tokyo, Tokyo, Japan}

\author{Wen-Ping Chen}
\affiliation{Graduate Institute of Astronomy, National Central University, 300 Jhongda Road, Zhongli, Taoyuan, Taiwan}


\author{Miju Kang}
\affiliation{Korea Astronomy and Space Science Institute, 776 Daedeokdae-ro, Yuseong-gu, Daejeon 34055, Republic of Korea}

\author{Sung-Ju Kang}
\affiliation{Korea Astronomy and Space Science Institute, 776 Daedeokdae-ro, Yuseong-gu, Daejeon 34055, Republic of Korea}


\author{Jeong-Eun Lee}
\affiliation{School of Space Research, Kyung Hee University, 1732, Deogyeong-Daero, Giheung-gu Yongin-shi, Gyunggi-do 17104, Korea}

\author{Oscar Morata}
\affiliation{Academia Sinica Institute of Astronomy and Astrophysics, 11F of AS/NTU Astronomy-Mathematics Building, No.1, Sec. 4, Roosevelt Rd, Taipei 10617, Taiwan}

\author{Andy Pon}
\affiliation{Department of Physics and Astronomy, The University of Western Ontario, 1151 Richmond Street, London, ON, N6A 3K7, Canada}

\author{Aleks Scholz}
\affiliation{SUPA, School of Physics \& Astronomy, North Haugh, St Andrews, KY16 9SS, United Kingdom}

\author{Satoko Takahashi}
\affiliation{Joint ALMA Observatory, Alonso de C\'ordova 3107, Vitacura, Santiago, Chile}
\affiliation{National Astronomical Observatory of Japa, 2-21-1 Osawa, Mitaka, Tokyo 181-8588, Japan}

\author{Hyunju Yoo}
\affiliation{School of Space Research, Kyung Hee University, 1732, Deogyeong-Daero, Giheung-gu Yongin-shi, Gyunggi-do 17104, Korea}
\affiliation{Department of Astronomy and Space Science, Chungnam National University, 99 Daehak-ro, Yuseong-gu, Daejeon 34134, Republic of Korea}


\author{The JCMT Transient Team}

\begin{abstract}

Investigating variability at the earliest stages of low-mass star formation is fundamental in  
understanding how a protostar assembles mass.
While many simulations of protostellar disks predict non-steady accretion onto protostars, deeper investigation requires robust observational constraints on the frequency and amplitude of variability events characterised across the observable SED. 
In this study, we develop methods to robustly analyse repeated observations of an area of the sky for submillimetre variability in order to determine constraints on the magnitude and frequency of deeply embedded protostars. We compare \mbox{850 $\mu$m} JCMT Transient Survey data with archival JCMT Gould Belt Survey data to investigate variability over 2-4 year timescales. Out of 175 bright, independent emission sources identified in the overlapping fields, 
we find 7 variable candidates, 5 of which we classify as \textit{Strong} and the remaining 2 as \textit{Extended} to indicate the latter are associated with larger-scale structure. For the \textit{Strong} variable candidates, we find an average fractional peak brightness change per year of $\abs{4.0}\%\mathrm{\:yr}^{-1}$ with a standard deviation of $2.7\%\mathrm{\:yr}^{-1}$. 
In total, 7\% of the protostars associated with \mbox{850 $\mu$m} emission in our sample show signs of variability. Four of the five \textit{Strong} sources are associated with a known protostar. The remaining source is a good follow-up target for an object that is anticipated to contain an enshrouded, deeply embedded protostar. In addition, we estimate the \mbox{850 $\mu$m} periodicity of the submillimetre variable source, EC 53, to be \mbox{567 $\pm$ 32 days} based on the archival Gould Belt Survey data. 

\end{abstract}


\section{Introduction}
\label{introductionsec}
  
 The accretion history of low-mass stars has been under investigation for many years \citep[see, for examples, ][]{kenyon1990,bell1994}. 
 Despite recent advances \citep{vorobyov2005,rice2010,mckeellp,dunham2012, bae2014,vorobyov2015}, few observational constraints exist regarding the dominant mass transfer processes occurring during the earliest phases of star formation. The presence of a disk around a protostar complicates 
 how matter is moved from the nascent envelope to the central source. Through a combination of rotation and magnetic fields, anisotropies develop in the otherwise symmetric infalling material predicted in 
 the seminal model of \cite{shu1977}, causing this material to first accrete onto the disk before being transported to the protostar  \citep[for a review on pre-main sequence accretion, see ][]{hartmann2016}. The rate 
 of this mass transport likely varies on both long and short timescales due to gravitational \citep{vorobyov2005,vorobyov2006}, magnetorotational \citep{armitage2001,zhu2009}, and spiral wave \citep{bae2016} instabilities. Material that builds up in the disk can compress the magnetosphere of a forming star, leading gas to rapidly flow onto the star through directed funnels (\citealt{romanova2003}; see also \citealt{cody2017}). In addition, the formation of giant planets 
 \citep{nayakshin2012} and stellar encounters in binary systems \citep{forgan2010,hodapp2012} can also affect the rate at which a protostar gains mass. 
  
 Episodic accretion (short bursts of mass accretion separated by long, quiescent periods) caused by these instabilities in the disk has been gaining considerable attention in the literature \citep[see][and references therein]{audard2014}. These events may be related to known outbursts observed in FU Orionis \citep{herbig1977,hartmann1985,reipurth1990} and EX Lupi \citep{herbig1989} objects. 
 In addition to these large outbursts, Classical T Tauri Stars (CTTS) have been known to exhibit lower levels of variability \citep[e.g.][]{johns1995,alencar2001,bouvier2007,donati2013,blinova2016}. Accretion variability events with analogous intensities and timescales may occur during the youngest stages of a forming star's evolution. A non-steady accretion rate is also one solution to ``the luminosity problem'': the order of magnitude discrepancy between the fainter, observed median protostellar luminosity and the predicted protostellar luminosity assuming constant, steady state accretion \citep{kenyon1990}. Other solutions to this problem, such as longer timescales over which low-mass stars form, have also been posed \citep[e.g.][]{mckeellp}. A recent study of 3000 Young Stellar Objects (YSOs) present within 18 different molecular clouds conducted by \cite{dunham2015} using data obtained by the \textit{Spitzer Space Telescope}, however, has provided evidence that Class 0+I protostellar lifetimes are relatively short (\mbox{0.46-0.72 Myr}; see also, \citealt{evans2009,heiderman2015,ribas2015,fischer2017}). 
 
Observing the changes in accretion rate around a forming star is fundamental to understanding the earliest stages of star formation. This is possible by measuring significant changes in the protostar's peak brightness over time since the protostar's luminosity is generated primarily by the accreting material. \cite{johnstone2013} modelled the spectral energy distribution (SED) of a deeply embedded protostar undergoing an outburst due to an increased mass accretion rate and concluded that while a stronger signal will be detected at the peak of the SED (mid- and far-infrared wavelengths), the outburst also would be observable in the submillimetre regime. The submillimetre luminosity arises from cascade reprocessing of the stellar radiation by the surrounding dust. The negligible heat capacity for interstellar dust causes the time lag between the burst and the observation at submillimetre wavelengths to be dominated by the light crossing time of the protostellar envelope, which \cite{johnstone2013} calculated to be several weeks to months long.

The benefit of observing variability at submillimetre wavelengths is that ground-based instruments such as the James Clerk Maxwell Telescope (JCMT) are available to monitor large areas of the sky at a regular cadence. The JCMT Transient Survey \citep{herczeg2017arxiv} is obtaining simultaneous  continuum data at \mbox{450 $\mu$m} and \mbox{850 $\mu$m} on 8 nearby (<500 pc) star-forming molecular clouds at approximately 28 day intervals, when the targets are observable using the Submillimetre Common-User Bolometre Array 2 (SCUBA-2). These eight, 30$\arcmin$ in diameter, regions were originally selected from the JCMT Gould Belt Survey (GBS; \citealt{wardthompson2007}) and shifted on sky to optimise the number of Class 0+I and Class II YSOs present in each field. The GBS data that overlap with the Transient Survey data were collected between 2012 and 2014 while the Transient Survey data collection began in December, 2015. 
Therefore, these two surveys can be compared to identify and investigate variable signals over 2-4 year timescales. Here, we consider Transient Survey data obtained prior to March 1$^{\mathrm{st}}$, 2017. In the present work, we only use \mbox{850 $\mu$m} data, which has higher signal to noise ratio than the \mbox{450 $\mu$m} data (see Section \ref{obssec}).


The GBS data were not obtained with the goal of detecting protostellar variability, so the observations are not regularly spaced in time and they often occur over only one or two nights for a given field. Therefore, in this study, we measure the flux-calibrated peak brightnesses of extracted sources in the co-added GBS images and compare them to the same sources in the co-added Transient Survey images. In this way, we robustly characterise the properties of all identified sources and become more sensitive to long-term changes. Furthering our investigation, we also correlate the identified \mbox{850 $\mu$m} emission sources with the positions of known YSOs \citep{megeath2012,stutz2013,dunham2015}. This is the first study that systematically analyses two consistent submillimetre datasets separated in time in order to determine constraints on the magnitude and the frequency of deeply embedded, variable protostars.

This paper is organised as follows: in Section \ref{obssec}, we summarise the Transient Survey, GBS, and YSO catalogue observations we use throughout this work. In Section \ref{drcalsec}, we give an overview of our data reduction, image alignment, and relative flux calibration procedures. In Section \ref{GBStransientresultssec}, we present the results of the comparison between the GBS and Transient Survey. This includes the identification of variable candidates, a description of the quality of those candidates, and the calculated fractional peak brightness changes per year. In Section \ref{GBStransientdiscusssec}, we discuss the results, highlight previously known variable sources in our sample, and construct a light curve for the variable source EC 53 (\citealt{hodapp2012,yoo2017arxiv}). 
Finally, in Section \ref{GBStransientconclusionsec}, we summarise our main results and provide concluding remarks.

 
\section{Observations}
\label{obssec}


\begin{table*}
\centering
\caption{A summary of the observed fields and their co-added noise at 850 $\mu$m. JCMT Transient Survey Fields are in bold and associated JCMT Gould Belt Survey fields are listed below each Transient Survey field.}
\label{fieldstable}
\begin{tabular}{|c|c|c|c|c|c|c|c|c|}
\hline
Tile Name$^{a}$ & \multicolumn{1}{|p{1.75cm}|}{\centering Central R.A. \\ (J2000)} & \multicolumn{1}{|p{1.5cm}|}{\centering Central Dec \\ (J2000)} & Start Date$^{b}$ & End Date$^{b}$ & $\Delta t^{c}$ & \multicolumn{1}{|p{2.25cm}|}{\centering \mbox{850 $\mu$m} Noise$^{d}$\\(Jy beam$^{-1}$)} & \# Obs.$^{e}$ &  \multicolumn{1}{|p{2.6cm}|}{\centering SCUBA-2 Programme\\ Identification}\\
\hline\hline
\textbf{IC348} & 03:44:18.3 & 32:05:16.0 & 20151222 & 20170209 & -- & 0.0043 & 9 & M16AL001\\
IC348-E & 03:44:24.4 & 32:02:08.6 & 20120816 & 20130205 & 3.5 & 0.0055 & 4 & MJLSG38\\
\textbf{NGC 1333} & 03:28:54.5 & 31:17:09.0 & 20151222 & 20170206 & -- & 0.0039 & 10 & M16AL001\\
NGC1333-N & 03:29:06.3 & 31:22:26.7 & 20120702 & 20120703 & 4.0 & 0.0052 & 4 & MJLSG38\\
\textbf{OMC 2-3} & 5:35:33.2 & -5:00:32 & 20151226 & 20170206 & -- & 0.0042 & 9 & M16AL001\\
OMC1 tile4 & 5:35:49.4 & -4:46:23 & 20120817 & 20130826 & 3.2 & 0.0045 & 4 & MJLSG31\\
\textbf{NGC 2068} & 5:46:13.0 & -0:06:05 & 20151226 & 20170206 & -- & 0.0039 & 10 & M16AL001\\
OrionBN\_450\_S & 5:46:13.0 & -0:06:05 & 20141116 & 20141122 & 1.6 & 0.0046 & 6 & MJLSG41\\
\textbf{NGC 2024} & 5:41:41.0 & -1:53:51 & 20151226 & 20170206 & -- & 0.0043 & 11 & M16AL001\\
OrionBS\_450\_E & 5:42:48.0 & -1:54:36 & 20141027 & 20141109 & 1.6 & 0.0045  & 6 & MJLSG41\\
OrionBS\_450\_W & 5:40:32.2 & -1:48:00 & 20130212 & 20130303 & 3.3 & 0.0045  & 4 & MJLSG41\\
\textbf{Oph Core} & 16:27:03.2 & -24:32:46.5 & 20160115 & 20170206 & -- & 0.0050 & 8 & M16AL001\\
L1688-1 & 16:27:02.9 & -24:41:44.5 & 20120506 & 20120508 & 4.0 & 0.0053 & 4 & MJLSG32 \\
L1688-2 & 16:27:15.1 & -24:10:09.7 & 20120518 & 20120520 & 4.0 & 0.0051 & 4 & MJLSG32\\
\textbf{Serpens Main} & 18:29:48.7 & 01:15:39.5 & 20160202 & 20170222 & -- & 0.0045 & 9 & M16AL001\\
SerpensMain1 & 18:29:59.8 & 01:14:46.9 & 20120518 & 20120519 & 4.1 & 0.0058 & 4 & MJLSG33 \\
\textbf{Serpens South} & 18:30:02.2 & -02:02:23.0 & 20160202 & 20170222 & -- & 0.0047 & 9 & M16AL001\\
SerpensS-NE & 18:31:35.4 & -01:53:50.3 & 20120421 & 20120503 & 4.2 & 0.0049 & 4 & MJLSG33 \\
SerpensS-NW & 18:29:30.8 & -01:47:07.3 & 20120503 & 20120505 & 4.2 & 0.0048 & 5 & MJLSG33 \\
\hline
\end{tabular}
\begin{flushleft}
$^{a}$ The Tile Name corresponds to the target identifier in the JCMT archive.\\ 
$^{b}$ The start and end dates refer to the date of the first observation taken by each survey and the last observation taken before March $1^{\mathrm{st}}$, 2017 (yyyymmdd). \\
$^{c}$ The time between the average GBS date and the average Transient Survey date (years).\\
$^{d}$ These measurements of the  \mbox{850 $\mu$m} noise levels are based on a point source detection in each field's co-added image after smoothing with a 6$\arcsec$ FWHM Gaussian kernel. The effective beam size after smoothing is 15.8$\arcsec$.\\
$^{e}$ The number of observations included in the co-add.
%
%
%
%
\end{flushleft}
\end{table*}

In this analysis, we combine JCMT Transient Survey \citep{herczeg2017arxiv} observations with archival data of the same regions observed by the JCMT Gould Belt Survey (hereafter, GBS; \citealt{wardthompson2007}). All observations in both surveys are performed using the Submillimetre Common-User Bolometer Array 2 (SCUBA-2) instrument \citep{holland2013} at the James Clerk Maxwell Telescope (JCMT). SCUBA-2 provides continuum coverage at both \mbox{450 $\mu$m} and \mbox{850 $\mu$m} simultaneously with half-power bandwidths of \mbox{32 $\mu$m} and \mbox{85 $\mu$m} \citep{holland2013} and effective beam sizes of 9.8$\arcsec$ and 14.6$\arcsec$, respectively \citep{dempsey2013}.  While we expect a stronger variability signal toward the peak of the spectral energy distribution, typically the mid to far-IR \citep{johnstone2013}, the practical data reduction and analysis of SCUBA-2 \mbox{450 $\mu$m} images is complicated by a strong atmospheric water vapour dependency in addition to a less stable beam profile due to dish deformation and focus errors \citep{dempsey2013}. These effects dramatically influence the signal to noise ratio (SNR) and require further investigation before careful corrections are possible.
In this paper we focus only on the \mbox{850 $\mu$m} data. All of the observations were taken in the PONG1800 mapping mode \citep{kackley2010}, yielding circular maps (``PONGs'') $\sim$0.5\textdegree  \hspace{0.3mm} in diameter. 

In total, there are eight Transient Survey fields \citep{herczeg2017arxiv}: three in the Orion A and B Molecular Clouds (OMC 2-3, NGC 2068, and NGC 2024), two in the Perseus Molecular Cloud (IC348 and NGC 1333), two in the Serpens Molecular Cloud (Serpens Main and Serpens South), and one in the Ophiuchus Molecular Cloud (Oph Core). Eleven Gould Belt Survey fields include bright, compact sources within areas of significant overlap of submillimetre emission with the Transient Survey data (central locations, observation dates, \mbox{850 $\mu$m} noise, and programme identification numbers for all observations are summarised in Table \ref{fieldstable}). Figures \ref{IC348mosfig} through \ref{SERPENSSOUTHmosfig} in Appendix \ref{discussextended} show the archival GBS data mosaicked with the JCMT Transient Survey data.

\subsection{The JCMT Transient Survey}
 
JCMT Transient Survey \citep{herczeg2017arxiv} observations began on December 26$^{th}$, 2015 and have continued at an approximate cadence of  28 days whenever a given field is observable at the JCMT. In this paper, we address all observations obtained prior to March 1$^{\mathrm{st}}$, 2017. 
All of the observations performed in this survey were taken in either band 1, 2, or 3 weather, where the zenith opacity at \mbox{225 GHz}, $\tau_{225\mathrm{\:GHz}}$ is less than 0.12 (corresponding to a precipitable water vapour of less than 2.58 mm). For more details on the
individual observations included in this work, see \cite{mairs2017}. 

\subsection{The JCMT Gould Belt Survey}

GBS data were obtained from 2012 to 2014 and are publicly available on the JCMT archive. The observed fields, however, are not necessarily centred on the same locations as the JCMT Transient Survey fields. Thus, multiple fields may overlap to cover the same area of the sky (see Table \ref{fieldstable} and Figures \ref{IC348mosfig} through \ref{SERPENSSOUTHmosfig}). All GBS observations were designed to reach a uniform depth across a wide area of each star-forming cloud. The data were collected in weather bands 1, 2, or 3 with 4 to 6 repeats such that a consistent sensitivity was achieved across the different atmospheric conditions. The GBS observations were not originally intended for studying protostellar variability, so they were not taken at a regular cadence. Often, all integrations of a given field were obtained within 1-2 nights. These data, however, are useful to compare with our recent observations as they provide brightness measurements of our identified sources across longer time separations (see $\Delta$t in Table \ref{fieldstable}). 

Additional GBS fields that overlap with the Transient Survey fields are not included in this study. Fields are excluded if a self-consistent relative flux calibration could not be performed for the data (see Section \ref{drcalsec}) or if there are no significantly bright or compact sources in the region of overlap between the two survey coverages. Often, these cases occur if a GBS field has a significant amount of extended structure (complicating the disentangling of compact structures from background emission as in the case of OMC 2-3) or compact structure that is very near the edge of the map where the noise is higher. In total, 24 GBS fields have some overlapping area with the Transient fields, of which 11 produced self-consistent flux calibration using bright, compact peaks that are also observed in the Transient fields (see Appendix \ref{discussextended}).

\newpage

\subsection{\textit{Spitzer Space Telescope} and \textit{Herschel Space Observatory} YSO Catalogues}
\label{ysorefsec}

In order to associate 850 $\mu$m emission sources with known Young Stellar Objects, we cross-match the \textit{Spitzer Space Telescope} catalogues of \cite{megeath2012} and \cite{dunham2015}, and the \textit{Herschel Space Observatory} YSO catalogue presented by \cite{stutz2013}. \cite{megeath2012} and \cite{stutz2013} focus on the Orion A and B Molecular Clouds while the \cite{dunham2015} catalogue provides information for the remaining regions addressed in this paper. We adopt the YSO classifications of \cite{megeath2012} throughout the area of their survey. In the catalogue, the authors denote Class 0+I and Flat spectrum YSOs as ``P'', for protostars, and Class II YSOs as ``D'', for disks. They also include protostellar candidate designations ``FP'', for faint protostar candidate, and ``RP'', for red protostar candidate. We make no attempt to further differentiate these four classes. In the case of YSOs discovered by \cite{stutz2013}, we only include the objects labelled by the authors as reliable protostars (Flag 1) in our analysis and generically refer to them as ``'protostars'' throughout this work. Finally, \cite{dunham2015} provides the extinction-corrected infrared spectral index, $\alpha^{\prime}$, for each YSO and the standard classification scheme to differentiate them \citep{greene1994,dunham2015}:

\vspace{4mm}

\begin{itemize}[leftmargin=*]

  \item[] Class 0+I: $\alpha^{\prime}\geq0.3$
  
  \vspace{2mm}
  
  \item[] Flat Spectrum: $-0.3\leq\alpha^{\prime}<0.3$
  
  \vspace{2mm}
  
  \item[] Class II: $-1.6\leq\alpha^{\prime}<-0.3$

  

\end{itemize}

\vspace{4mm}

\noindent Where possible (all regions except those in the Orion A and B Molecular Clouds), we differentiate Flat Spectrum sources from protostars and refer to Class II sources as ``disks'' throughout the rest of this paper.

\section{Data Reduction and Image Calibration}
\label{drcalsec}

We largely follow the three step data reduction and calibration methodology adopted by the JCMT Transient Survey, which is described in detail by \cite{mairs2017}. We develop further methods, however, described in Section \ref{postredcalsec} to perform a relative flux calibration between the two independent datasets (the GBS and the Transient Survey). We first construct robust images from the raw SCUBA-2 data for all observations. Then, we perform a spatial alignment for each image using a reference field. Finally, we perform a relative flux calibration to bring all observations into agreement with the mean, co-added image (for the GBS and Transient surveys separately).

\subsection{Data Reduction}
\label{GBStransientdatareducsec}

Following \cite{mairs2017}, we first perform the data reduction using the iterative map-making software {\sc{makemap}} (described in detail by \citealt{chapin2013}) in the {\sc{SMURF}} package (\citealt{jenness2013}) found within the {\sc{Starlink}} software \citep{currie2014}. We grid each map to 3$\arcsec$ pixels and define convergence of the iterative solution when the difference in individual pixels changed on average by <0.1\% of the rms noise present in the map. For both the GBS and Transient data independently, we first run an ``auto-mask'' reduction where the map-making software identifies regions of significant emission and separates this signal from the atmospheric noise
that is eventually subtracted from the final output image.  We co-add the first 4 observations for each survey individually in order to create corresponding ``external masks'' with boundaries defined by a signal to noise ratio of at least 3. The external masks applied to the GBS and Transient surveys have negligible differences. Then, we perform a second round of data reduction using these masks to define areas of robust astronomical emission as an additional constraint in {\sc{makemap}}'s solution. In this way, we are able to simultaneously recover faint, extended structures as well as isolated and embedded compact sources. The final mosaic produced is in units of picowatts (pW), which we initially convert to \mbox{Jy beam$^{-1}$} using the \mbox{850 $\mu$m} JCMT default aperture flux conversion factor \mbox{565 Jy pW$^{-1}$ beam$^{-1}$}, based on a beam FWHM of 14.6$\arcsec$ 
\citep{dempsey2013}. 

The constructed SCUBA-2 maps are not sensitive to large-scale structures because these are filtered out during the data reduction process \citep{chapin2013}. For detecting submillimetre variability of compact sources, \cite{mairs2017} determined that the most robust results can be obtained by filtering out information on scales $>200\arcsec$ (reduction \textit{R3} in \citealt{mairs2017}). In this way, the peak fluxes of compact objects are well recovered and bright, compact source extraction is less confused by extended emission\footnote{For an overview of the effect of spatial filtering on SCUBA-2 data see \cite{mairs2015,mairs2017}.}. The \mbox{CO(J=3-2)} emission line contributes to the flux measured in these \mbox{850 $\mu$m} continuum observations (\citealt{johnstone1999}, \citealt{drabek2012}, \citealt{coude2016}). \cite{mairs2016} show, however, that the peak brightnesses of compact sources are not significantly affected by the removal of this line. 

\subsection{Post-Reduction Alignment and Flux Calibration}
\label{postredcalsec}

Nominally, the JCMT has a pointing error of 2$\arcsec$-6$\arcsec$ and a flux calibration uncertainty of  $\sim5\%-10\%$ \citep{dempsey2013,mairs2017}. Using the methods developed by \cite{mairs2017}, we achieve an alignment of $<1\arcsec$ and a flux calibration uncertainty (in a typical measurement) of 2\%-3\%. Since each image has a similar flux calibration uncertainty, co-adding the data further reduces the uncertainty in a given measurement by the square root of the number of observations included in the co-add. Therefore, a source's peak brightness may have an uncertainty as low as 1\% or less (see Tables \ref{fieldstable} and \ref{potvartable}). Briefly, the image alignment and flux calibration procedures rely on comparing the properties of bright, compact sources detected across multiple observations of the same area of the sky. To identify these sources, subtract larger-scale background flux, and extract properties such as the central position, size, and peak brightness, we employ the algorithm {\sc{Gaussclumps}} \citep{stutzki1990}, which models each compact object with a Gaussian profile. Specifically, we use the {\sc{starlink}} software \citep{currie2014} implementation of {\sc{Gaussclumps}} found within the {\sc{cupid}} \citep{berry2007} package. We expect the bright, point-like sources in an image to resemble Gaussian structures based on the shape of the JCMT beam. In order to mitigate spurious noise features, before running the source extraction algorithm we first smooth the maps with a 6$\arcsec$ (FWHM) Gaussian kernel \citep[see Appendix B of ][]{mairs2017}. 

As described by \cite{mairs2017}, the image alignment is performed by selecting the robust, Gaussian sources that have a peak brightness of at least 200 \mbox{mJy beam$^{-1}$} (\mbox{SNR $\sim$ 10} for an individual Transient Survey observation) and a maximum effective radius of 10$\arcsec$ (the effective radius is $r = \sqrt{\mathrm{FWHM_{\mathrm{1}}}\times \mathrm{FWHM}_{\mathrm{2}}}/2$ where the FWHM$_{N}$ terms are the full widths at half maximum of the major and minor axes of the fitted two dimensional Gaussian), comparing their central locations in each observation to a reference image, and then correcting for that offset. For both the Gould Belt and Transient Surveys, the chosen reference field for each region is the first Transient Survey observation of that region. The absolute position of each source is of little importance for our goals, as the most critical measurement for understanding the variability of a given point source is the relative peak brightness and the alignment uncertainty is small enough that we are able to confidently associate known protostars with the emission peaks.  

\cite{mairs2017} describe the procedure to identify and use robust calibrator sources in each field to self-consistently perform a relative flux calibration. The chosen calibrator sources  are referred to as \textit{Family members}. A given \textit{Family Member} in the Transient Survey is selected by measuring the peak brightness of all sources that are brighter than \mbox{500 mJy beam$^{-1}$} with effective radii $<10\arcsec$ normalised to their average peak brightness and comparing that value to all of the other sources in a given image with these properties. The largest set of sources that display a low amount of scatter from observation to observation with respect to one another (defined by a threshold of 6\% in the standard deviation) is selected to be a \textit{Family}. The fact that these sources agree well with one another over time suggests that none of them are intrinsically varying to the level we can detect and that they are tracking the flux uncertainty of the telescope. In this manner, we determine if each observation, as a whole, is slightly brighter or fainter than the mean image allowing each epoch (image) to be corrected by a constant multiplicative factor. 

We perform the relative flux calibration individually for the GBS data and the Transient data. As described in Section \ref{obssec}, the GBS data were often taken over a short time frame (one or two days) while the Transient data were taken over more than one year. Therefore, sources that do not appear to vary across the GBS observations could show signs of variability in the Transient data. In addition, a slowly varying source may look constant in both datasets individually, but given the long separation in time between the two surveys (see Table \ref{fieldstable}), the peak brightness could be dramatically different when cross-compared (see Section \ref{GBStransientresultssec}). 

\begin{figure*} 	
\centering
\subfloat{\label{}\includegraphics[width=9cm,height=7.8cm]{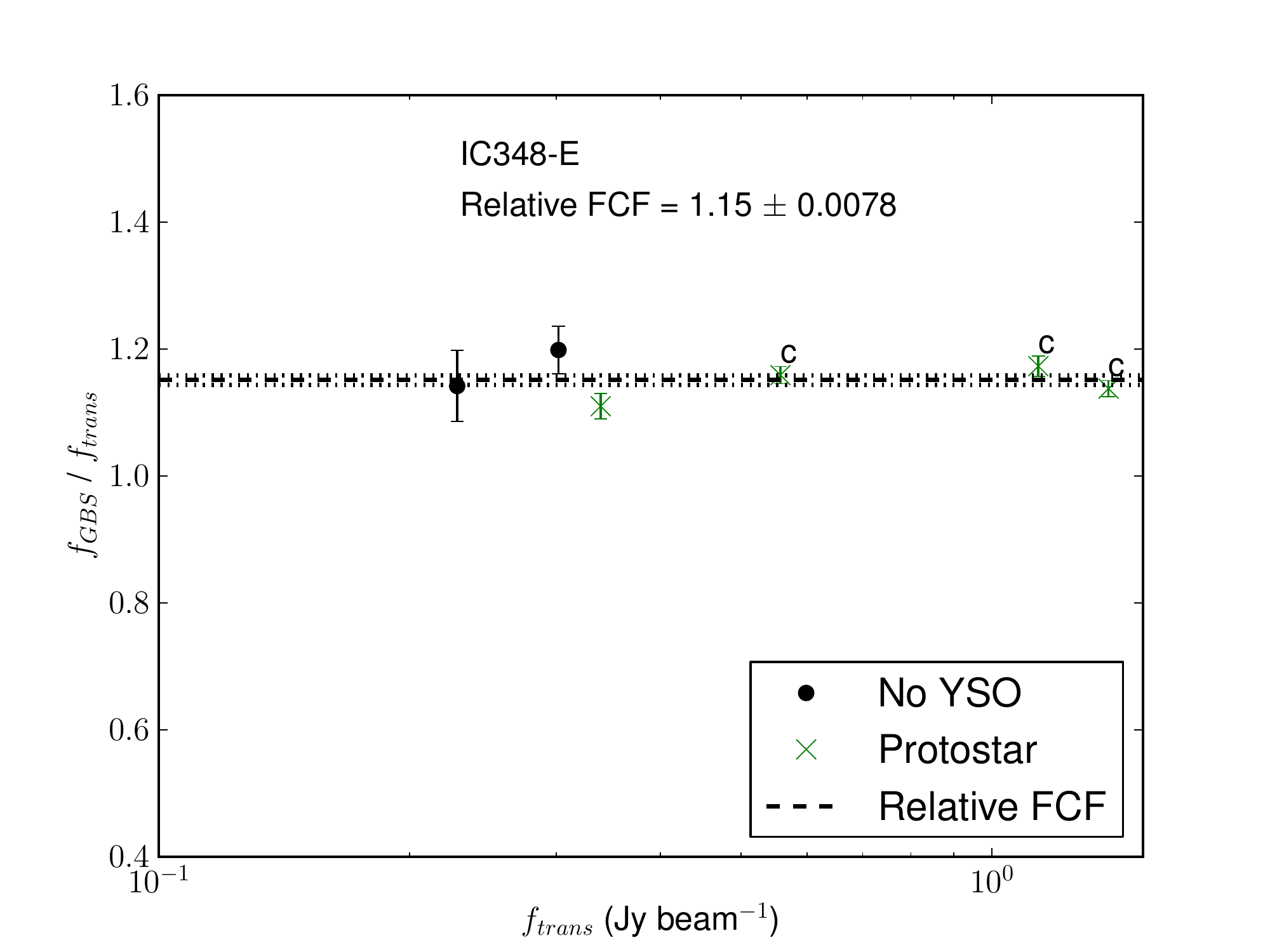}}
\subfloat{\label{}\includegraphics[width=9cm,height=7.8cm]{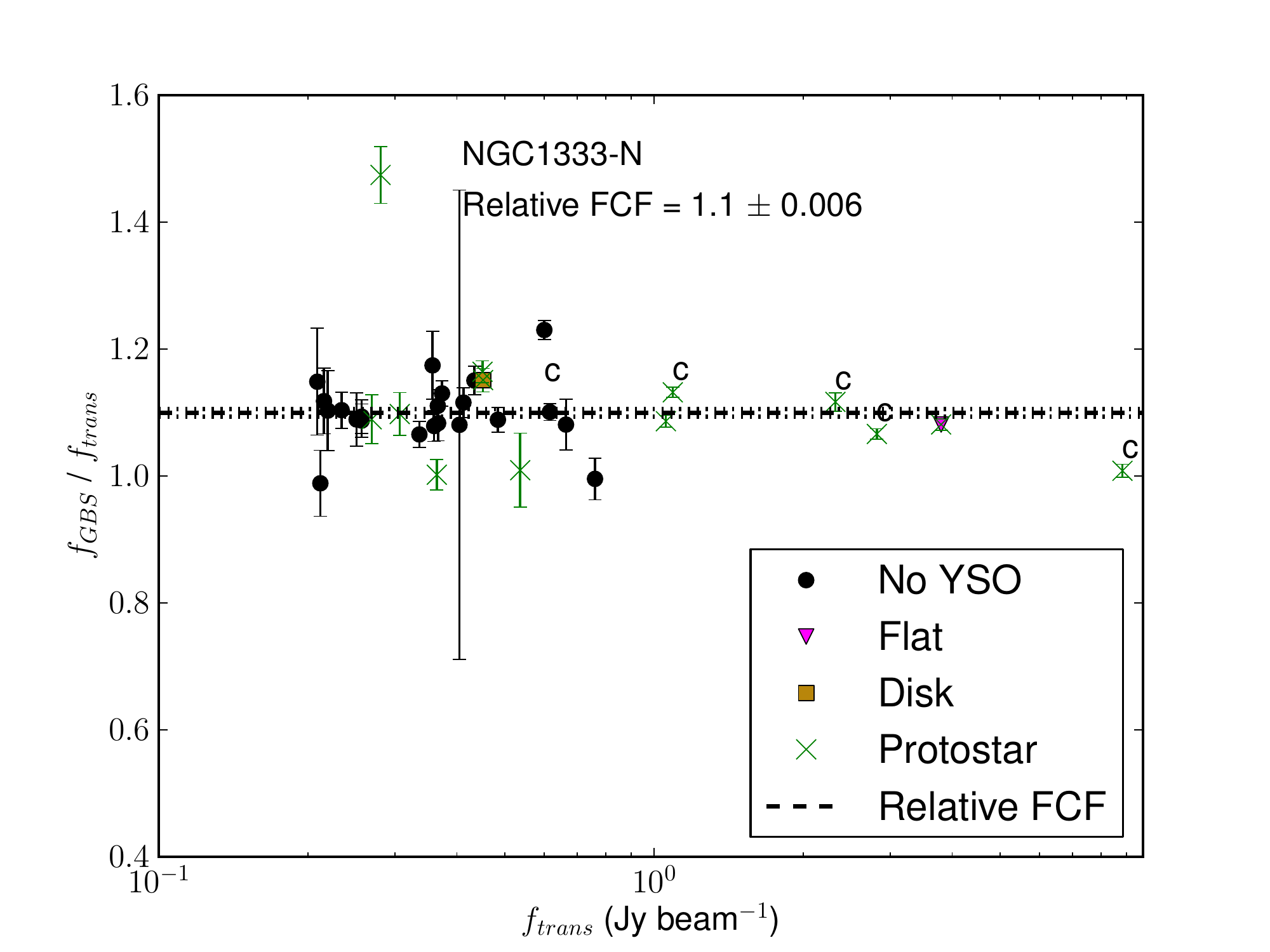}}\\
\subfloat{\label{}\includegraphics[width=9cm,height=7.8cm]{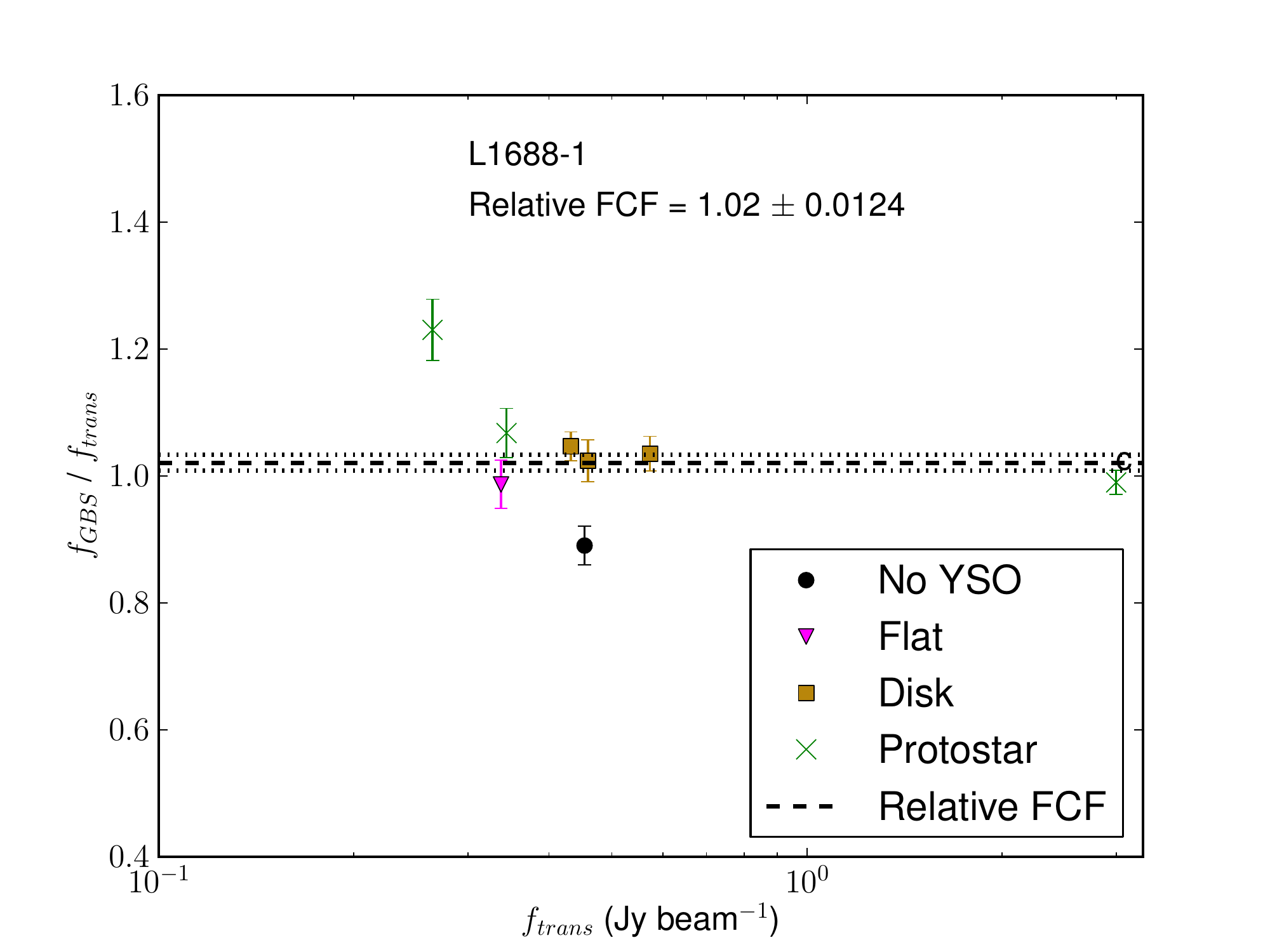}}
\subfloat{\label{}\includegraphics[width=9cm,height=7.8cm]{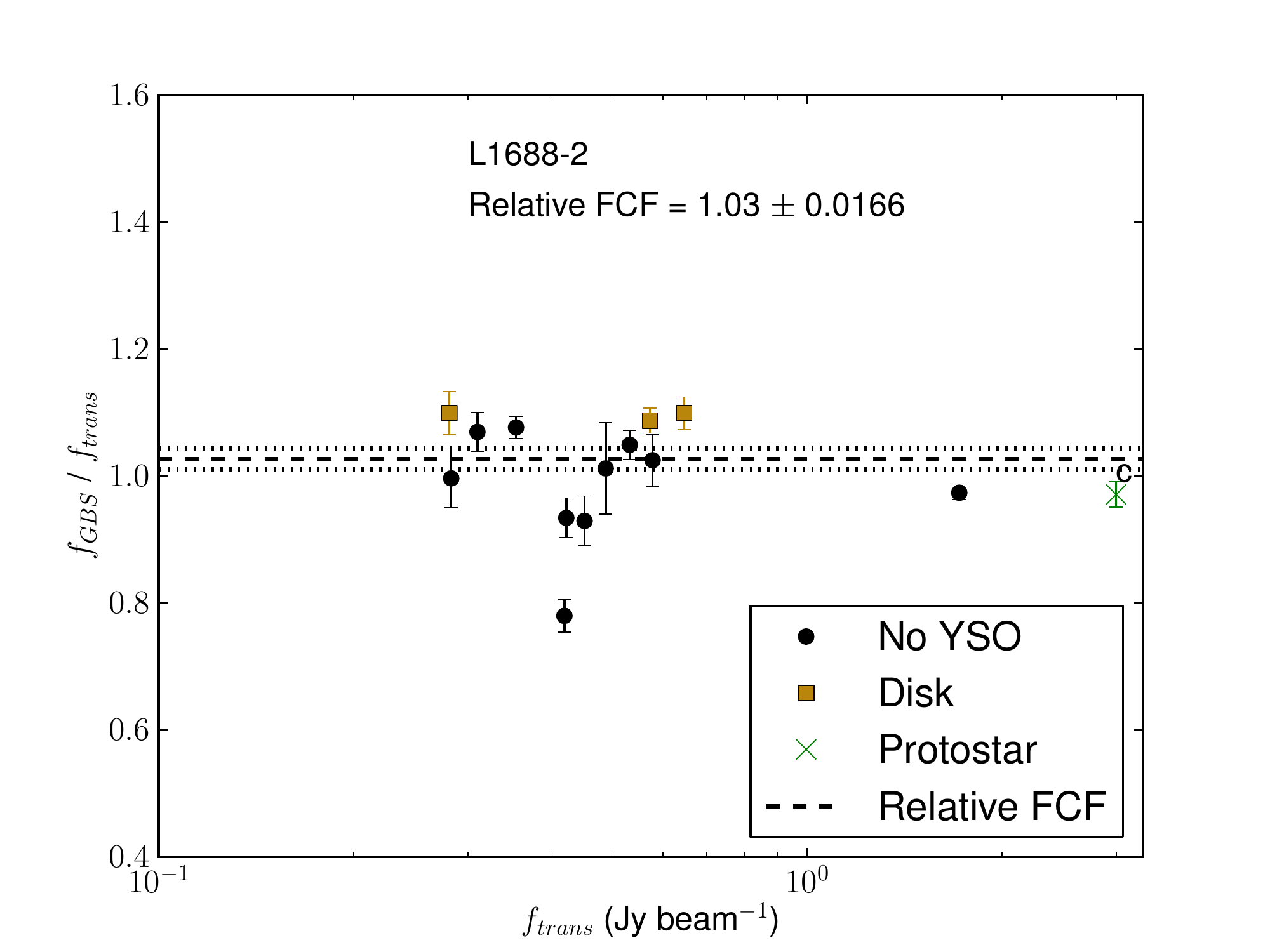}}
\caption{The mean GBS peak brightness divided by the mean Transient Survey peak brightness for all sources brighter than 200 mJy beam$^{-1}$ with radii less than $10\arcsec$ in the Perseus (top) and Ophiuchus (bottom) Molecular Cloud fields. The ratios are plotted against their mean peak brightnesses as measured across the Transient Survey. Points labeled with a ``c'' are chosen to be calibrators (\textit{Family members}) in both the GBS and Transient Survey data independently. Each point is coloured according to its association with YSOs (see text and legend). The error bars represent the combination of the uncertainty in the rescaled GBS peak 
brightness measurements and the uncertainty 
in the Transient Survey peak brightness measurements. The dashed line represents the derived relative flux calibration factor between the GBS data and the Transient Survey data (the number by which to divide to bring the GBS data into relative calibration with the Transient Survey data). The dotted lines represent the FCF uncertainty. }
\label{firstweightedmeanfig}
\end{figure*}

\begin{figure*} 	
\centering
\subfloat{\label{}\includegraphics[width=9cm,height=7.8cm]{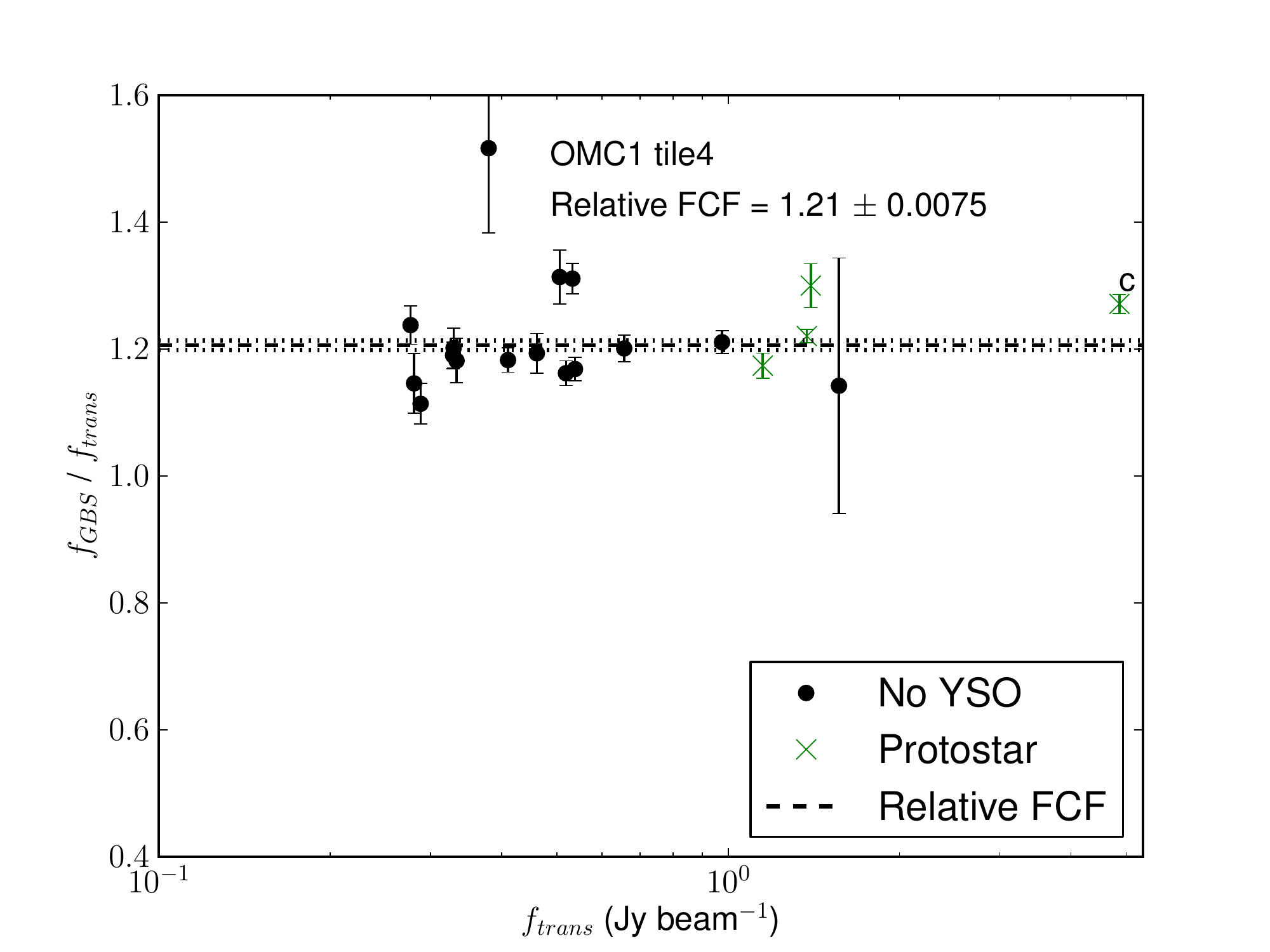}}
\subfloat{\label{}\includegraphics[width=9cm,height=7.8cm]{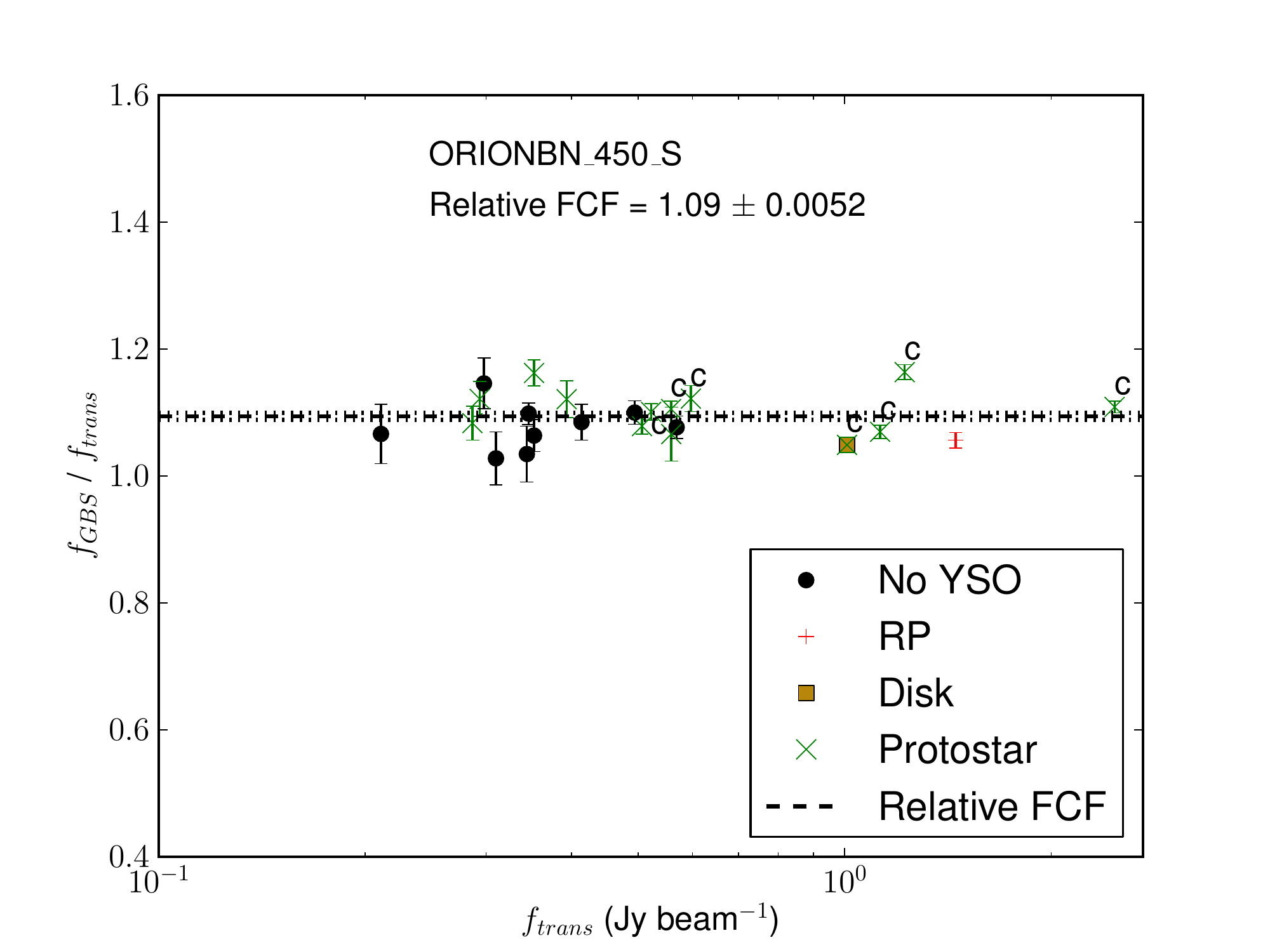}}\\
\subfloat{\label{}\includegraphics[width=9cm,height=7.8cm]{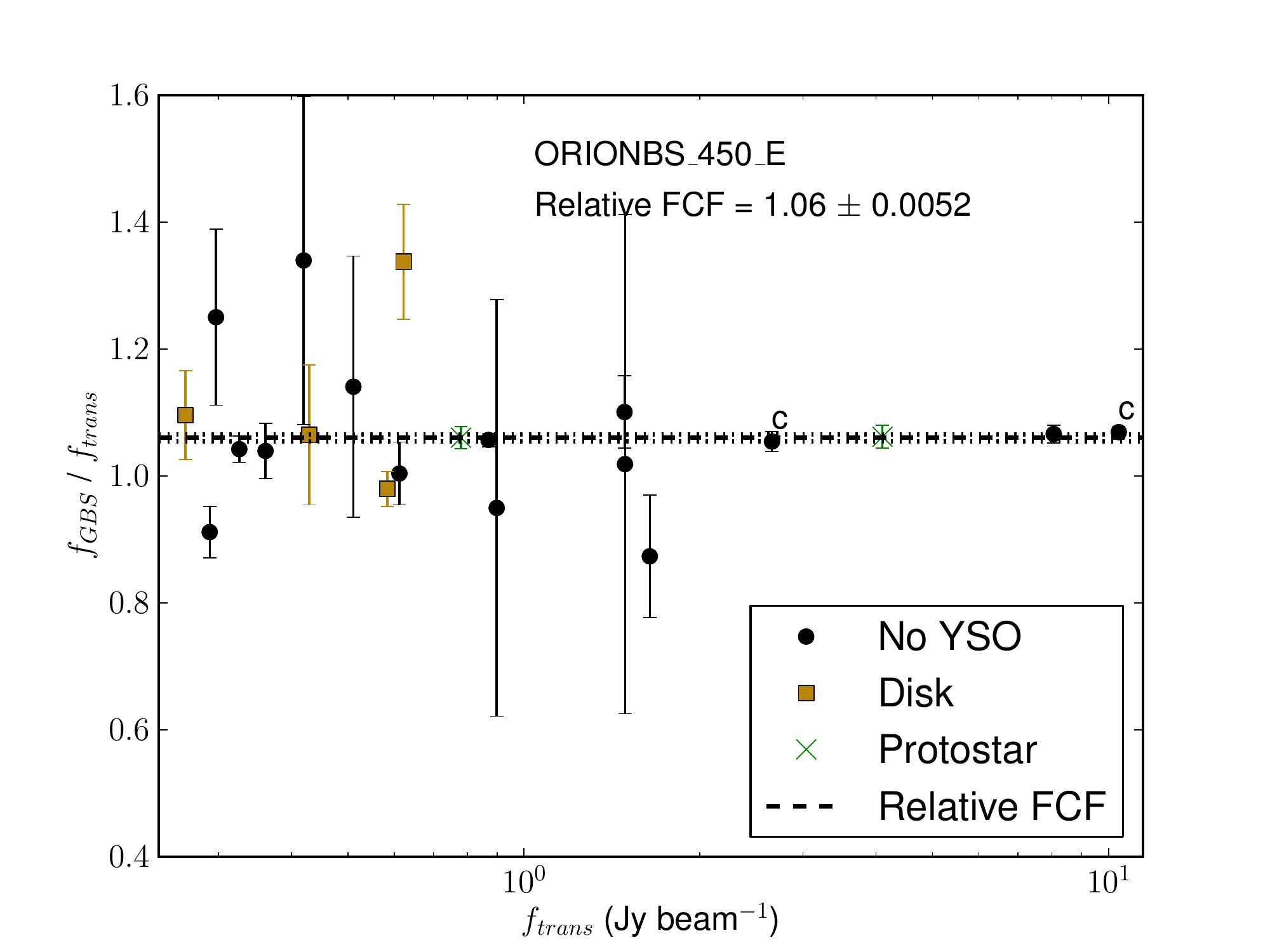}}
\subfloat{\label{}\includegraphics[width=9cm,height=7.8cm]{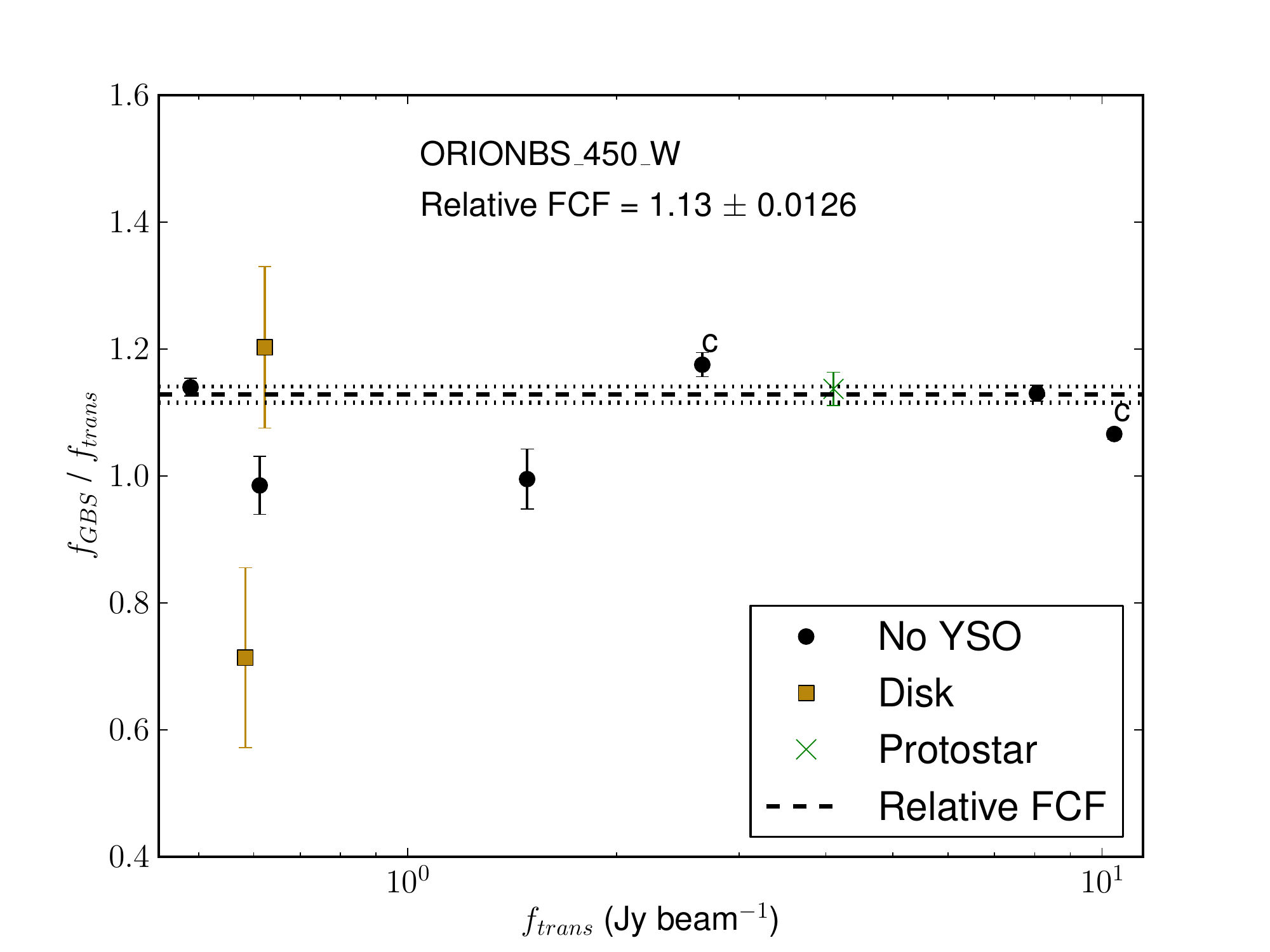}}
\caption{Same as Figure \ref{firstweightedmeanfig} for the Orion A and B Molecular Cloud fields.}
\label{secondweightedmeanfig}
\end{figure*}

\begin{figure*} 	
\centering
\subfloat{\label{}\includegraphics[width=9cm,height=7.8cm]{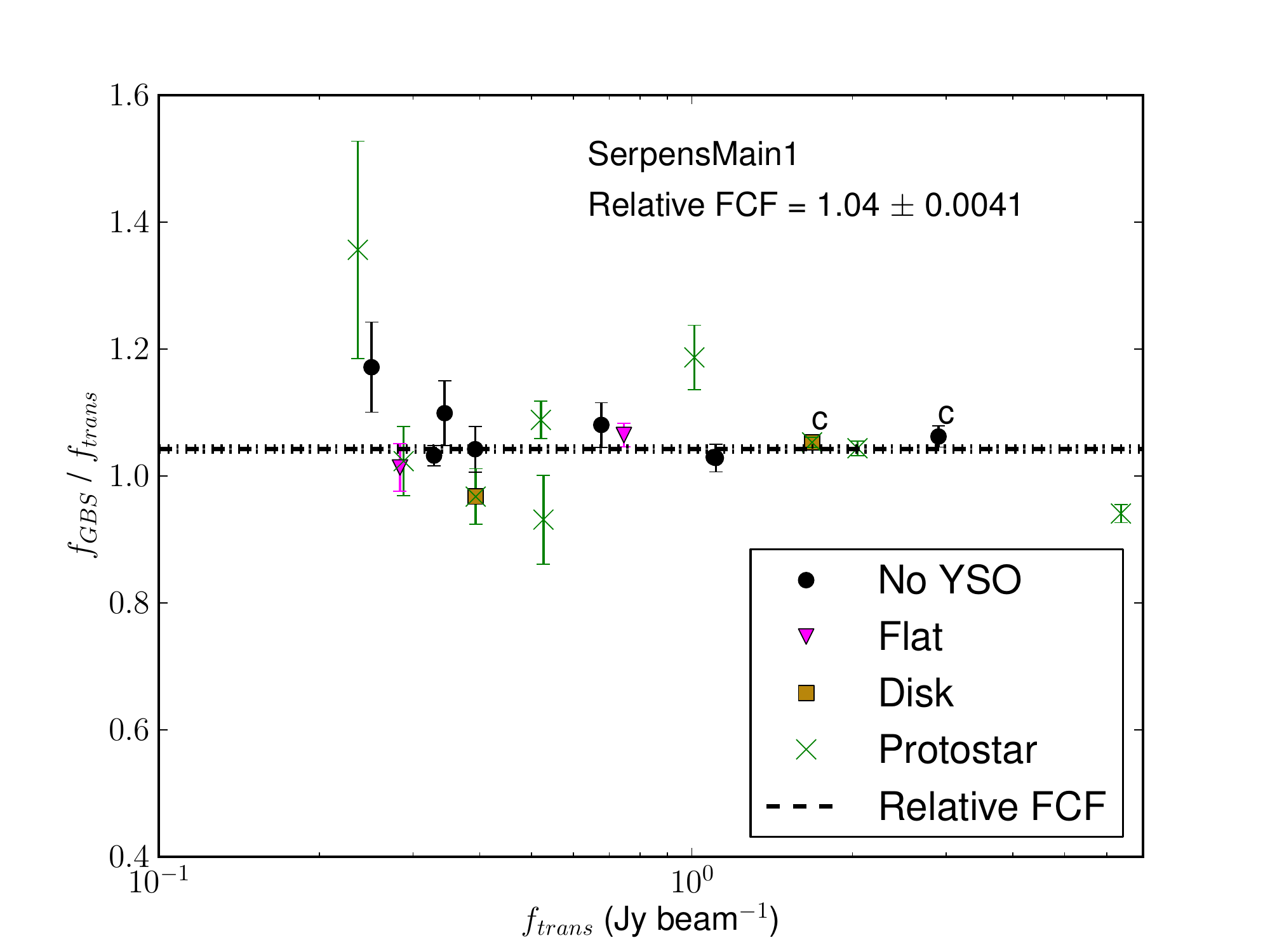}}
\subfloat{\label{}\includegraphics[width=9cm,height=7.8cm]{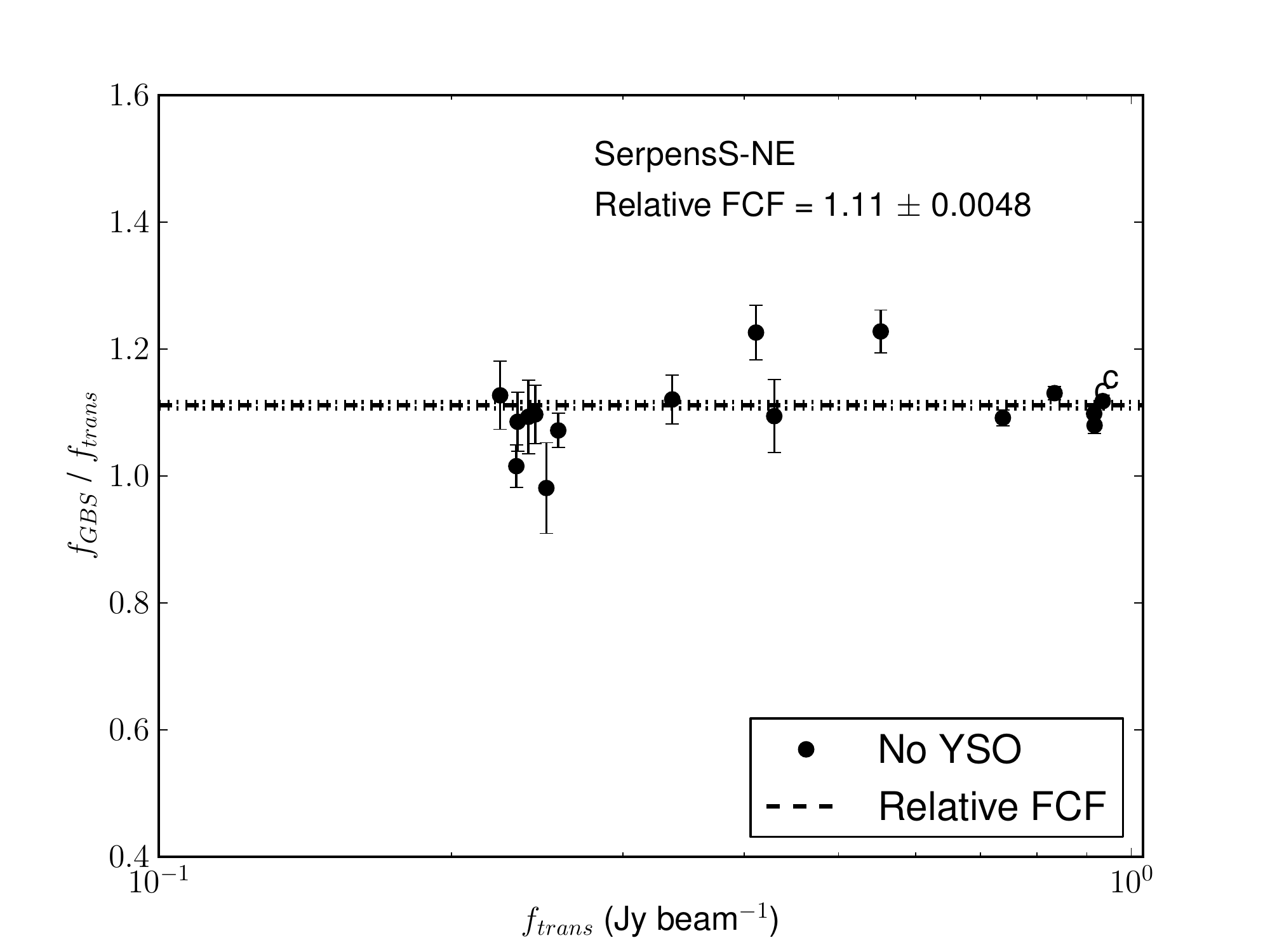}}\\
\subfloat{\label{}\includegraphics[width=9cm,height=7.8cm]{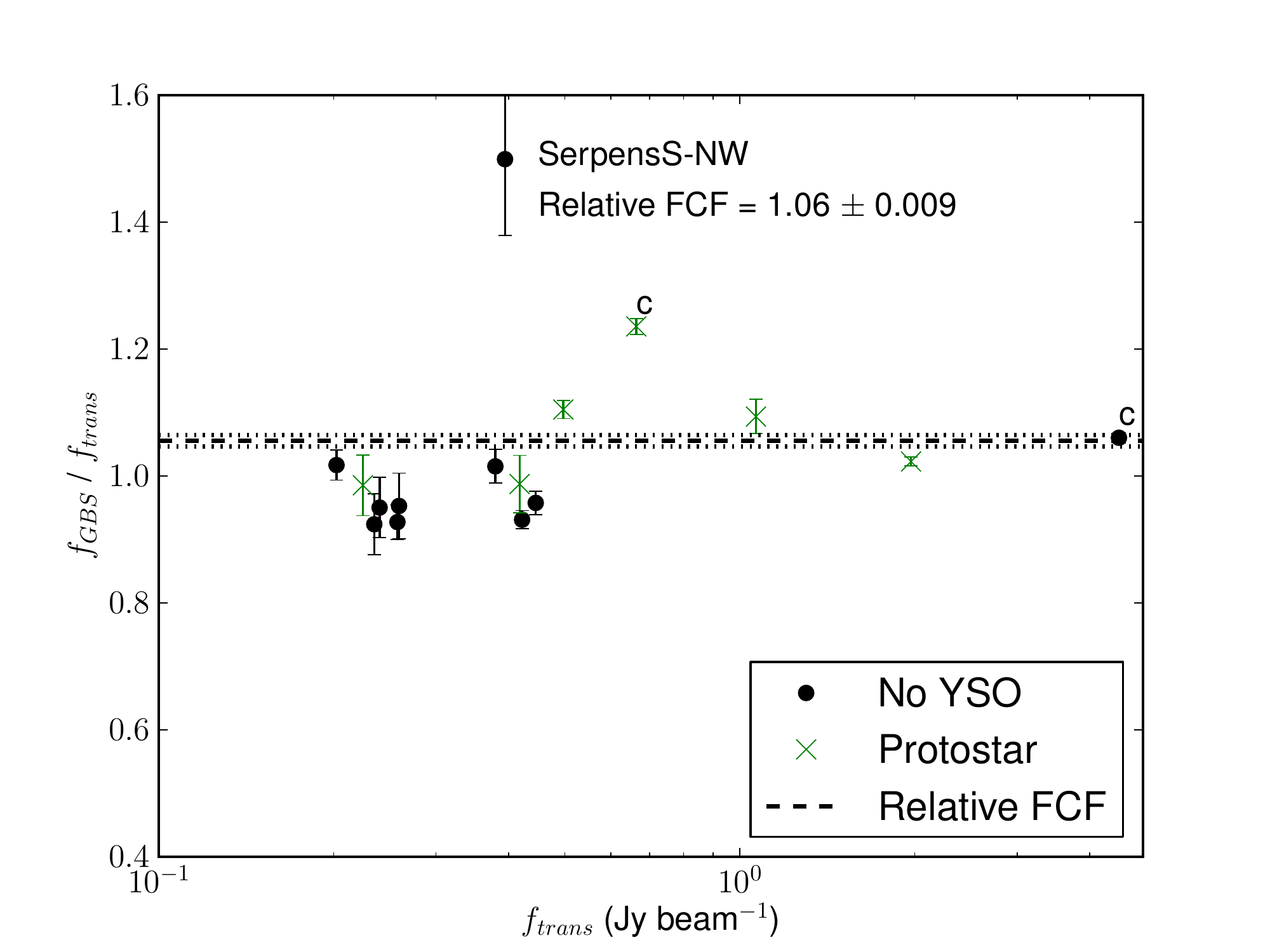}}
\caption{Same as Figure \ref{firstweightedmeanfig} for the Serpens Molecular Cloud fields.}
\label{lastweightedmeanfig}
\end{figure*}

\newpage

In order to bring the GBS data and the Transient Survey data into relative flux calibration with one another, we first identify sources common to both surveys. 
The sources we select 
have peak brightnesses 
larger than $200 \mathrm{\:mJy\:beam}^{-1}$ and effective radii less than $10\arcsec$. For each source, we measure the average peak brightness ($f_{\mathrm{GBS}}$ and $f_{\mathrm{trans}}$) and derive the associated uncertainties ($\sigma_{f_{\mathrm{GBS}}}$ and $\sigma_{f_{\mathrm{trans}}}$)
by calculating the measured standard deviation across all observations within each survey individually and 
dividing the result by the square root of the number of observations that were included in the calculation (see Table \ref{potvartable}). 

We calculate a calibration factor using the measured source brightnesses by dividing the $f_{\mathrm{GBS}}$ values by their corresponding $f_{\mathrm{trans}}$ values. For the $i^{th}$ source, we label this ratio $R_{i}$ and we propagate the uncertainties, $\sigma_{R,i}$, in order to calculate the weighted mean, $\bar{R}$,

 \begin{equation}
\label{weightedmeaneq}
\bar{R} = \frac{\sum_{i=1}^{n}R_{i}w_{i}}{\Sigma_{i=1}^{n}w_{i}},
\end{equation}

\noindent where $n$ is the number of sources and the weights are the inverse square of the propagated uncertainties,

\begin{equation}
w_{i} = \sigma_{R,i}^{-2}.
\end{equation}

The uncertainty in the weighted mean, $\sigma_{\bar{R}}$, is given by

\begin{equation}
\label{uncweightedmeaneq}
\sigma_{\bar{R}} = \sqrt{\frac{1}{\Sigma_{i=1}^{n}w_{i}}}.
\end{equation}

\noindent Next, we calculate the difference, $\delta$, between each peak brightness ratio and the weighted mean by subtracting the latter from the former and dividing by the uncertainty associated with that source added in quadrature to the uncertainty in the weighted mean, 

\begin{equation}
\sigma_{\mathrm{tot},i} =  \sqrt{\sigma_{R,i}^{2}+\sigma_{\bar{R}}^{2}}
\end{equation}

\begin{equation}
\label{outliereq}
\delta_{i} = \frac{R_{i}-\bar{R}}{\sigma_{\mathrm{tot},i}}. 
\end{equation}

\noindent We consider all sources with $\abs{\delta}>3$ to be deviant outliers that might skew the weighted mean. Therefore, we remove these sources from the calculation and recompute the weighted mean. This refined weighted mean (that  excludes outliers) represents the initial approximation of the relative flux calibration factor between overlapping average GBS and Transient Survey images. 

To explore how the peak brightness uncertainties in each data set ($\sigma_{f_{\mathrm{GBS}}}$ and $\sigma_{f_{\mathrm{trans}}}$) affect the refined weighted mean, we use a Monte Carlo analysis. We fix $\sigma_{f_{\mathrm{GBS}}}$ and $\sigma_{f_{\mathrm{trans}}}$ for each source and randomly draw new peak brightness measurements from normal distributions with mean values equal to $f_{\mathrm{GBS}}$ and 
$f_{\mathrm{trans}}$ and standard deviations equal to $\sigma_{f_{{\mathrm{GBS}}}}$ and $\sigma_{f_{\mathrm{trans}}}$. In this way, we calculate a new peak brightness ratio for each source, $R_{i}$, that is within the derived measurement uncertainties. Then, we calculate a new weighted mean (Equation \ref{weightedmeaneq}) based on these values, discard outliers in the same manner as before, and compute a refined weighted mean. We repeat this process 10,000 times.

Of these 10,000 refined weighted mean values, we adopt the average as the relative flux calibration factor, or, \textit{relative FCF}. This is the number by which we divide the GBS image to bring it into relative calibration with the Transient Survey image. The standard deviation in refined weighted mean values is the uncertainty in the relative FCF, $\sigma_{\mathrm{FCF}}$. 

We plot the results in Figures \ref{firstweightedmeanfig} through \ref{lastweightedmeanfig}. The colours represent a source's association with known YSOs. In these figures, to be associated with a protostar, flat spectrum source, protostellar candidate, or disk, the peak position of the source must be within 10$\arcsec$ of the YSO location (to match the radius of the largest compact source we consider). Points labelled with a ``c'' are \textit{Family Members} in both the GBS and Transient Surveys, i.e. they appear to have stable peak brightnesses in each dataset independently but not necessarily when compared over several year timescales. 
The FCF values are indicated by the dashed lines and $\sigma_{\mathrm{FCF}}$ is indicated by the dotted lines. As the noise is relatively constant across a map and large scale modes have been mostly removed in the data reduction procedure, we expect the relative FCF value to remain constant across the full field of any single image.

The GBS images were all originally calibrated at a level that is slightly brighter than their respective Transient Survey images obtained prior to March 1$^{st}$, 2017. This is the result of several factors that affect the nominal flux calibration performed at the JCMT, before we apply the relative flux calibration presented above. For instance, the nominal flux calibration values are based on data obtained between 2011 and 2012 \citep{dempsey2013}, the water vapour monitor was replaced in 2015, and SCUBA-2 had a filter upgrade in late 2016. The combination of these factors have led to differences in the original calibration of the data obtained in the GBS and the Transient Survey eras. These effects are corrected for by our relative flux calibration. 

\section{Results}
\label{GBStransientresultssec}

\begin{figure*} 	
\centering
\subfloat{\label{}\includegraphics[width=9cm,height=7.8cm]{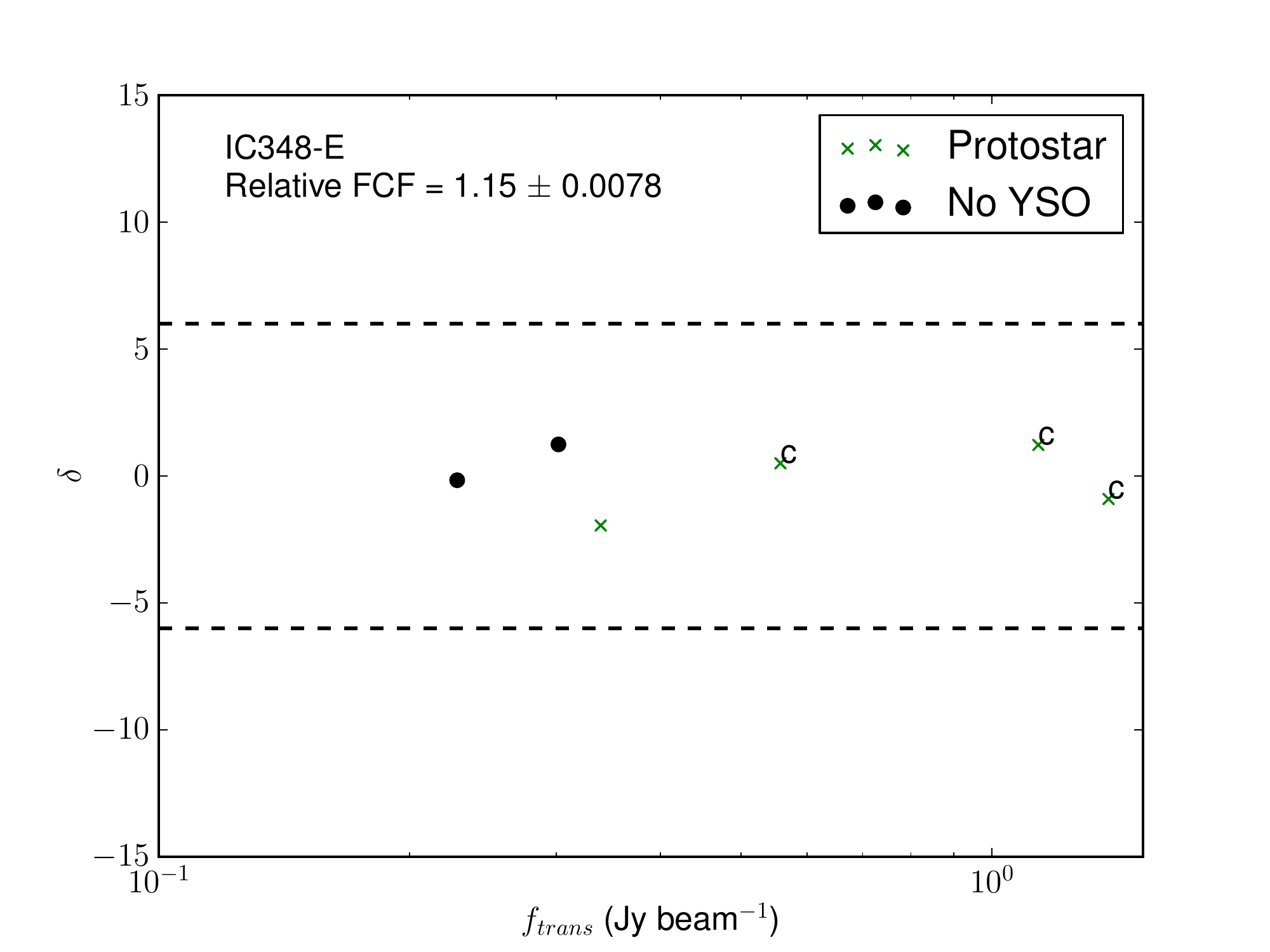}}
\subfloat{\label{}\includegraphics[width=9cm,height=7.8cm]{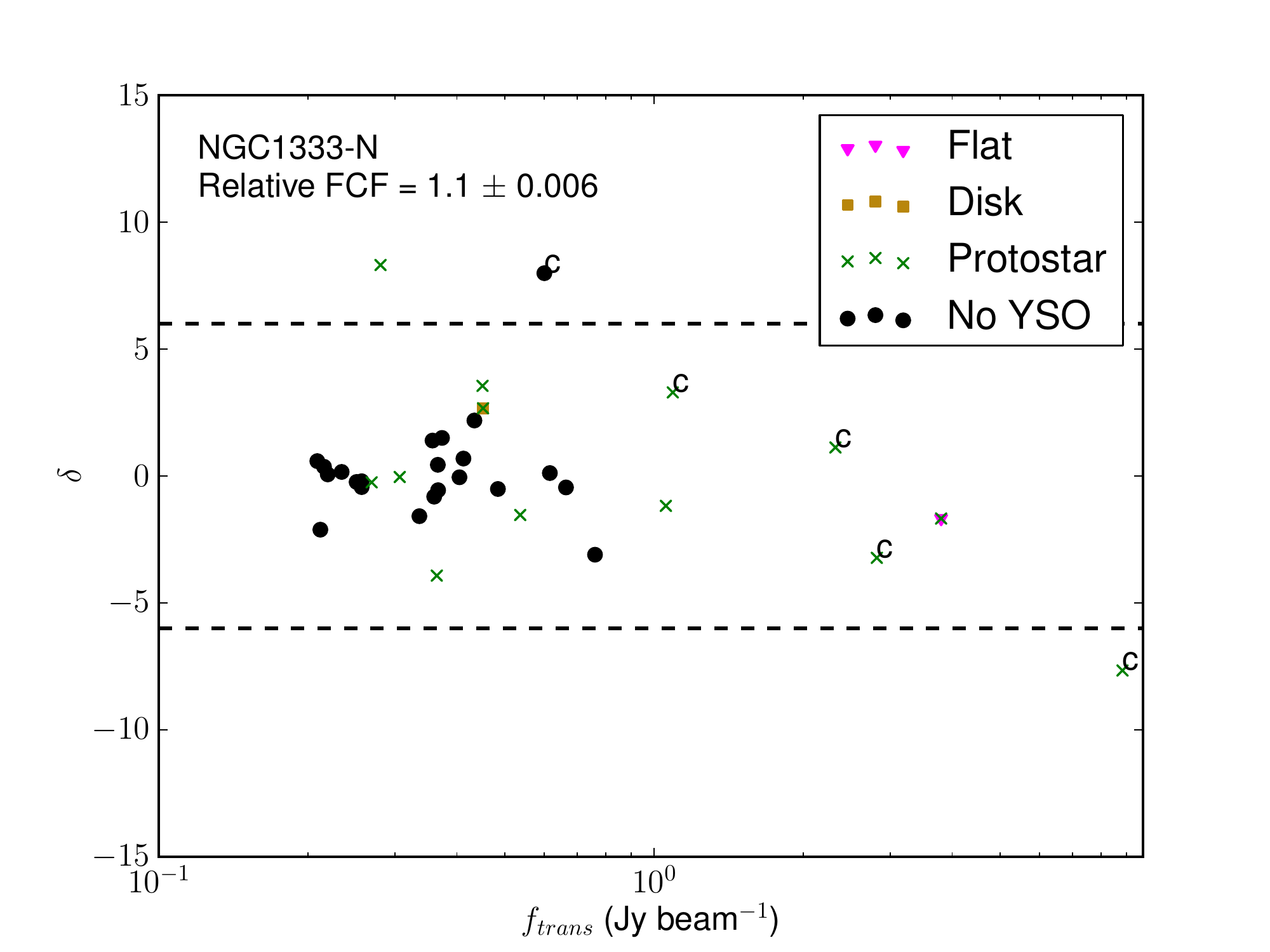}}\\
\subfloat{\label{}\includegraphics[width=9cm,height=7.8cm]{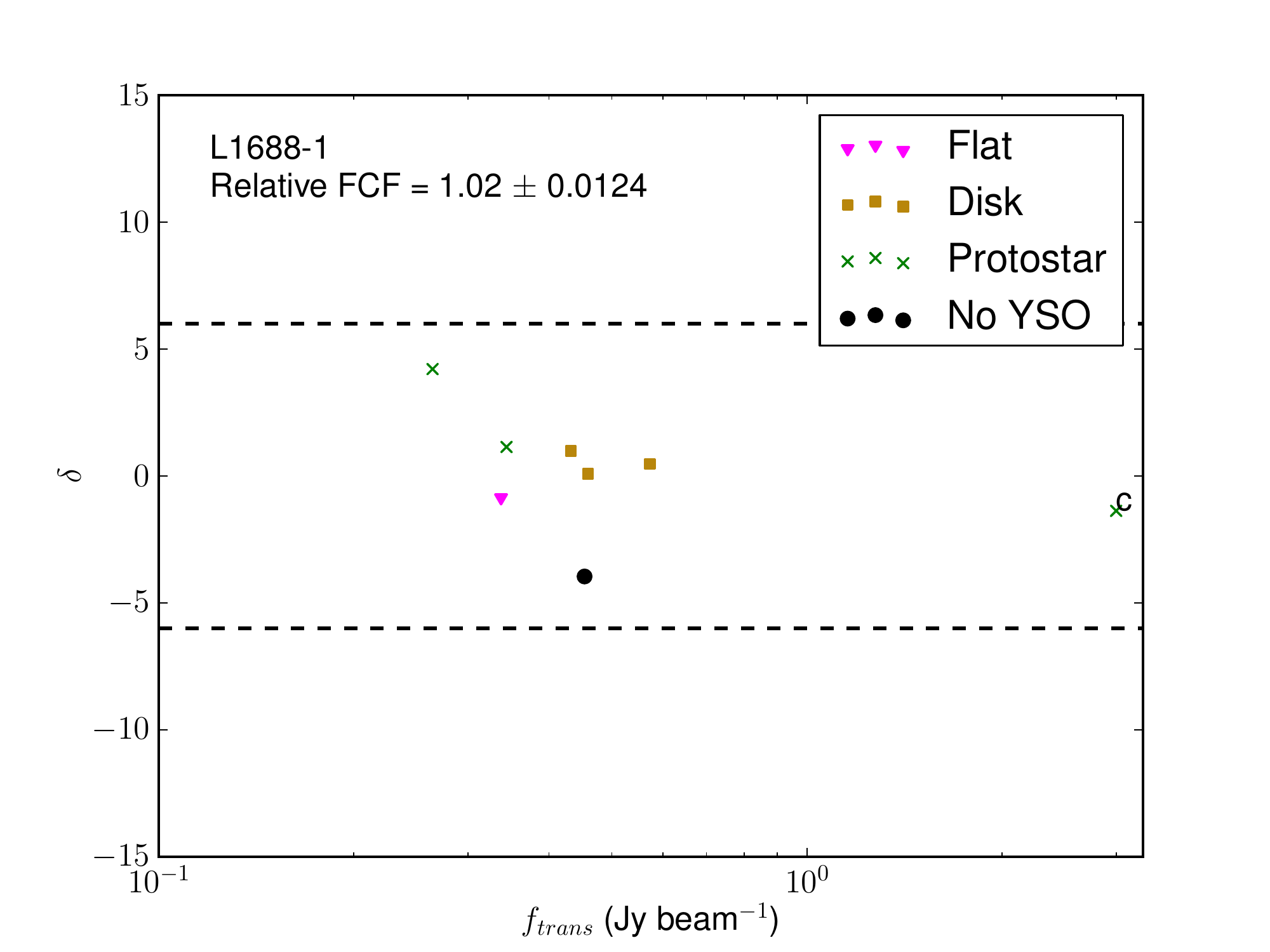}}
\subfloat{\label{}\includegraphics[width=9cm,height=7.8cm]{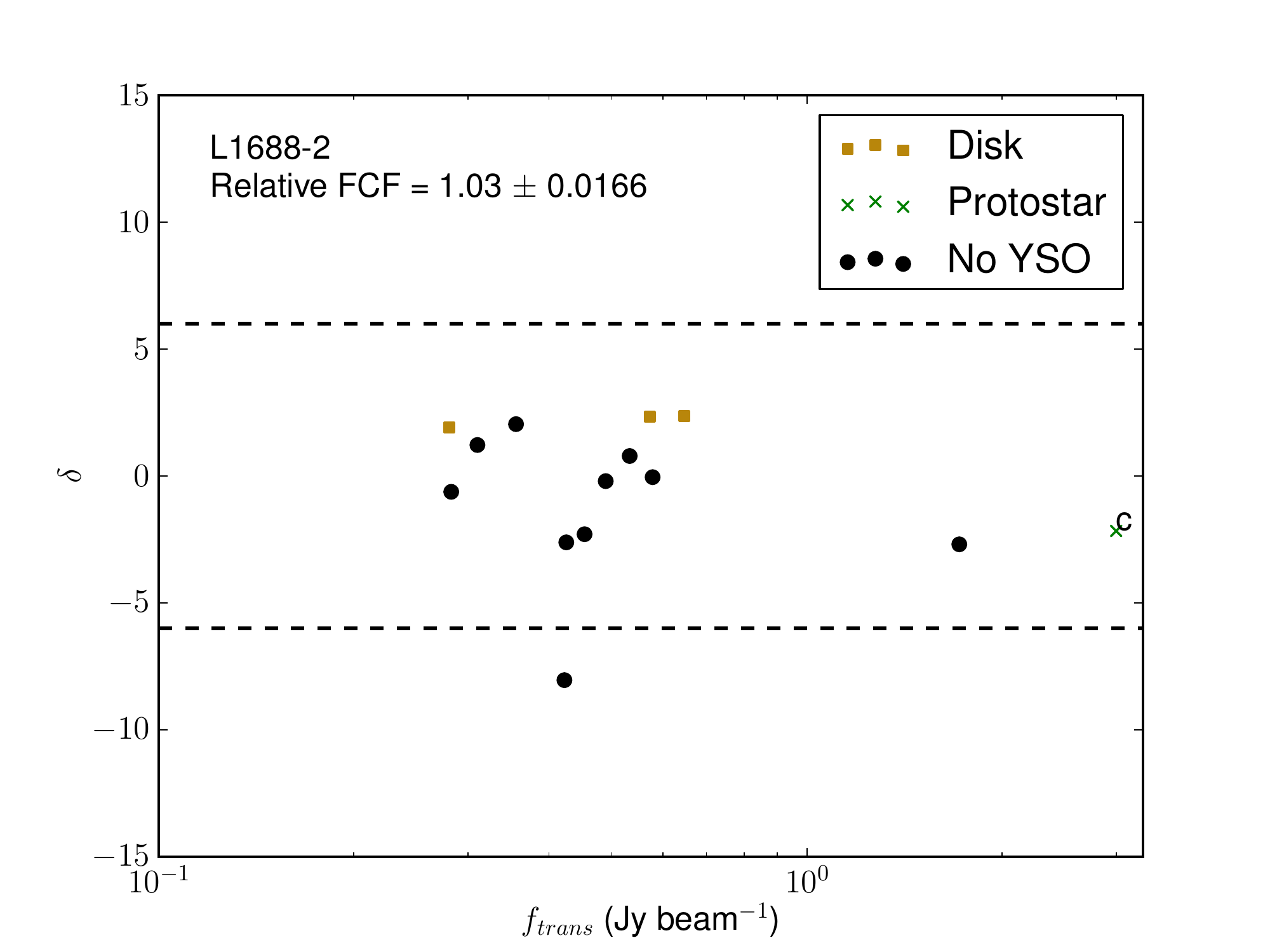}}
\caption{The deviation from the FCF for all sources brighter than 200 mJy beam$^{-1}$ with radii $<10\arcsec$ in the Perseus (top) and Ophiuchus (bottom) Molecular Cloud fields. The ratios are plotted against their mean peak brightness as measured across the Transient Survey. Points labeled with a ``c'' are chosen to be calibrators (\textit{Family members}) in both the GBS and Transient Survey data independently. Each point is coloured according to its association with YSOs (see text and legend). Dashed lines are drawn at $\pm4$ to highlight sources defined to be significant outliers.}
\label{firstsigmafig}
\end{figure*}

\begin{figure*} 	
\centering
\subfloat{\label{}\includegraphics[width=9cm,height=7.8cm]{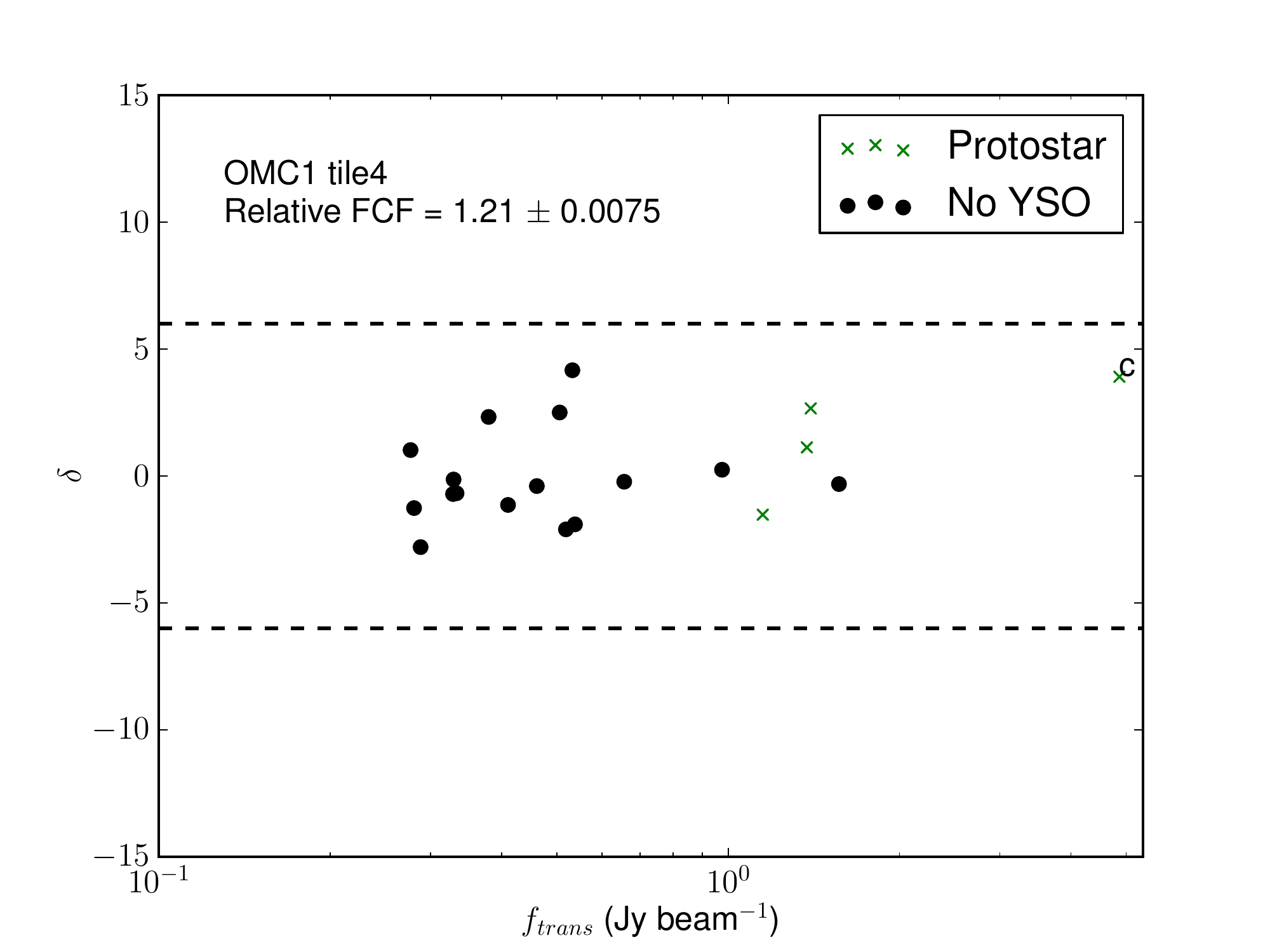}}
\subfloat{\label{}\includegraphics[width=9cm,height=7.8cm]{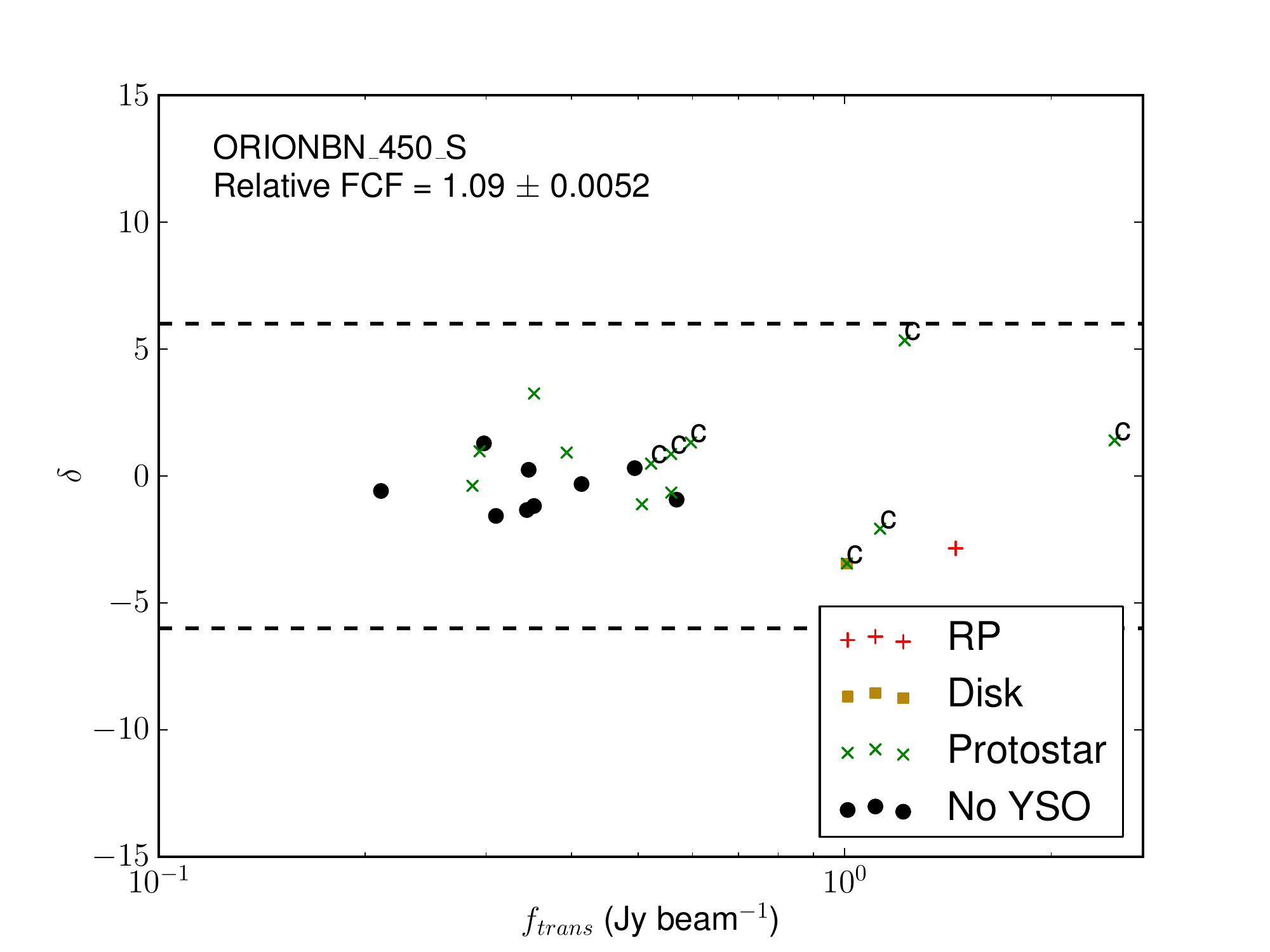}}\\
\subfloat{\label{}\includegraphics[width=9cm,height=7.8cm]{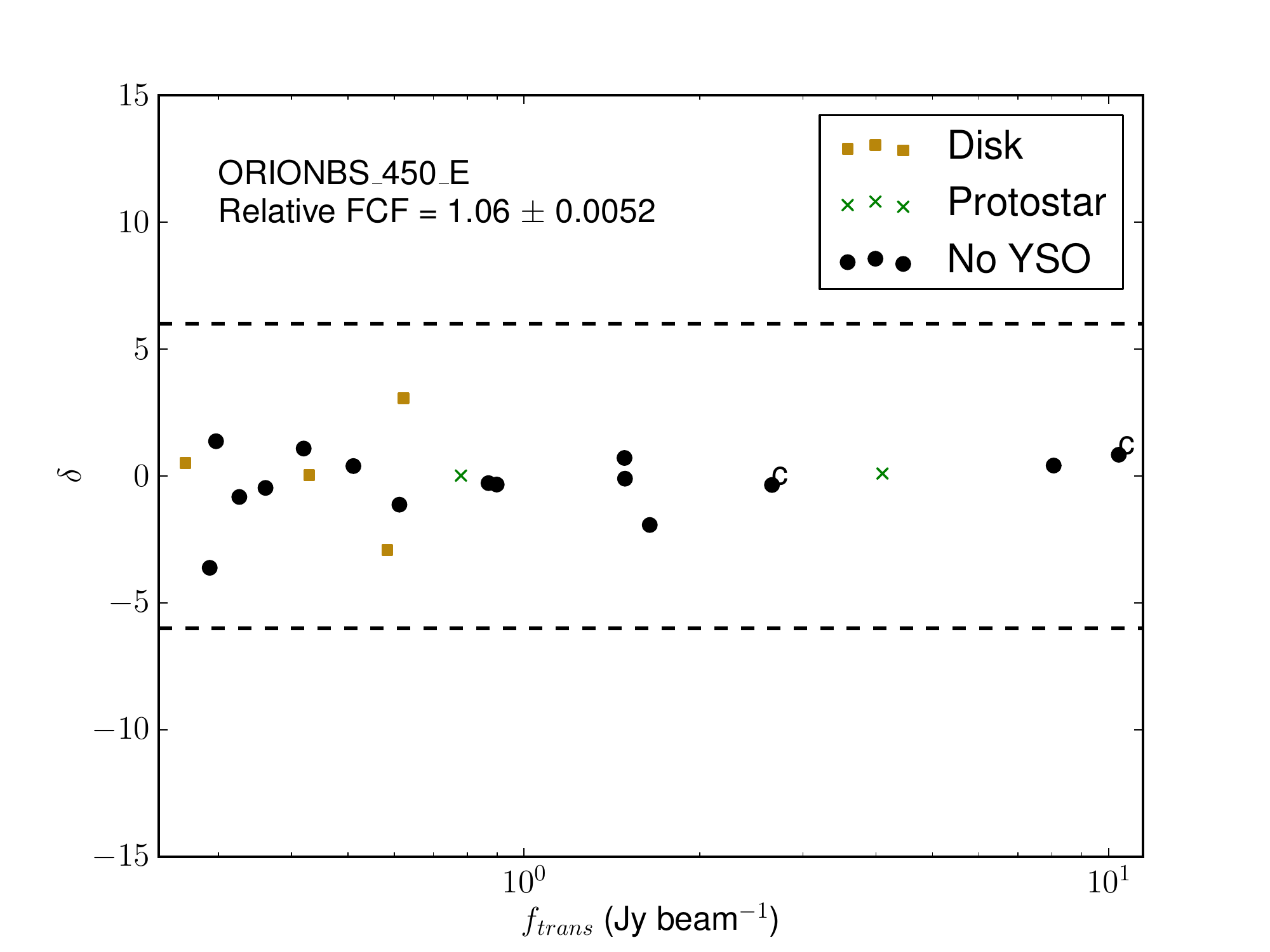}}
\subfloat{\label{}\includegraphics[width=9cm,height=7.8cm]{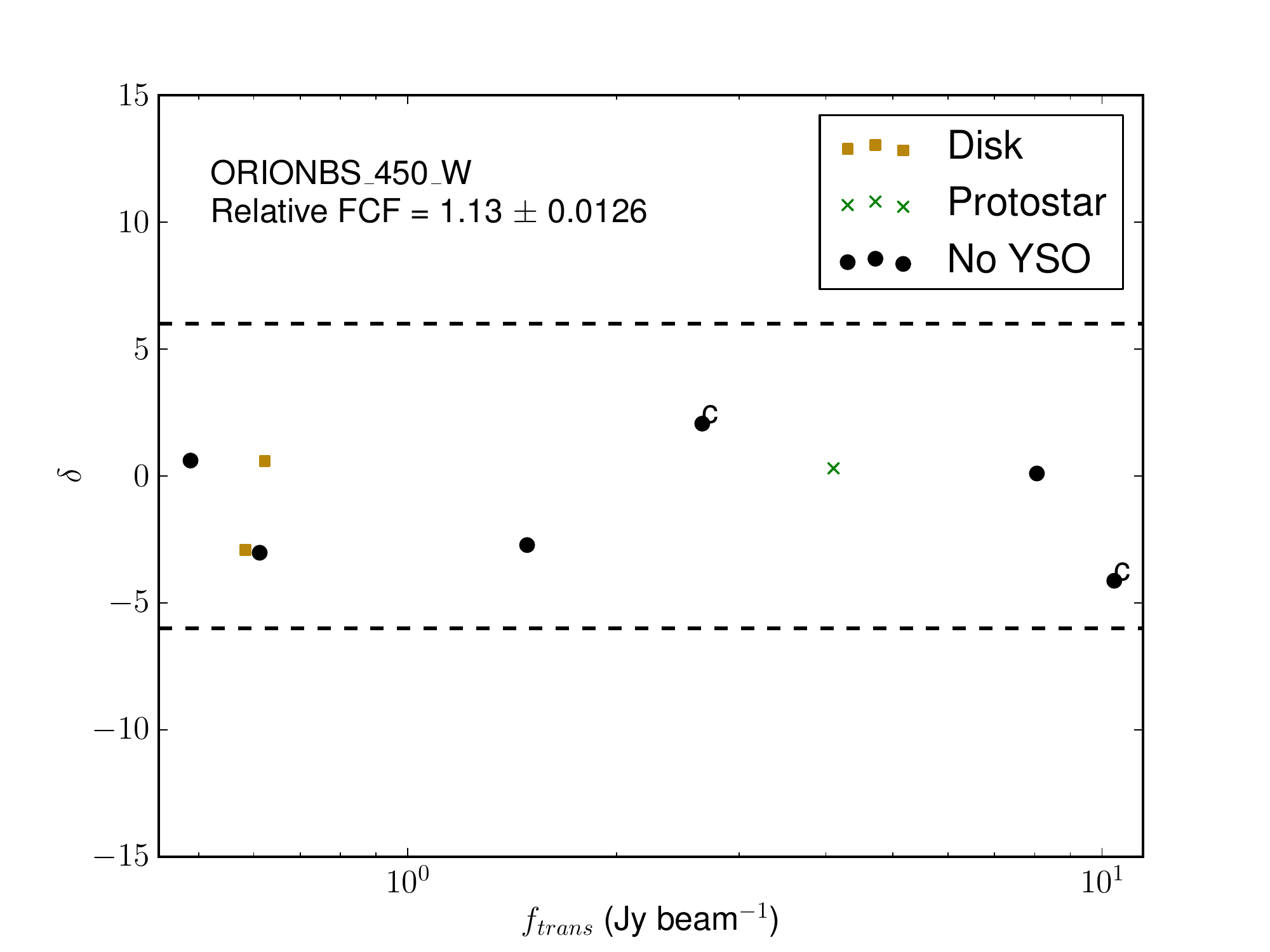}}
\caption{Same as Figure \ref{firstsigmafig} for the Orion A and B Molecular Cloud fields.}
\label{secondsigmafig}
\end{figure*}

\begin{figure*} 	
\centering
\subfloat{\label{}\includegraphics[width=9cm,height=7.8cm]{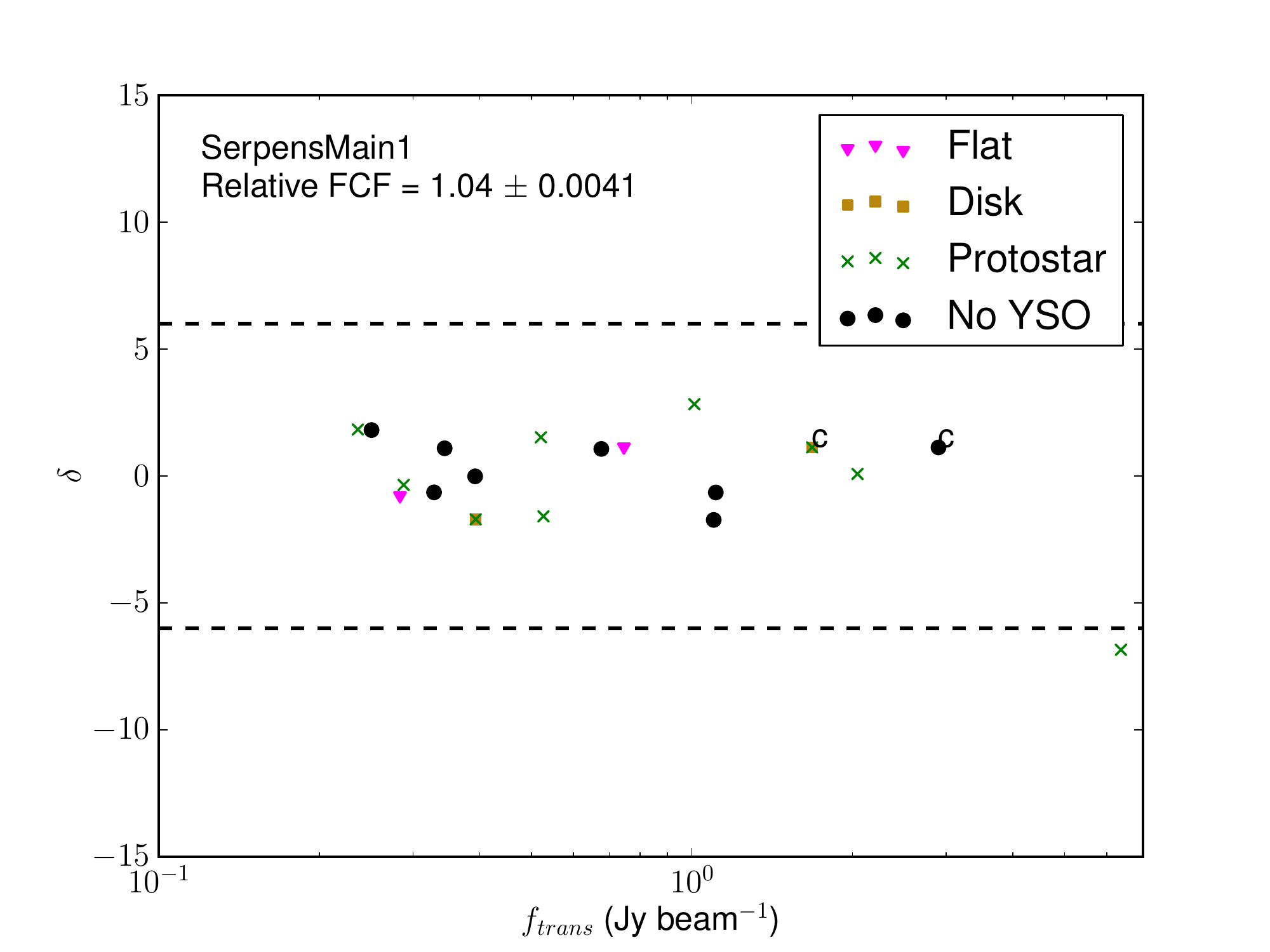}}
\subfloat{\label{}\includegraphics[width=9cm,height=7.8cm]{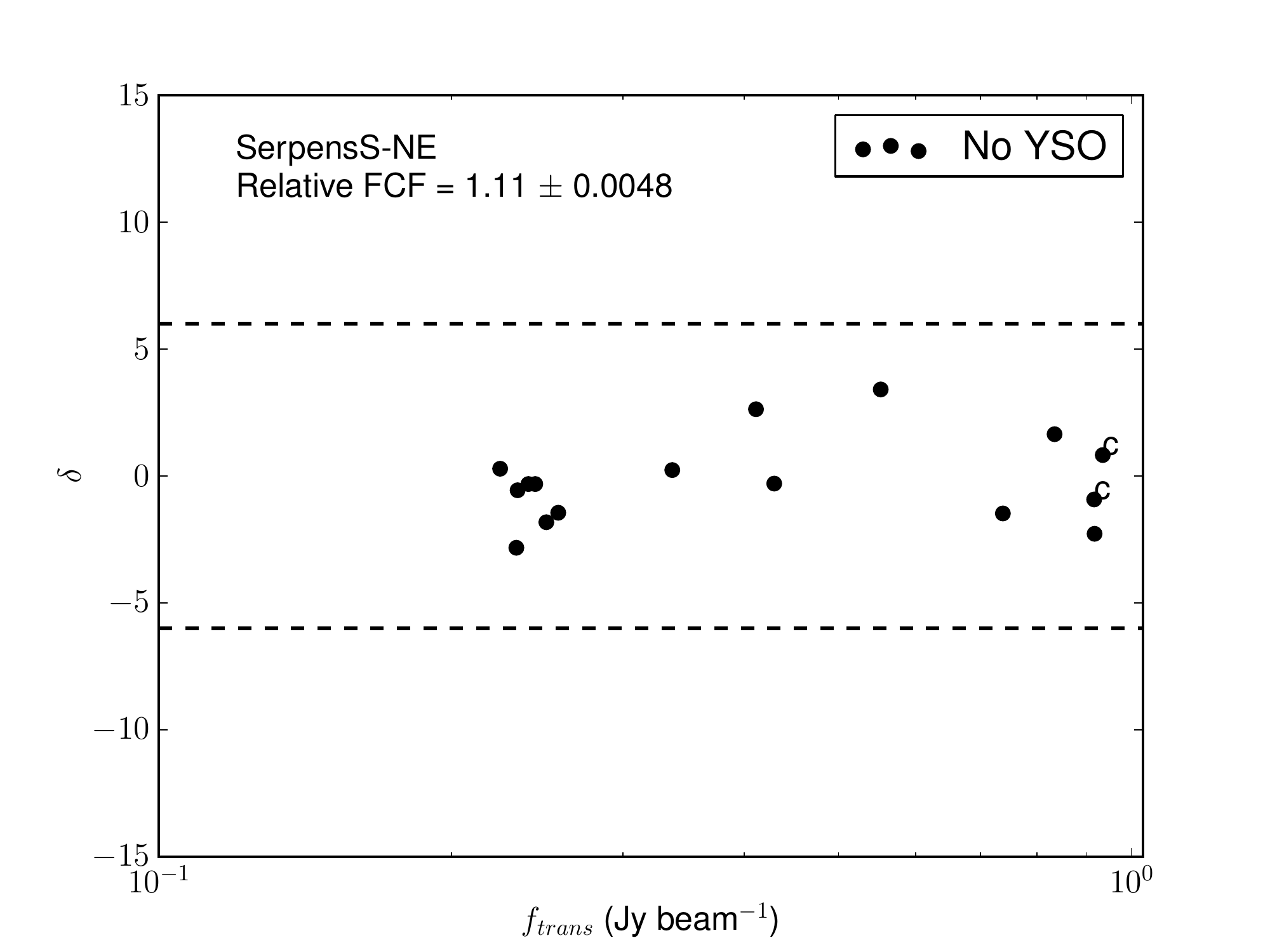}}\\
\subfloat{\label{}\includegraphics[width=9cm,height=7.8cm]{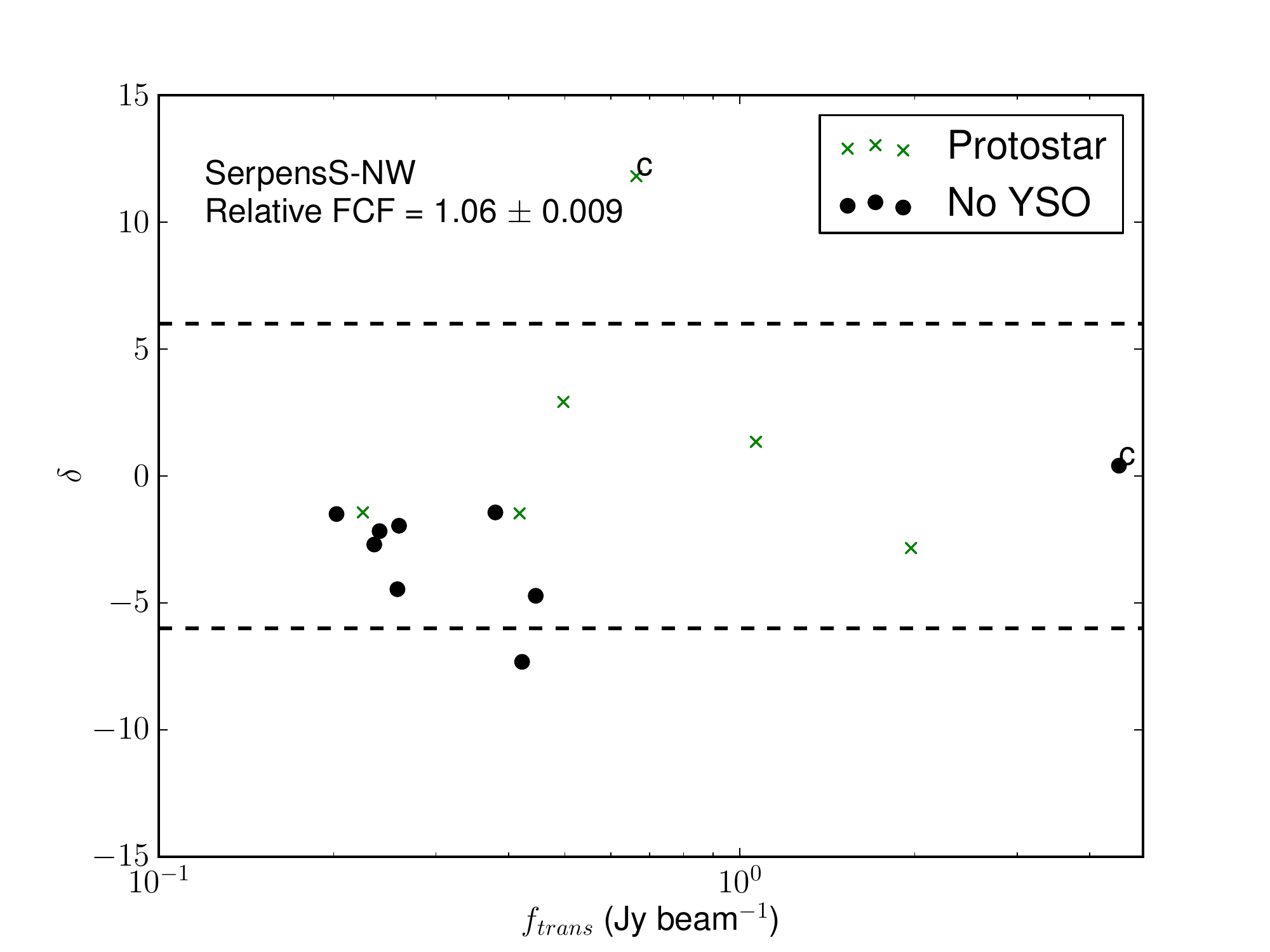}}
\caption{Same as Figure \ref{firstsigmafig} for the Serpens Molecular Cloud fields.}
\label{lastsigmafig}
\end{figure*}

\begin{figure*} 	
\centering
\includegraphics[width=14cm,height=11cm]{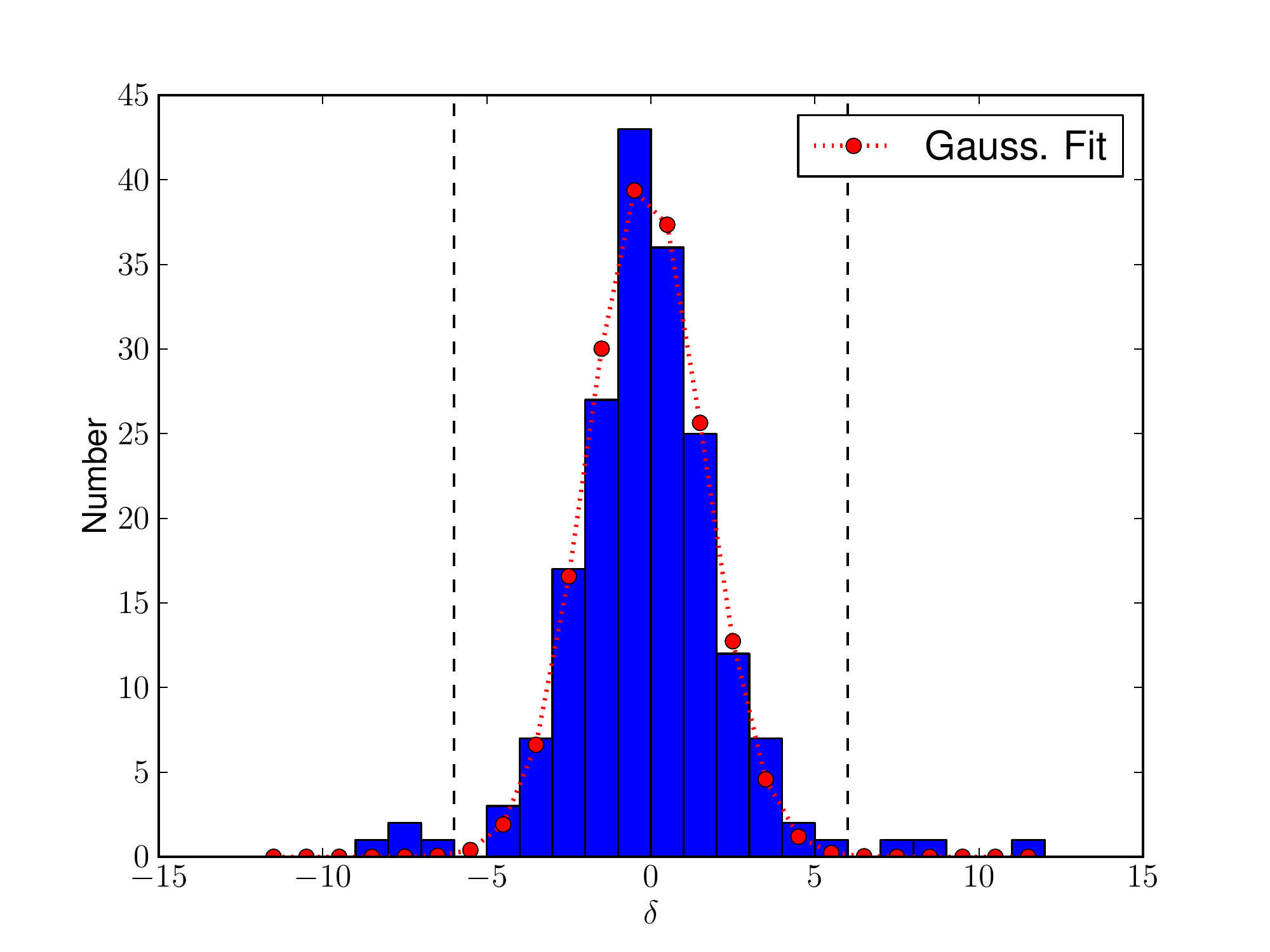}
\caption{
The distribution of $\delta$ values for all sources. The red points represent a Gaussian fit to the histogram. The vertical dashed lines indicate the threshold for a significant detection of a variable candidate.}
\label{deltadist}
\end{figure*}

 
\begin{sidewaystable}
\centering
\caption{Summary of the variable candidate source properties.}
\label{potvartable}
\begin{tabular}{|c|c|c|c|c|c|c|c|c|c|c|c|c|}
\hline
GBS Field & ID & Other Name\footnote{Reference name from literature. YSO name where possible.}  & R.A. (J2000)  & Dec (J2000) & $f_{\mathrm{trans}}$\footnote{$f_{\mathrm{x}}$ is the mean source peak brightness measured across the Transient Survey ($x=trans$) or GBS ($x=GBS$) in Jy beam$^{-1}$.} & $\frac{\sigma_{f_{\mathrm{trans}}}}{f_{\mathrm{trans}}}$\footnote{The standard deviation in $f$ divided by the square root of the number of observations, normalised by $f$. In units of \%.} & $f_{\mathrm{GBS}}$$^{\mathrm{b}}$ & $\frac{\sigma_{f_{\mathrm{GBS}}}}{f_{\mathrm{GBS}}}$$^{\mathrm{c}}$ & $\abs{\delta}$ & $\frac{\dot{f}}{f_{\mathrm{trans}}}$ (\% yr$^{-1}$) & $\sigma_{\dot{f}/f}$ & Category\\
\hline\hline 
NGC1333-N & 			PER-1   & IRAS4A\footnote{\cite{jennings1987}}		 & 3:29:10.42 & 	31:13:30.63 	& 	8.83 & 	1.0 	& 8.10 	& 0.4 & 7.66 	& 2.09 	& 0.28 & S\\
NGC1333-N & 			PER-10 & Bolo 40\footnote{\cite{enoch2006}}			 & 3:28:59.86 & 	31:21:33.09 	& 	0.60 & 	1.0 	& 0.67 	& 0.7 & 7.99 	& -3.03 	& 0.36 & S\\
NGC1333-N & 			PER-34 & [LAL96] 213\footnote{\cite{lada1996}}		 & 3:29:07.66 & 	31:21:54.05 	& 	0.28 & 	2.8 	& 0.38 	& 1.2 & 8.31 	& -8.64 	& 0.87 & S\\ 
SerpensMain1 & 		SER-1   & [KOB2004] 258b\footnote{\cite{kaas2004}}	 &18:29:49.80 & 	1:15:19.33 	& 	6.38 & 	1.5 	& 5.76 	& 0.4 & 6.85 	& 2.37 	& 0.39 & S\\
SerpensS-NW & 	SER-10 & IRAS 18270-0153\footnote{\cite{connelley2007}}&18:29:37.99 & 	-1:51:04.66 	& 	0.66 & 	0.8 	& 0.78 	& 0.6 & 11.81 	& -4.05 	& 0.35 & S\\\hline
L1688-2 & 			OPH-14 & --	& 								    16:26:24.95 & 	-24:24:23.92 	& 	0.42 & 	1.9 	& 0.32 	& 2.7 & 8.04 	& 6.10 	& 0.78 & E\\
SerpensS-NW & 	SER-21 & SerpS-MM15\footnote{\label{maurynote}\cite{maury2011}}& 18:30:02.69 & -2:01:09.33 & 	0.42 & 	0.9 	& 0.37 	& 1.3 & 7.32 	& 2.82 	& 0.38 & E\\\hline
OMC1 tile4 & 			ORA-36 & HOPS 383\footnote{\cite{safron2015}}		 & 5:35:29.67 & 	-4:59:37.25 	& 	0.53 & 	0.9 	& 0.58	& 1.6 & 4.17 	& -2.66 	& 0.64 & Pos.\\
\hline
\end{tabular}
\end{sidewaystable}

In Section \ref{postredcalsec}, above, we described how we bring the mean GBS data and the mean Transient Survey data into relative flux calibration with one another by using the bright, compact sources in each overlapping field. Once the relative FCF has been computed, we repeat our calculation to search for significant outliers by replacing the individually calculated weighted mean of the source peak ratios, $\bar{R}$, with the mean FCF and $\sigma_{\bar{R}}$ with $\sigma_{\mathrm{FCF}}$ in Equation \ref{outliereq}. Here, we define the significance of outliers based on the distribution of these newly calculated $\delta$ values.

Figures \ref{firstsigmafig} to \ref{lastsigmafig} show the $\delta$ value of every source. In Figure \ref{deltadist}, we show the distribution of $\delta$ values for all sources fit with a Gaussian curve. The fitted Gaussian has a standard deviation of $\sigma_{GaussFit}=1.76$.  
The measured $\delta$ values significantly deviate from the fitted Gaussian profile beyond a threshold of $\delta=\pm6$. Therefore, we define sources that have $\abs{\delta} > 6$ as noteworthy outliers. The probability of detecting a source with $\abs{\delta}>6$ from the Gaussian distribution of values is 0.06\%.    
Over all 11 GBS fields that overlap with the Transient Survey data, however, we find that 7 out of 175 independent sources\footnote{Sources that appear in multiple GBS fields as well as a Transient Survey field are only counted once.} brighter than \mbox{200 mJy beam$^{-1}$} with radii \mbox{<10\arcsec} exceed this threshold. 
Thus, 96\% of the identified sources show no sign of variation between the average GBS data and the average Transient data to our sensitivity, while the remaining 4\% are considered variable candidates and form the basis of all further investigation. We summarise the properties of the variable candidates in Table \ref{potvartable}.

 In order to discern the quality of each of the variable candidates, we construct difference maps, subtracting the GBS co-add from the Transient co-add for each region, and perform a visual analysis on each map to note sources with extended features in confused regions as well as larger-scale differences between the maps which can complicate our measurements (see Figures \ref{difffig1} and \ref{difffig2} in Appendix \ref{discussextended}).

\begin{table*}
\centering
\caption{Associations between variable candidate and YSOs.}
\label{potvarysotable}
\begin{tabular}{|c|c|c|c|c|c|}
\hline
GBS Field & ID & Other Name & N$_{\mathrm{proto}}$\footnote{The number of protostars within 10$\arcsec$ of the source peak.}  & Dist$_{\mathrm{proto}}$\footnote{The distance, In units of $\arcsec$, between the source peak and the nearest protostar within 15$\arcsec$. -- indicates no nearby objects.}  & Category\\
\hline\hline 
NGC1333-N & 			PER-1 & 	IRAS4A			& 2 & 2.64 &  S\\
NGC1333-N & 			PER-10 &  Bolo 40			& 0 & -- &  S\\
NGC1333-N & 			PER-34 & [LAL96] 213		& 1 & 2.56 &  S\\
SerpensMain1 & 		SER-1 & 	[KOB2004] 258b	& 1 & 3.16 &  S\\
SerpensS-NW & 	SER-10 & IRAS 18270-0153	& 1 & 6.22 & S\\\hline
L1688-2 & 			OPH-14 & --				& 0 & 11.31 &  E\\
SerpensS-NW & 	SER-21 & SerpS-MM15		& 0 & -- & E\\\hline
OMC1 tile4 & 			ORA-36 & HOPS 383		& 0 & 13.49\footnote{The extracted 2D Gaussian traces an extended structure, slightly shifting the peak of this emission source away from HOPS 383 but a brightness change associated with a compact feature containing the Class 0 protostar is apparent in the constructed difference maps (see Section \ref{prevknownvar} and Figure \ref{difffig1}).} & Pos.\\
\hline
\end{tabular}
\end{table*}
  
  \begin{table*}[h]
\centering
\caption{Associations between 850 $\mu$m emission sources and YSOs.}
\label{ysotable}
\begin{tabular}{|c|c|c|c|c|}
\hline
& \# \mbox{850 $\mu$m} Sources\footnote{Number of robust, bright, compact sources extracted using the {\sc{Gaussclumps}} algorithm.} & \# With Proto (10$\arcsec$)\footnote{Number of sources with a protostar within 10$\arcsec$ of their peak location (also expressed as a percentage of the total number of sources in this category).} & \# With Disk (10$\arcsec$)\footnote{Number of sources with a disk (Class II YSO) within 10$\arcsec$ of their peak location (also expressed as a percentage of the total number of sources in this category).} &  \# With no known YSO \\
\hline\hline 
Total & 175 & 54 & 13 & 108 \\
Strong Var. Can. & 5 (3\%) & 4 (7\%) &  0 (0\%) & 1 (1\%) \\
All Var. Can.  & 7 (4\%) & 4 (7\%) &  0 (0\%) & 3 (3\%) \\
\hline
\end{tabular}
\end{table*}

While investigating the difference maps, we searched for any indication of compact sources that were present only in the GBS data and not in the Transient Survey data (or vice versa) in case an object had varied such that it fell below the detection threshold (200 \mbox{mJy beam$^{-1}$}) in one dataset but not in the other. 
No significant objects of this type were identified.

Through our visual inspection of each source in their respective difference maps, we define 2 categories of variable candidate:

\newpage

\begin{itemize}[leftmargin=*]

  \item[] \textit{Strong Candidates} (S): The source exceeds the significance threshold ($\abs{\delta}>6$), has an average Transient Survey peak brightness measurement of $f_{\mathrm{trans}}>200\mathrm{\:mJy\:beam}^{-1}$, and has a radius less than 10$\arcsec$. These sources have an obvious indication of significant compact structure at the location of the source in the difference map.
  
  \vspace{2mm}
  
  \item[] \textit{Extended Candidates} (E): The source exceeds the significance threshold ($\abs{\delta}>6$), has an average Transient Survey peak brightness measurement of $f_{\mathrm{trans}}>200\mathrm{\:mJy\:beam}^{-1}$, and a radius less than 10$\arcsec$. There is little indication of compact structure present at the location of the source in the difference map; surrounding extended structures complicate the measurements.
  
\end{itemize}

\noindent We find 5 \textit{Strong} sources and 2 \textit{Extended} sources. 
\textit{Extended} variable candidates display very little evidence of real compact emission changing brightness between the GBS and Transient Survey datasets. It is likely that we are underestimating the uncertainty for these objects due to the uncertainty in the background subtraction computed by the {\sc{Gaussclumps}} algorithm in confused areas. 
Both of these remain interesting sources, though they are less robust candidates for variability than those classified as \textit{Strong}. The \textit{Extended} source OPH-14 traces part of the larger scale structure around the prototypical Class 0 source VLA 1623 \citep{andre1993}, though we see no significant evidence that the deeply embedded protostar itself is undergoing a significant brightness change at \mbox{850 $\mu$m}. In addition to the \textit{Strong} and \textit{Extended} candidates, one \textit{Possible} candidate (HOPS 383; \citealt{safron2015}) is also identified in Tables \ref{potvartable} and \ref{potvarysotable} and is discussed further in Section \ref{prevknownvar}.

In general, we expect the presence of a disk to influence the accretion rate of material onto the central protostar. Of the 5 \textit{Strong} variable candidates, 4 are associated with a known protostar (and likely, therefore, a young circumstellar disk)  and none are associated with a known evolved disk object (see N$_{\mathrm{proto}}$ in Table \ref{potvarysotable}; see also Table \ref{ysotable}). 
To be associated with a protostar or a disk, the peak position of the source must be within 10$\arcsec$ of the YSO location. 
Recall that the presence of a dusty envelope increases the likelihood of detecting variability at submillimetre wavelengths due to the reprocessing of the light emitted by a burst event. Generally, the envelope is faint or non-existant for known, more evolved disk candidates seen at infrared wavelengths. 

In total, out of the 175 independent sources identified across all 8 Transient Survey fields and their 11 associated GBS fields, there are 54 protostars and 13 disk objects within 10$\arcsec$ of source peaks. 
Therefore, to the sensitivity achieved across these surveys, approximately 7\% of known protostars associated with an \mbox{850 $\mu$m} emission source to be potentially varying. Only 3\% of the sources not known to be associated with a YSO show signs of variability.  One of the \textit{Strong} variable candidates, IRAS4A (PER-1), has two protostars associated with its peak. Another, SER-1, displays an elongated structure that may indicate multiple embedded sources or some outflow activity. Excluding the \textit{Extended} sources, which are less robust detections, only one source without a known protostar or disk shows signs of variability. We do not expect starless cores to be variable. Therefore, the \textit{Strong} variable candidate without a known embedded YSO, Bolo 40 \citep{enoch2006}, is a good target for follow-up studies to identify signatures of very faint, deeply embedded protostars that have yet to be detected.  


\begin{figure*} 	
\centering
\includegraphics[width=14cm,height=11cm]{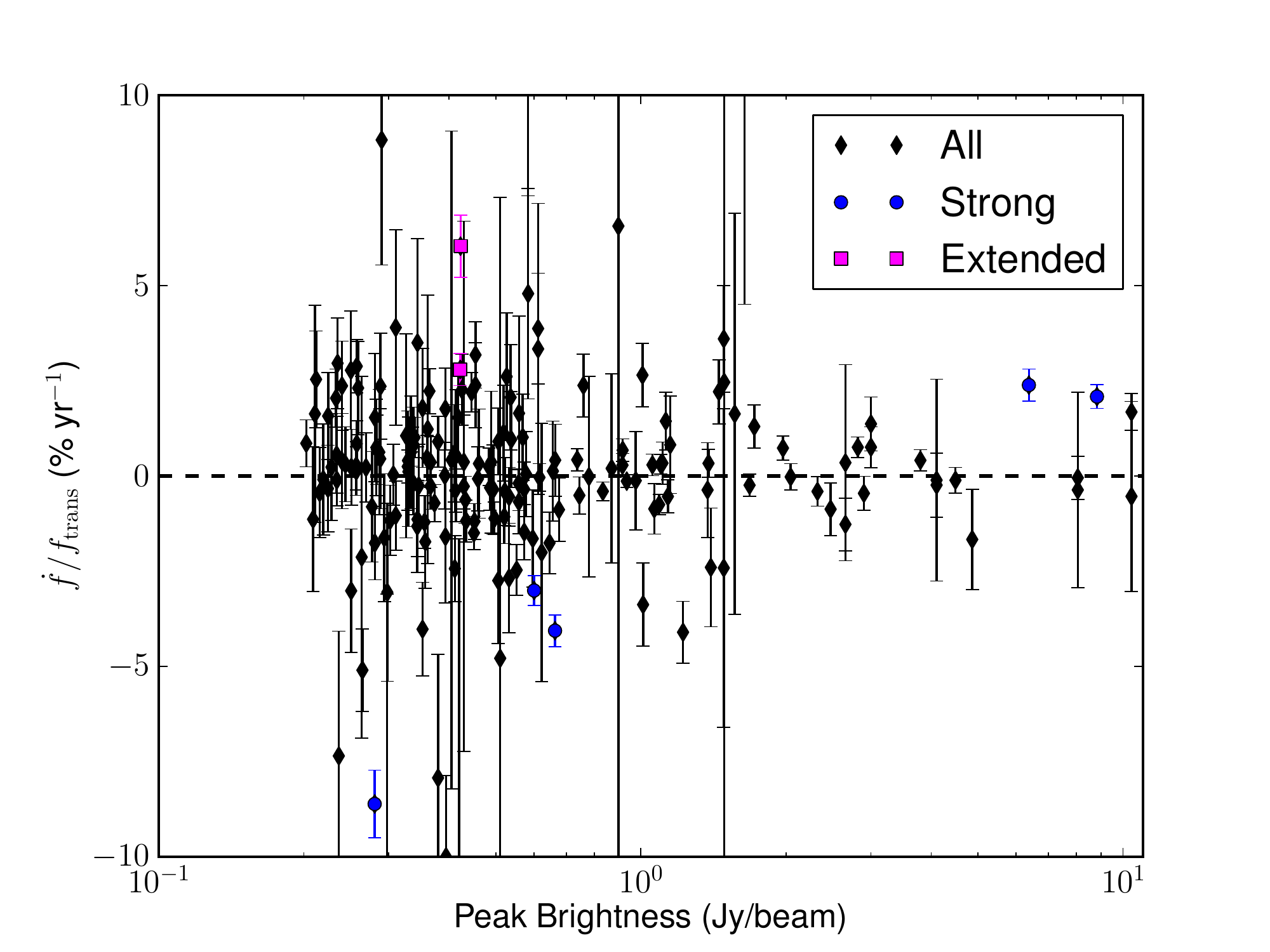}
\caption{The change in peak brightness divided by the difference between the average GBS and Transient Survey observation dates, normalised to the average Transient Survey peak brightness (($\dot{f}/f_{\mathrm{trans}}$), Equation \ref{deltafeq}). \textit{Strong} variable candidates are indicated by blue circles. \textit{Extended} variable candidates are indicated by magenta squares. All other sources are indicated by black diamonds. Variable candidates are intermixed with non-variable sources as the detection sensitivity varies from field to field (see Table \ref{deltafsensitivitytab}).}
\label{deltaffig}
\end{figure*}

The GBS data and the Transient Survey data were obtained 2-4 years apart, depending on the field (see Table \ref{fieldstable}), which allows us to characterise the apparent, significant submillimetre brightness changes in terms of a fractional change per year (averaged over both datasets) assuming a linear change throughout the GBS/Transient Survey time lag, 

\begin{equation}
\label{deltafeq}
\frac{\dot{f}}{f_{\mathrm{trans}}} = \frac{f_{\mathrm{trans}}-f_{\mathrm{GBS}}}{f_{\mathrm{trans}}}\times\left(\frac{1}{\bar{t}_{\mathrm{trans}}-\bar{t}_{\mathrm{GBS}}}\right),
\end{equation}

\noindent where $\bar{t}_{\mathrm{GBS}}$ is the mean GBS observation date for that field and $\bar{t}_{\mathrm{trans}}$ is the mean Transient Survey observation date for that field (the parenthetical term in the denominator is the same as $\Delta t$ in Table \ref{fieldstable}). We choose to normalise the result to the mean Transient Survey observation to express the brightness change as a percentage of the (near) current source peak brightness. The associated error, $\sigma_{\dot{f}/f}$, is the combination of the Transient Survey average peak brightness error (see Section \ref{postredcalsec} and Table \ref{potvartable}), the GBS peak brightness error, and the error in the weighted mean.

\begin{table*}[h]
\centering
\caption{Current $\dot{f}/f_{\mathrm{trans}}$ sensitivity limits.}
\label{deltafsensitivitytab}
\begin{tabular}{|c|c|c|}
\hline
GBS Field &  $\Delta f$ Sensitivity Limit (\%yr$^{-1}$) & Typical $\Delta f$ Sensitivity (\%yr$^{-1}$)\\
\hline\hline
IC348-E & 2.2 & 2.9\\
L1688-1 & 3.3 & 5.0\\
L1688-2 & 2.9 & 4.8\\
NGC1333-N & 1.4 & 3.6\\
OMC1 tile4 & 2.0 & 4.3\\
OrionBN\_450\_S & 3.7 & 7.4\\
OrionBS\_450\_E & 3.8 & 17.7\footnote{The large uncertainties in OrionBS\_450\_E are due to complication in background structure subtraction executed by the {\sc{GaussClumps}} algorithm (see Appendix \ref{discussextended}). These are, however, bright and isolated sources in this field that are well recovered and better represented by the $\abs{\dot{f}/f_{\mathrm{trans}}}$ sensitivity limit in the adjacent column.}\\
OrionBS\_450\_W & 2.4 & 4.7\\
SerpensMain1 & 1.1 & 5.0\\
SerpensS-NE & 1.0 & 4.6\\
SerpensS-NW & 1.5 & 3.8\\
\hline
\end{tabular}
\end{table*}


We plot $\dot{f}/f_{\mathrm{trans}}$ as a function of the Transient Survey average peak brightness in Figure \ref{deltaffig}, coloured according to category. A typical, \textit{Strong} variable candidate has an average brightness change per year of $\abs{4.0}\%\mathrm{\:yr}^{-1}$ with a standard deviation of $2.7\%\mathrm{\:yr}^{-1}$. The \textit{Extended} variable candidates are grouped in a similar area of parameter space. This grouping is primarily the result of poor background flux subtraction performed by the {\sc{Gaussclumps}} algorithm in these more confused regions. This strengthens the notion that the uncertainties in these measurements are likely underestimated. Every calibrated GBS field includes a distribution of sources that appear slightly brighter and slightly fainter than their Transient Survey counterparts. The brightest \textit{Strong} candidates all display a positive $\dot{f}/f_{\mathrm{trans}}$, while the fainter \textit{Strong} candidates all show significant brightness decreases between the GBS and Transient Survey eras. This may indicate an intrinsic difference in the underlying accretion process for different types of protostars. This trend, however, is only based on small number statistics.

Three factors  contribute to the sensitivity of $\abs{\dot{f}/f_{\mathrm{trans}}}$: the uncertainties in the GBS and Transient Survey peak brightness measurements, the uncertainty in the calibration, and the time lag between the two data sets for 
each field (see $\Delta t$ in Table \ref{fieldstable}). In Table \ref{deltafsensitivitytab}, we present both the sensitivity limit and the typical sensitivity in $\abs{\dot{f}/f_{\mathrm{trans}}}$ for each field. To calculate the sensitivity limit for each field, 
we set Equation \ref{outliereq} (replacing $\bar{R}$ with the mean FCF and $\sigma_{R}$ with $\sigma_{\mathrm{FCF}}$) equal to 6 (the threshold at which we define a source to be a variable candidate) and solve for the GBS peak brightness 
in terms of the Transient Survey peak brightness. Then, we substitute the result into Equation \ref{deltafeq} for each source and find the minimum allowed $\abs{\dot{f}/f_{\mathrm{trans}}}$ for a variable candidate in each field. For the typical $
\abs{\dot{f}/f_{\mathrm{trans}}}$ sensitivity, we report the median $\abs{\dot{f}/f_{\mathrm{trans}}}$ value calculated in this fashion. As the Transient Survey continues and we are able to observe longer time separations, our sensitivity will 
improve.


\section{Discussion}
\label{GBStransientdiscusssec}


The relative flux calibration agreement between GBS fields associated with the same Transient Survey field is robust. There are 13 sources in total that overlap two GBS fields and a Transient Survey field. There are no cases where a source is considered to be a variable candidate in one GBS field and not the other. In total, we find 7 sources with significantly discrepant GBS to Transient Survey peak brightness ratios, indicating they are candidates for variability. 
 These sources are distributed around $\dot{f}/f_{\mathrm{trans}} = 0 \% \mathrm{\:yr}^{-1}$ (Figure \ref{deltaffig}). 
Though we do not necessarily expect linear changes over 2-4 year timescales, assuming linearity, the average $\abs{\dot{f}/f_{\mathrm{trans}}}$ of the \textit{Strong} variable candidates over several year timescales using co-added, well calibrated maps is $\abs{4.0}\%\mathrm{\:yr}^{-1}$ with a standard deviation of $2.7\%\mathrm{\:yr}^{-1}$ (see Table \ref{potvartable} and Figure \ref{deltaffig}). We are not sensitive to short timescale, small fluctuations that may average out the signal over time (see Section \ref{EC53sec} for a further example of the limitations of time averaging), so we expect there to be more variable candidates in these fields that can be identified at higher time resolutions and sensitivities. An analysis of variability within the Transient Survey data alone is addressed by Johnstone et al., (in preparation). 

The observed difference in the submillimetre flux of an embedded protostar undergoing a change in accretion rate is determined by the heating or cooling of the dusty envelope. As long as the dust temperature remains above \mbox{$\sim$25 K}, 
the sub-mm response will be approximately linear to this change in temperature. When the dust temperature is lower, the submillimetre response becomes much stronger. In equilibrium, the dust temperature, $T_{\mathrm{dust}}$, in most of the envelope is expected to vary with accretion luminosity, $L_{\mathrm{acc}}$,  as $T_{\mathrm{dust}} \propto L_{\mathrm{acc}}^{1/4}$ \citep{johnstone2013}. As \cite{johnstone2013} note, however, this relationship will break down in the very outer envelope where significant heating comes from the external radiation field and therefore the temperature of the dust remains relatively fixed.

For small fractional changes in temperature, the accretion luminosity is approximated by
\begin{equation}
\label{lumvsaccreq}
\frac{\Delta L_{\mathrm{p}}}{L_{\mathrm{p}}} \propto 4\times\frac{\Delta T_{\mathrm{dust}}}{T}.
\end{equation} 
Assuming the submillimetre response is roughly proportional to the dust temperature which itself is determined primarily by the accretion luminosity, a 4\% change in the observed submillimetre flux corresponds to a $\sim$16\% change in the accretion luminosity (accretion rate) of the central protostar. If the envelope temperature drops below \mbox{$\sim$25 K}, then the submillimetre response will be closer to one-to-one (i.e. a 4\% change in the observed submillimetre flux corresponds to $\sim$4\% change in accretion luminosity).  Much work is still required to fully understand the relationship between the observed submillimetre flux and the accretion luminosity (see, for example, \citealt{johnstone2013}).

We find 7\% of the known protostars in our observed fields display a typical $16\%$ accretion variability over $\sim$3 years (using Equation \ref{lumvsaccreq}). In Figure 3 of \cite{herczeg2017arxiv}, the authors summarise expectation values of accretion variability over specific timescales, taking into consideration a model in which the variability is driven by large-scale gravitational instabilities in the disk \citep{vorobyov2010} and a model in which smaller-scale magneto-rotational instabilities are included \cite{bae2014}. The models of \cite{vorobyov2010} predict that $\sim$7\% of protostars will undergo a 7\%-8\% change in accretion luminosity over 3 years of observations. \cite{bae2014}, however, predict that $\sim7\%$ of protostars will undergo a 40\% change in accretion luminosity over this same timeframe.  Under the simple assumption that submillimetre brightness varies linearly as dust temperature, Equation \ref{lumvsaccreq} predicts our results to lie  between the two models. 
We note, however, that this result is tempered by uncertainties in the relationship between changes in the submillimetre flux due to the protostellar luminosity, the reliability expected from the models of protostellar episodic accretion over few year timsescales (as a detailed investigation relies on accurately tracing the physics of the inner disk), and the sensitivity of our source detection methods. 

Future surveys studying accretion variability onto deeply embedded protostars would benefit from working in the far infrared, near the peak of the protostellar envelope spectral energy distribution in order to have a more linear relationship between observed brightness changes and the underlying accretion luminosity variation \citep[see ][]{johnstone2013}. These surveys are likely to be undertaken by space observatories such as the \textit{Space Infrared Telescope for Cosmology and Astrophysics} (\textit{SPICA}; Roelfsema et al., submitted) or the \textit{Origins Space Telescope} (\textit{OST}; \citealt{meixner2016}) which will further benefit from the stability of space-based observations, provided that the calibration and dynamic range of the instruments is excellent. As these missions are still a decade or more away, an obvious first undertaking will be to use \textit{Herschel Space Observatory} observations of nearby star-forming regions as a previous epoch, yielding a multi-decade delta in time. As this paper shows, however, there are significant challenges in collating two disparate data sets and thus care will need to taken in order to reach relative uncertainties at the $1\%-2\%$ level.

\newpage

\subsection{Previously Known Signatures of Variability}
\label{prevknownvar}

Nearly all of the submillimetre emission sources in Table \ref{potvartable} are detected in previous surveys \citep[see, for example,][]{johnstone2000,johnstone2001,skrutskie2006,kirk2006,difrancesco2008}, and many have been associated with outflows. Two \textit{Strong} sources have previously-known indicators of variability from outflow knots or inferences from spectroscopic diagnostics: IRAS4A (PER-1) and IRAS 18270-0153 (SER-10). 

%


In the case of IRAS4A, the evidence is in the form of outflowing jets with compact knots \citep{choi2001}, a phenomenon that has been characterised around protostellar and Herbig-Haro objects for many years \citep[see, for examples, ][and references therein]{reipurth1989,cernicharo1996,reipurth2004}. These jets may be caused by episodic outbursts \citep{choi2001} that increase the density of the ejected material for a short period of time, leaving some indication of the history of activity around the source. Other evidence, however, points to jet precession as the source of the knots \citep{choi2006,santangelo2015}. 

IRAS 18270-0153 \citep{connelley2007} has been classified as an ``FU Orionis-like'' object, owing to its deep CO and water vapour absorption bands and lack of clearly defined photospheric absorption lines \citep{greene1996,connelley2010}. These features indicate the presence of a very hot, optically thick inner disk, which is a signpost for an ongoing FUor accretion outburst \citep{zhu2007}. 
In addition, this protostar has a notable bipolar outflow in H$_{\mathrm{2}}$ \citep{zhang2015}. 
We observe a decrease in brightness at the rate of \mbox{4.3\% yr$^{-1}$}, assuming a linear change. 

In addition to these \textit{Strong} candidates, we highlight one further source of interest, ORA-36 (HOPS 383; \citealt{safron2015}), which we list as a \textit{Possible} variable candidate in Tables \ref{potvartable} and \ref{potvarysotable}. HOPS 383 is a source that contains the youngest known Class 0 protostar that has shown evidence of an outburst in both infrared and submillimetre data \citep{safron2015}. Mid- and far-infrared photometric data indicate that the source underwent a strong outburst between 2004 and 2012. By comparing \mbox{450 $\mu$m} Submillimetre Common-User Bolometre Array (SCUBA) data from 1998 to \mbox{350 $\mu$m} Submillimetre APEX Bolometer Camera (SABOCA) data in 2011, \cite{safron2015} found that the source had doubled in brightness in the submillimetre. 
We expect that an accretion outburst would be much brighter in the infrared \citep{johnstone2013}. Here, we see a clear indication of a compact emission source closely associated with HOPS 383. We do not consider it a robust variable candidate as its $\delta$ value is less than 6 ($\delta=4.17$; see Table \ref{potvartable}), though it appears to be undergoing a brightness decrease between the GBS and Transient Survey data at the level of $-2.66\pm0.64\%\:\mathrm{yr}^{-1}$. \citealt{fischer2017atel} have also noted a recent brightness decrease in the $H$ and $K$-bands for this source. The extracted 2D Gaussian source traces an extended structure in which the compact feature is embedded (see Figure \ref{difffig1}), causing the peak of the identified source to artificially shift further from the protostar it contains. 


In addition to the variables, some non-varying sources are also of interest. Our observations coincide with 2 long-term radio (cm wavelength) variable sources in IC 348 (VLA 2 and VLA 3; \citealt{forbrich2011}), but neither are significantly variable at \mbox{850 $\mu$m}. 
VLA 2 was found to decrease in brightness from \mbox{0.09 mJy} in 2001 \citep{avila2001} to \mbox{0.04 mJy} in 2008,  
In this work, we have found no significant brightness change between the GBS and Transient Survey observations ($\delta=0.81$) for the associated \mbox{850 $\mu$m} emission source. 
Similarly, we find no \mbox{850 $\mu$m} emission source to coincide with VLA 3, which increased in radio brightness from 0.41 mJy in 2001 \cite{avila2001} to 0.542 mJy in 2008, 
It is important to recognise that radio and submillimetre variability arise from different phenomena. In the former case, magnetic flares and synchrotron radiation dominate while in the latter case we are tracing accretion events. 


\begin{figure*} 	
\centering
\subfloat{\label{}\includegraphics[width=14cm,height=11cm]{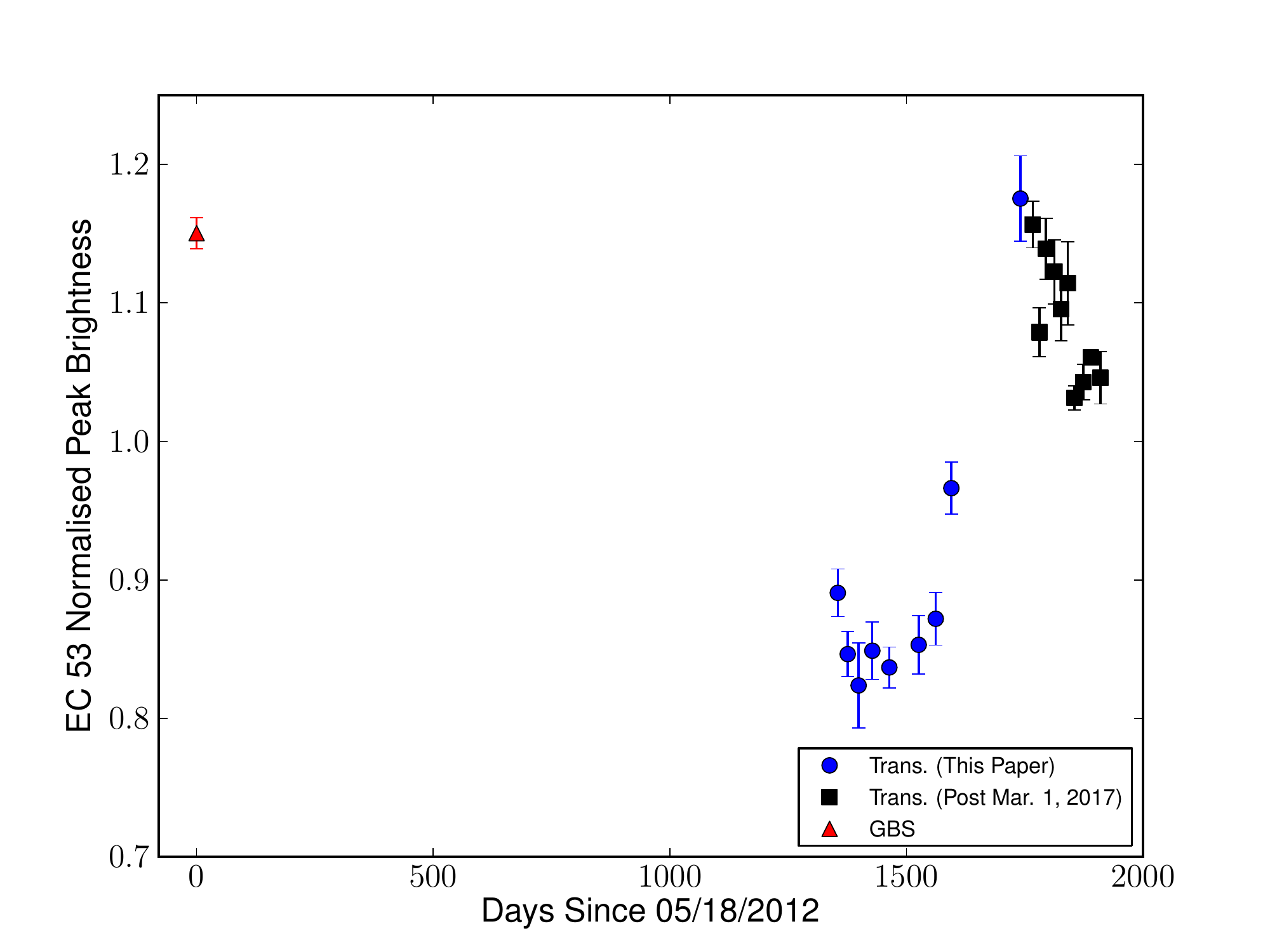}}\\
\subfloat{\label{}\includegraphics[width=14cm,height=11cm]{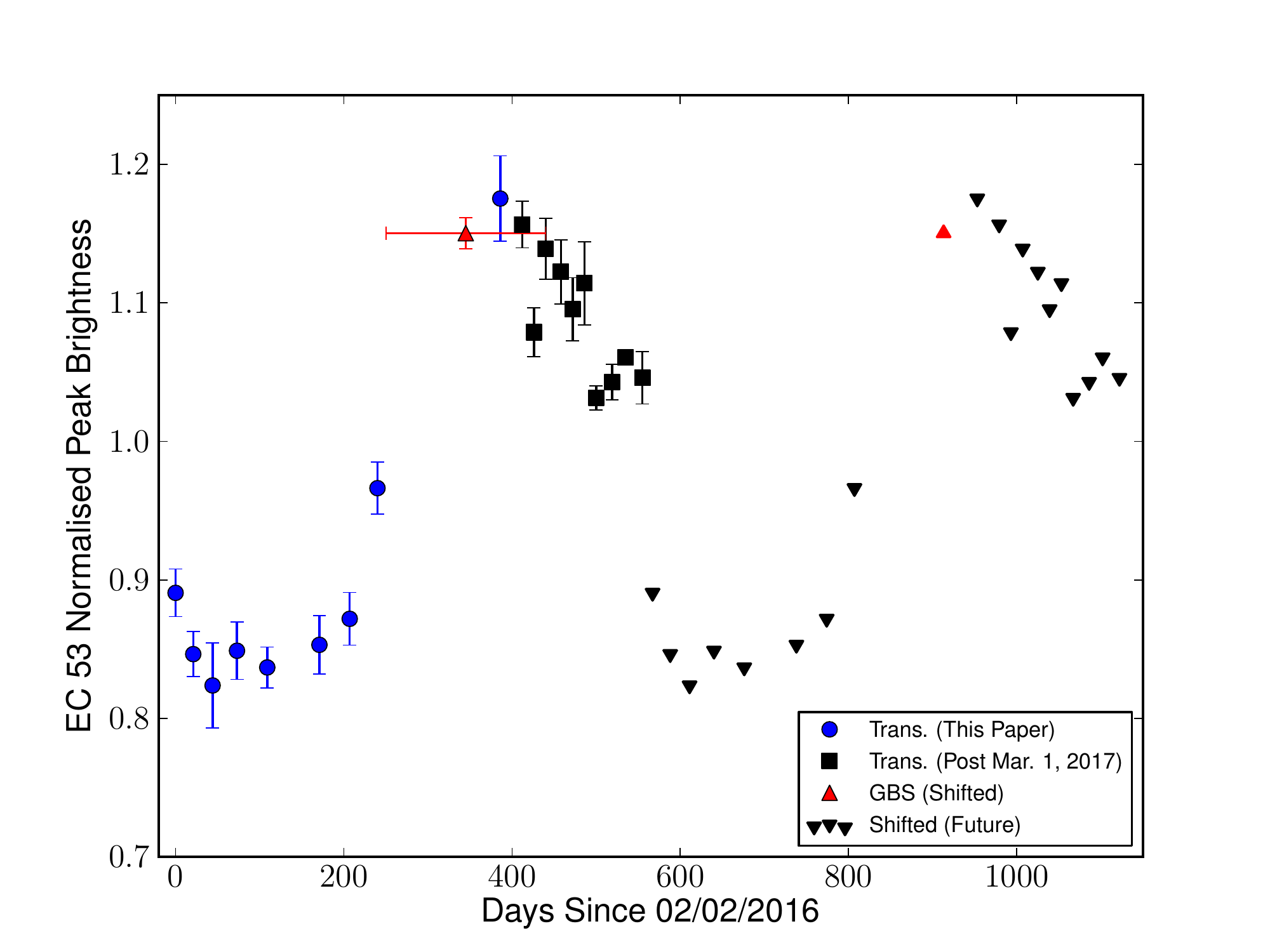}}
\caption{The 850 $\mu$m light curve of EC 53 \protect(see also, \citealt{yoo2017arxiv}). \textit{Top}: The red (upward) triangle represents the average, calibrated GBS data, the blue circles represent data analysed in this paper, and the black squares represent data that has been collected between March 1$^{st}$, 2017 and July 5$^{th}$, 2017. \textit{Bottom}: The black (downward) triangles represent all obtained Transient Survey data shifted one period (567 days) into the future. The GBS data presented in this Figure has been shifted three and then four increments of 567 days until it matched the current and next periodic cycles. The error bars in the GBS dates suggest a reasonable range of values that agree with the rise in the light curve.}
\label{EC53fig}
\end{figure*}

\subsection{The Period of the Submillimetre Variable EC 53}
\label{EC53sec}

One prominent source in the Serpens Main Transient Survey field, EC 53 \citep{eiroa1992}, is identified as a sub-mm variable in a companion paper \citep{yoo2017arxiv} and has previously been identified as a periodic variable in near-IR ($K$-band) photometry \citep{hodapp2012}. The periodicity is thought to be driven by accretion instabilities triggered by a nearby companion \citep{hodapp2012}. 
We do not detect it as a variable candidate in this paper ($\delta=2.83$) because EC 53's periodicity 
is timed such that the difference between the GBS measurement and the average Transient measurement is only moderate.  In addition, the uncertainty (standard deviation) in EC 53's time averaged brightness across the Transient Survey data is high, because the brightness change over one period is significant (see Johnstone et al., in prep. for an analysis performed at a higher time resolution). To make this more clear, the source's \mbox{850 $\mu$m} light curve presented in Figure \ref{EC53fig} (see also, \citealt{yoo2017arxiv}). In the top panel, the GBS and Transient peak brightness data are plotted against the dates of observation. The red (upward) triangle represents the average, calibrated GBS data, the blue circles represent Transient Survey data included in the investigation presented in this paper (obtained between February 2$^{nd}$, 2016 and February 22$^{nd}$, 2017), and the black squares represent Transient Survey data obtained between March 1$^{st}$, 2017 and July 5$^{th}$, 2017. The Transient Survey error bars represent the standard deviation of the normalised peak brightnesses across the \textit{Family} of calibrator sources for each epoch (whereas the uncertainty of the GBS data point includes all 4 GBS observations taken over two nights, see Table \ref{fieldstable}). 
 
 In the bottom panel, we shift the calibrated GBS data from its original observation date until it reasonably aligns with the Transient Survey observation dates. We find that a shift of 1700 days from the original observation date agrees well with the rise of the peak brightness (red, upward triangle) with a range of acceptable values from 1605 days to 1795 days (horizontal, red errorbars). 
A shift of \mbox{1700$\pm$95 days} corresponds to 3 full periods of \mbox{567 $\pm$ 32 days} at \mbox{850 $\mu$m}. 
To this accuracy, we independently confirm the \mbox{850 $\mu$m} periodicity is consistent with the 543 day $K$-band periodicity \citep{hodapp2012}.  In order to show the periodic nature of EC 53, we assume the cycle is continuously repeating and we display all of the obtained Transient Survey observations shifted by 567 days (black, downward triangles), once again including the average GBS data in this future cycle. 



\section{Summary and Conclusions}
\label{GBStransientconclusionsec}

In this paper, we investigate signatures of \mbox{850 $\mu$m} submillimetre variability by comparing archival GBS fields, observed between 2012 and 2014, to JCMT Transient Survey fields, observed between December 2015 and March 1$^{\mathrm{st}}$ 2017. We follow the data reduction and calibration procedures presented in \cite{mairs2017} to self-consistently align and calibrate each set of observations from the two surveys individually before bringing the co-added GBS images into relative flux calibration with the co-added Transient Survey images (see Section \ref{drcalsec} and Figures \ref{firstweightedmeanfig} through \ref{lastweightedmeanfig}). Using the source extraction algorithm {\sc{Gaussclumps}} \citep{stutzki1990}, we identify 175 independent bright (\mbox{>200 mJy beam$^{-1}$}), compact (effective radius <10$\arcsec$) compact emission objects that were well fit with Gaussian profiles, correlate them with known protostars (Class 0+I, Flat spectrum sources; see Section \ref{ysorefsec}) and disks (Class II sources), and identify objects that are significant outliers in their GBS to Transient Survey average peak brightness ratio with respect to the other sources in the field (see Figures \ref{firstsigmafig} through \ref{lastsigmafig}, Figure \ref{deltadist} and Equation \ref{outliereq}). Based on a visual inspection for compact structure present around the area of each outlying source in calibrated difference maps (the Transient Survey co-add subtracted from the GBS co-add), 
we define two categories for the quality of each variable candidate. 

Our main results are summarised as follows:

\begin{enumerate}[labelindent=0pt,labelwidth=\widthof{\ref{last-item}},label=\arabic*.,itemindent=1em,leftmargin=!]

\item We have developed methods to robustly analyze repeated observations of an area of the sky for signatures of submillimeter variability. We identify 11 archival GBS fields that can be self-consistently flux calibrated and have significant overlap with the 8 Transient Survey fields (see Appendix \ref{discussextended}).

\item Out of 175 independent compact, \mbox{850 $\mu$m} emission sources, we find a total of 7 variable candidates, 4 of which are associated with known protostars. 

\item We classify 5 of the variable candidates as \textit{Strong} and 2 as \textit{Extended} (see Section \ref{GBStransientresultssec}), and we highlight one additional source, HOPS 383, as a \textit{Possible} detection of variability. Two \textit{Strong} sources (IRAS4A, IRAS 18270-0153) along with HOPS 383 
have previously noted signatures of variability. There is one \textit{Strong} variable candidate without a known protostellar or disk association (Bolo 40/PER-10). This is a good target for follow-up studies to identify signatures of very faint, deeply embedded protostars (see Table \ref{potvartable}). 

\item 
The average flux change for the strong variable candidates is $\abs{4.0}\%\mathrm{\:yr}^{-1}$ with a standard deviation of $2.7\%\mathrm{\:yr}^{-1}$ over 2-4 year timescales. Assuming the heating of the envelope is responsible for the changing luminosity, this corresponds to a change in accretion rate of $\sim16\%$ (see Section \ref{GBStransientdiscusssec}). The observed changes in flux for all sources are distributed around $0\%\mathrm{\:yr}^{-1}$. 

\item Using the archival GBS data, we strengthen the detection of the submillimetre variable source EC 53 by adding a critical data point to its periodic light curve and determine its \mbox{850 $\mu$m} period to be \mbox{567 $\pm$ 32 days}. This value is consistent with the 543 day period perviously reported in the $K$-band (see Figure \ref{EC53fig}, \citealt{yoo2017arxiv}, and \citealt{hodapp2012}).

\end{enumerate}

Throughout this work, we have developed methods to robustly analyse repeated observations of an area of the sky for signatures of submillimetre variability. The JCMT Transient Survey will continue through at least January, 2019 and as more data are collected we will have the opportunity to continue this investigation to fainter sources and smaller levels of variability over longer timescales. Future directions include periodically co-adding sets of Transient Survey observations to construct very precise light curves (sacrificing temporal resolution for higher sensitivity), improving the flux calibration procedures such that more archival data can be included, comparing these observations with the simultaneous \mbox{450 $\mu$m} datasets once the latter can be robustly reduced and calibrated, and applying these observational constraints to current simulations of variable protostellar accretion. 

\section*{Acknowledgements}

Steve Mairs was partially supported by the Natural Sciences and
Engineering Research Council (NSERC) of Canada graduate scholarship
program. Doug Johnstone is supported by the National Research Council
of Canada and by an NSERC Discovery Grant. Gregory Herczeg is supported by general grant 11473005 awarded by the National Science Foundation of China. Andy Pon received partial salary support from  a Canadian Institute for Theoretical Astrophysics ( CITA)  National Fellowship. Miju Kang was supported by Basic Science Research Program through the
National Research Foundation of Korea (NRF) funded by the Ministry of
Science, ICT \& Future Planning (No. NRF-2015R1C1A1A01052160).

\vspace{3mm}

The authors wish to recognise and acknowledge the very significant cultural role and
reverence that the summit of Maunakea has always had within the indigenous Hawaiian
community. We are most fortunate to have the opportunity to conduct observations from
this mountain. The James Clerk Maxwell Telescope is operated by the East Asian Observatory on behalf of The National Astronomical Observatory of Japan, Academia Sinica Institute of Astronomy and Astrophysics, the Korea Astronomy and Space Science Institute, the National Astronomical Observatories of China and the Chinese Academy of Sciences (Grant No. XDB09000000), with additional funding support from the Science and Technology Facilities Council of the United Kingdom and participating universities in the United Kingdom and Canada. The James Clerk Maxwell Telescope has historically been operated by the Joint Astronomy Centre on behalf of the Science and Technology Facilities Council of the United Kingdom, the National Research Council of Canada and the Netherlands Organisation for Scientific Research. Additional funds for the construction of SCUBA-2 were provided by the Canada Foundation for Innovation.  The identification number for the JCMT Transient Survey 
data used in this paper is M16AL001. The identification numbers for the archival Gould Belt Survey data used in this paper are: MJLSG31, MJLSG32, MJLSG33, MJLSG38, and MJLSG41. The authors thank the JCMT staff for their support
of the data collection and reduction efforts. This research has made use of NASA's Astrophysics Data System and the facilities of the Canadian Astronomy Data Centre operated by the National Research Council of Canada with the support of the Canadian Space Agency. The authors would especially like to thank Chang Won Lee and Harriet Parsons for their useful insights and suggestions along with the extended JCMT Transient Team\footnote{Joanna Bulger, Subaru Telescope; Eun Jung Chung, KASI; Yuxin He,  Xinjiang Astronomical Observatory; Po-Chieh Huang, National Central University; Miju Kang, KASI; Gwanjeong Kim, KASI; Jongsoo Kim, KASI; Kyoung Hee Kim, KNU/KNUE; Mi-Ryang Kim, Chungbuk University; ShinYoung Kim, KASI/UST; Yi-Jehng Kuan, National Taiwan Normal University; Woojin Kwon, KASI/UST; Shih-Ping Lai, National Tsing Hua University; Bhavana Lalchand, National Central University; Chang Wong Lee, KASI; Feng Long, KIAA/Peking University; A-Ran Lyo, KASI; Harriet Parsons, East Asian Observatory;  Ramprasad Rao, ASIAA; Jonathan Rawlings, University College London; Manash Samal, National Central University;  Archana Soam, KASI; Dimitris Stamatellos, University of Central Lancashire; Wang Yiren, Peking University; Miaomiao Zhang, Max Planck Institute for Astrophysics; Jianjun Zhou, Xinjiang Astronomical Observatory} for their support. This research used the services
of the Canadian Advanced Network for Astronomy Research (CANFAR), which in turn is
supported by CANARIE, Compute Canada, University of Victoria, the National Research
Council of Canada, and the Canadian Space Agency. This research made use of {\sc{APLpy}}, an open-source plotting package for Python hosted at http://aplpy.github.com, and {\sc{matplotlib}}, a 2D plotting library for Python \citep{matplotlib}. 

\facility{JCMT (SCUBA-2) \citep{holland2013}}

\software{Starlink \citep{currie2014}, Astropy \citep{astropy}, Python version 3.5, APLpy \citep{aplpy}, Matplotlib \citep{matplotlib}}

\newpage

\appendix

\section{Source Extraction and Comparison}
\label{discussextended}

\begin{figure*}  	
\centering
\includegraphics[width=18cm,height=13.2cm]{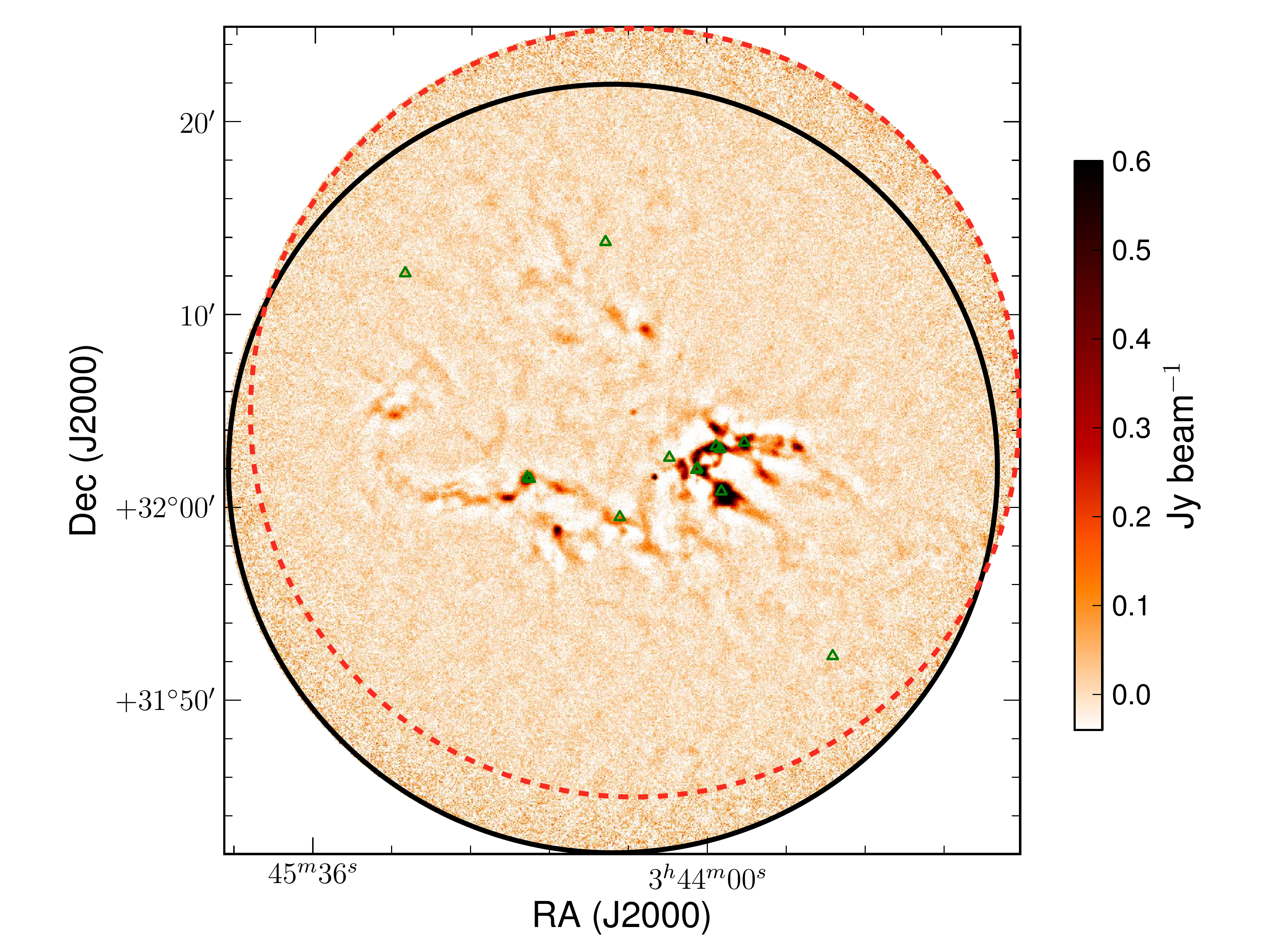}
\caption{The Transient Survey field IC348 mosaicked with its corresponding archival GBS fields at 850 $\mu$m. The area of each observed GBS and Transient Survey field included in the mosaic is bounded by a circle. The solid black circle is the Transient Survey field. The red (dashed) circle shows the boundary of the IC348-E GBS field. The green triangles represent the positions of known protostars taken from the \textit{Spitzer Space Telescope} catalogue of \protect\cite{dunham2015}.}
\label{IC348mosfig}
\end{figure*}

\begin{figure*}  	
\centering
\includegraphics[width=18cm,height=13.2cm]{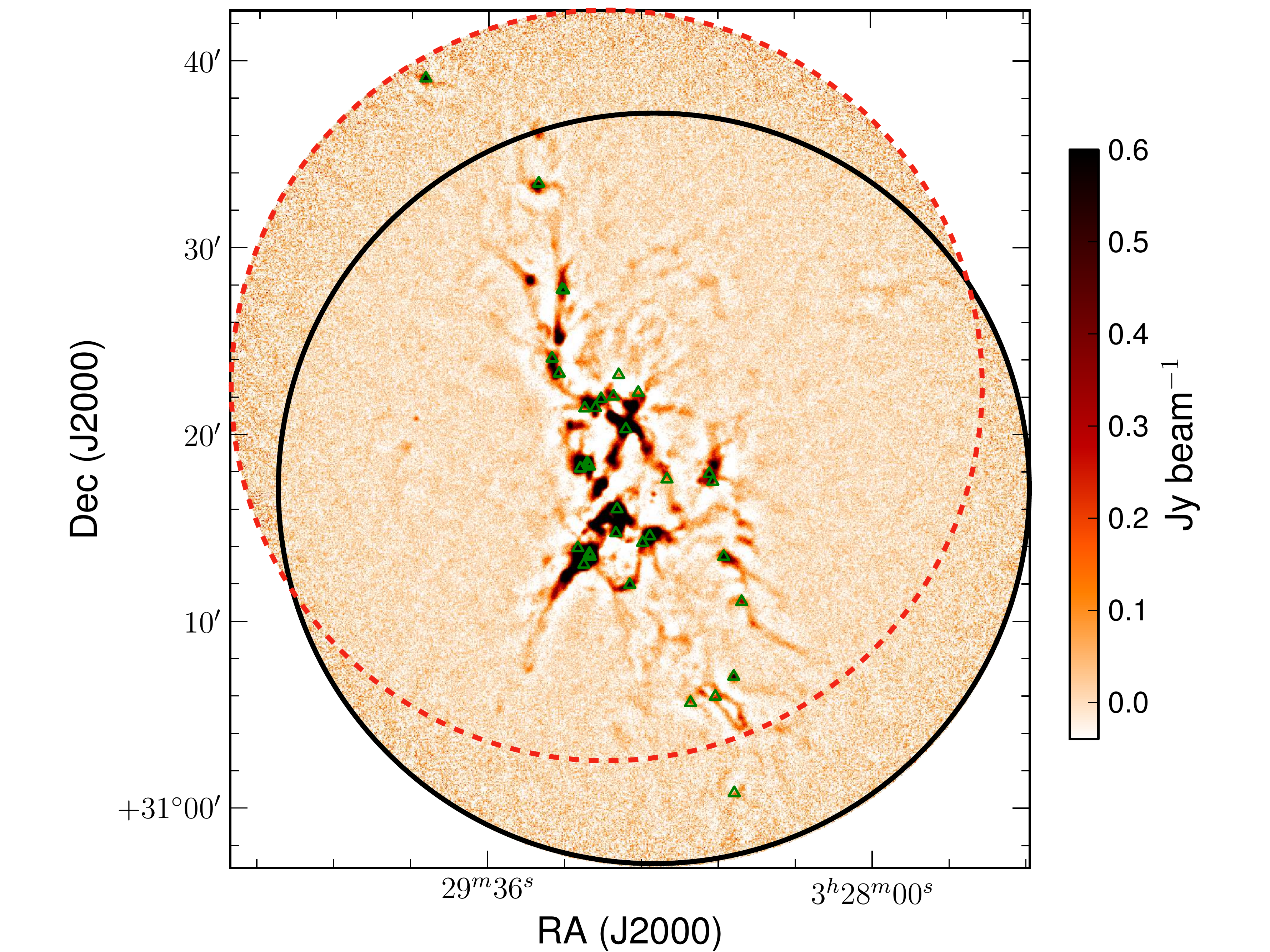}
\caption{Same as Figure \ref{IC348mosfig}, but showing the NGC1333 field with its corresponding archival GBS field. The red (dashed) circle shows the NGC1333-N GBS field.}
\label{NGC1333mosfig}
\end{figure*}

\begin{figure*}  	
\centering
\includegraphics[width=18cm,height=16.5cm]{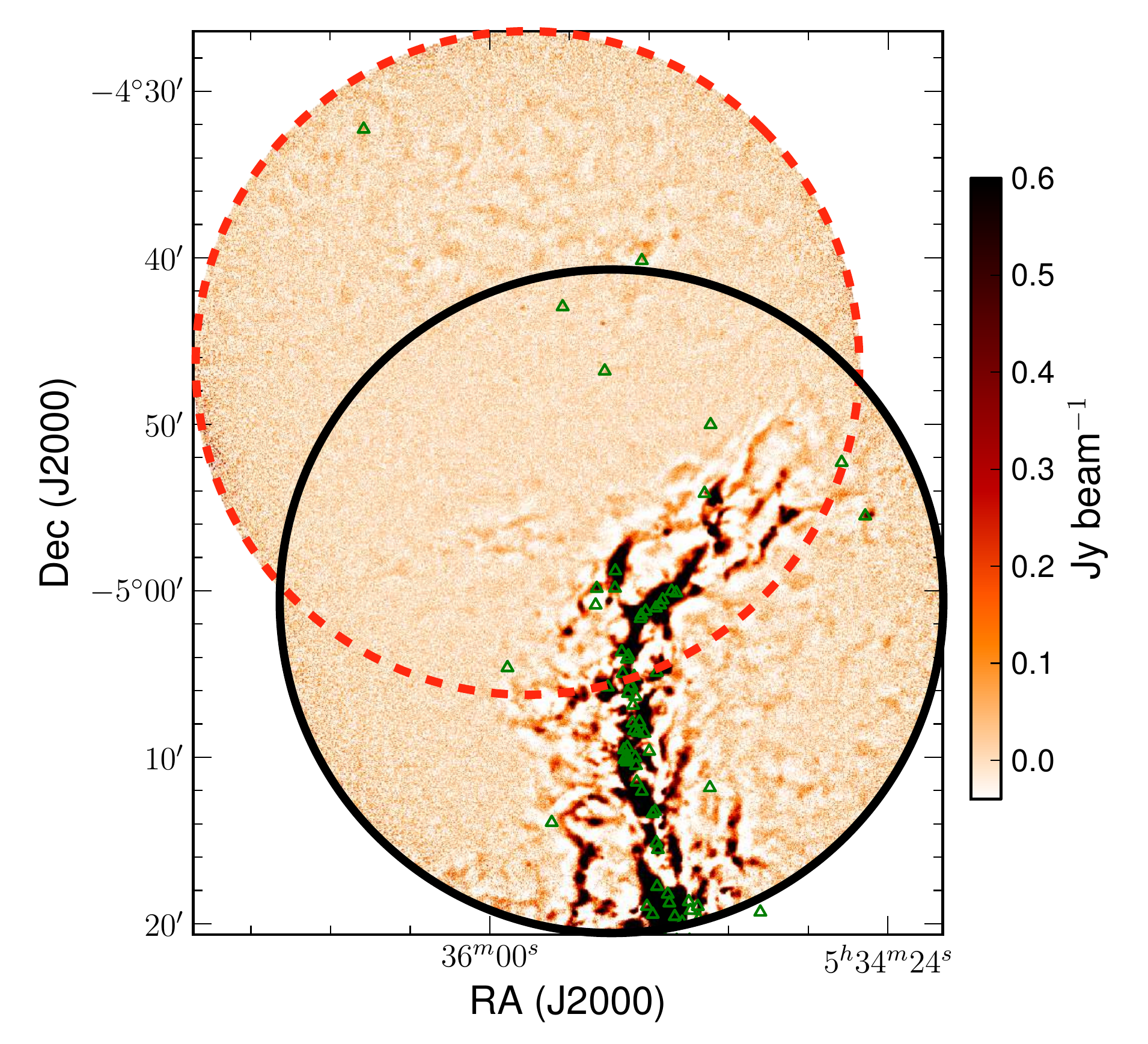}
\caption{Same as Figure \ref{IC348mosfig}, but showing the OMC 2-3 field with its corresponding archival GBS fields. The red (dashed) circle shows the OMC1 tile4 GBS field. The green triangles represent the positions of known protostars taken from the \textit{Spitzer Space Telescope} and \textit{Herschel Space Observatory} catalogues of \protect\cite{megeath2012} and \protect\cite{stutz2013}, respectively.}
\label{OMC23mosfig}
\end{figure*}

 \begin{figure*}  	
\centering
\includegraphics[width=18cm,height=13.2cm]{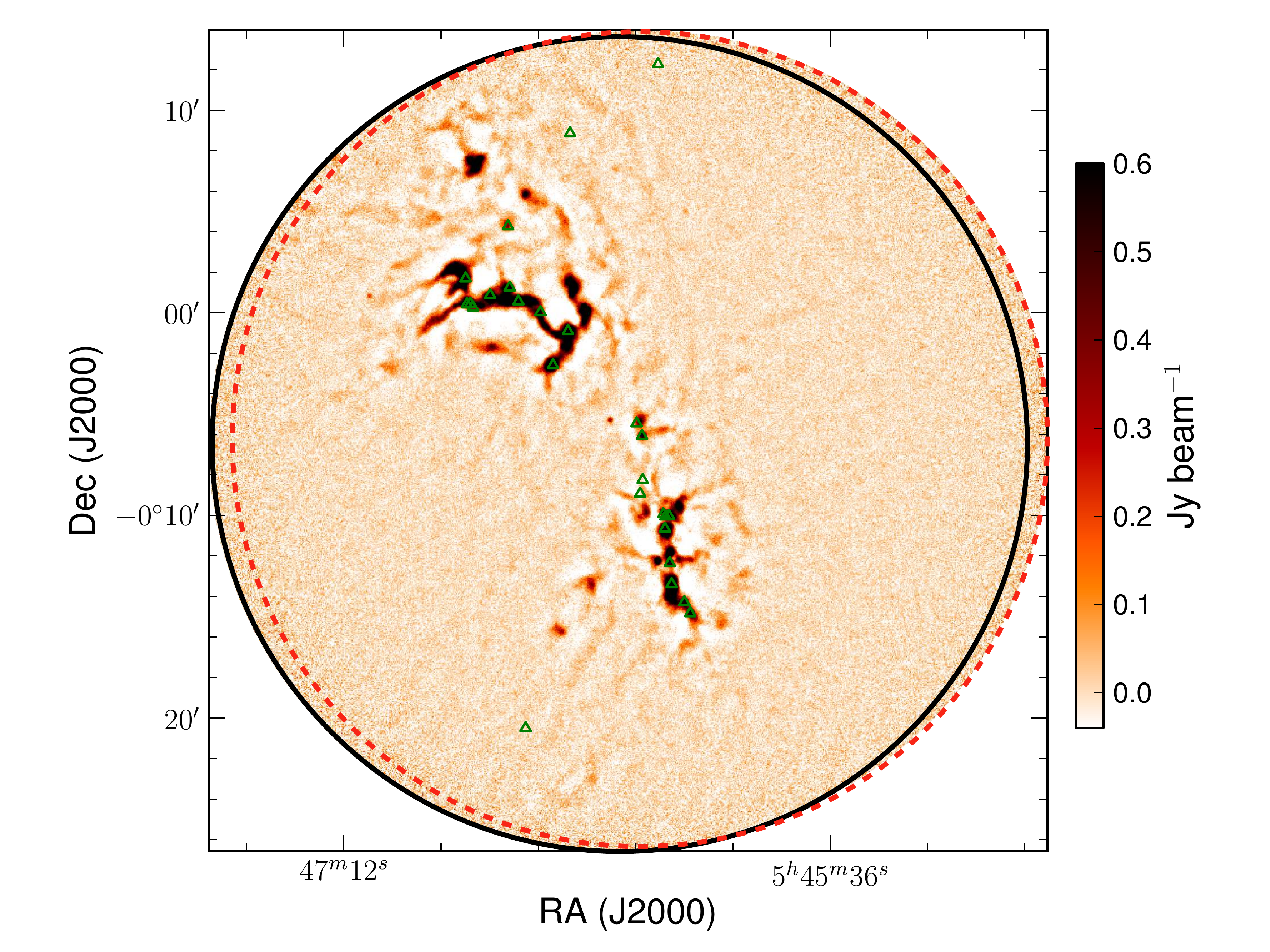}
\caption{Same as Figure \ref{IC348mosfig}, but showing the NGC2068 field with its corresponding archival GBS field. The red (dashed) circle shows the ORIONBN\_450\_S GBS field. The green triangles represent the positions of known protostars taken from the \textit{Spitzer Space Telescope} and \textit{Herschel Space Observatory} catalogues of \protect\cite{megeath2012} and \protect\cite{stutz2013}, respectively.}
\label{NGC2068mosfig}
\end{figure*}

\begin{figure*}  
\centering
\includegraphics[width=18cm,height=10cm]{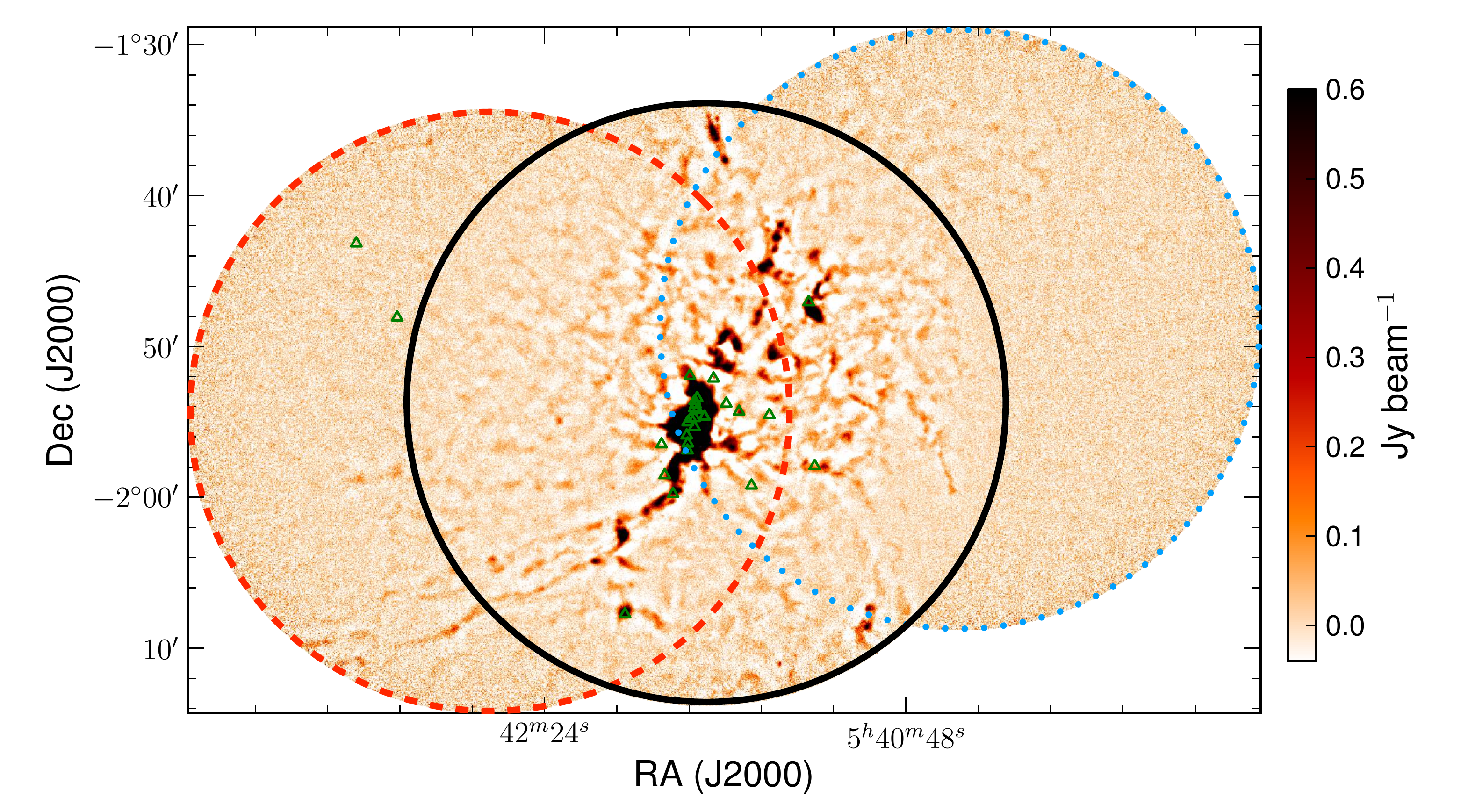}
\caption{Same as Figure \ref{IC348mosfig}, but showing the NGC2024 field with its corresponding archival GBS fields. The red (dashed) circle shows the ORIONBS\_450\_E GBS field while the blue (dotted) circle shows the ORIONBS\_450\_W GBS field. The green triangles represent the positions of known protostars taken from the \textit{Spitzer Space Telescope} and \textit{Herschel Space Observatory} catalogues of \protect\cite{megeath2012} and \protect\cite{stutz2013}, respectively.}
\label{NGC2024mosfig}
\end{figure*}

\begin{figure*}  	
\centering
\includegraphics[width=16.45cm,height=18cm]{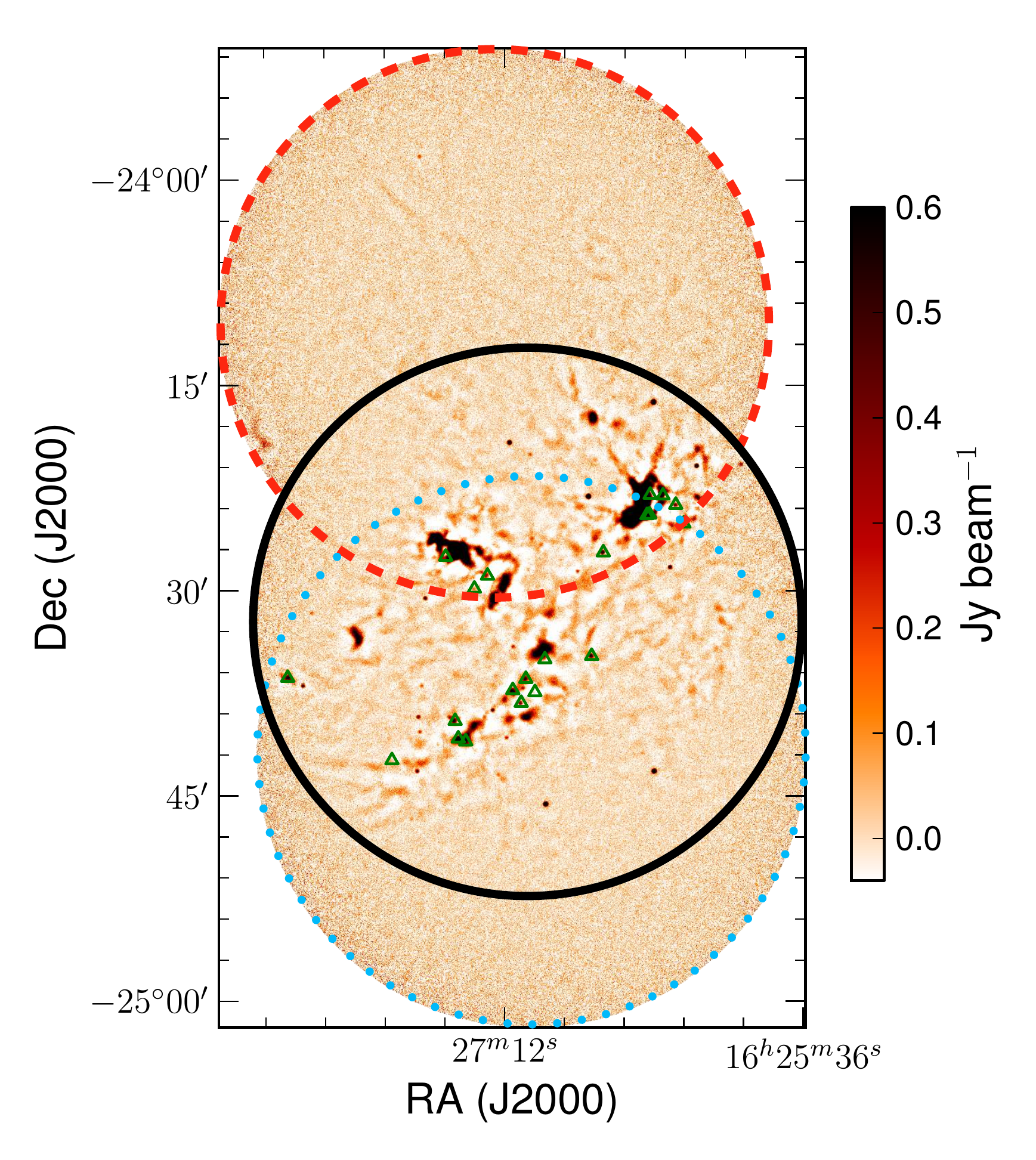}
\caption{Same as Figure \ref{IC348mosfig}, but showing the Oph Core field with its corresponding archival GBS field. The red (dashed) circle shows the L1688-2 GBS field while the blue (dotted) circle shows the L1688-1 GBS field.}
\label{OPHCOREmosfig}
\end{figure*}

\begin{figure*}  	
\centering
\includegraphics[width=18cm,height=13.2cm]{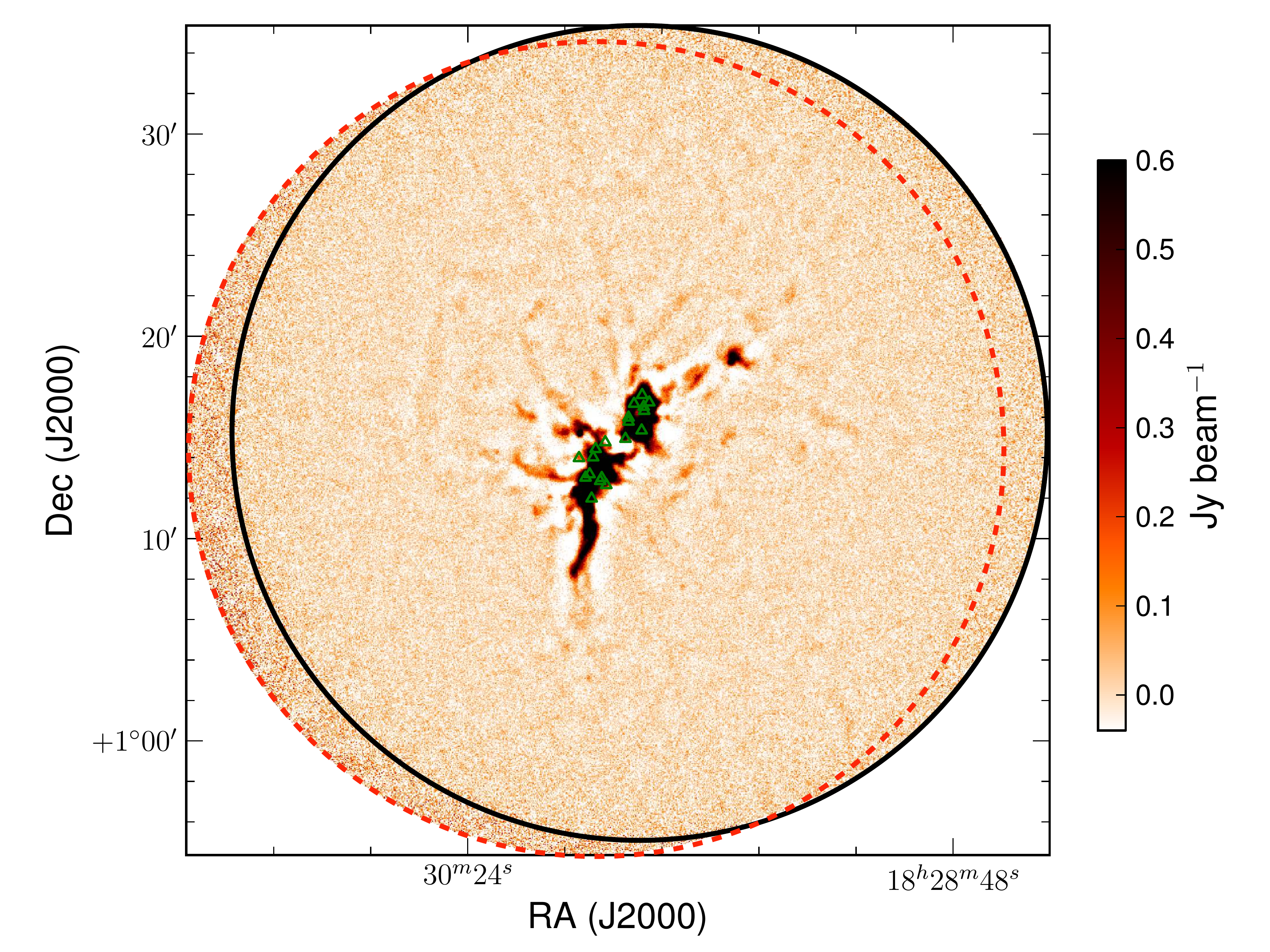}
\caption{Same as Figure \ref{IC348mosfig}, but showing the Serpens Main field with its corresponding archival GBS field. The red (dashed) circle shows the SerpensMain1 GBS field.}
\label{SERPENSMAINmosfig}
\end{figure*}

\begin{figure*}  	
\centering
\includegraphics[width=18cm,height=11.2cm]{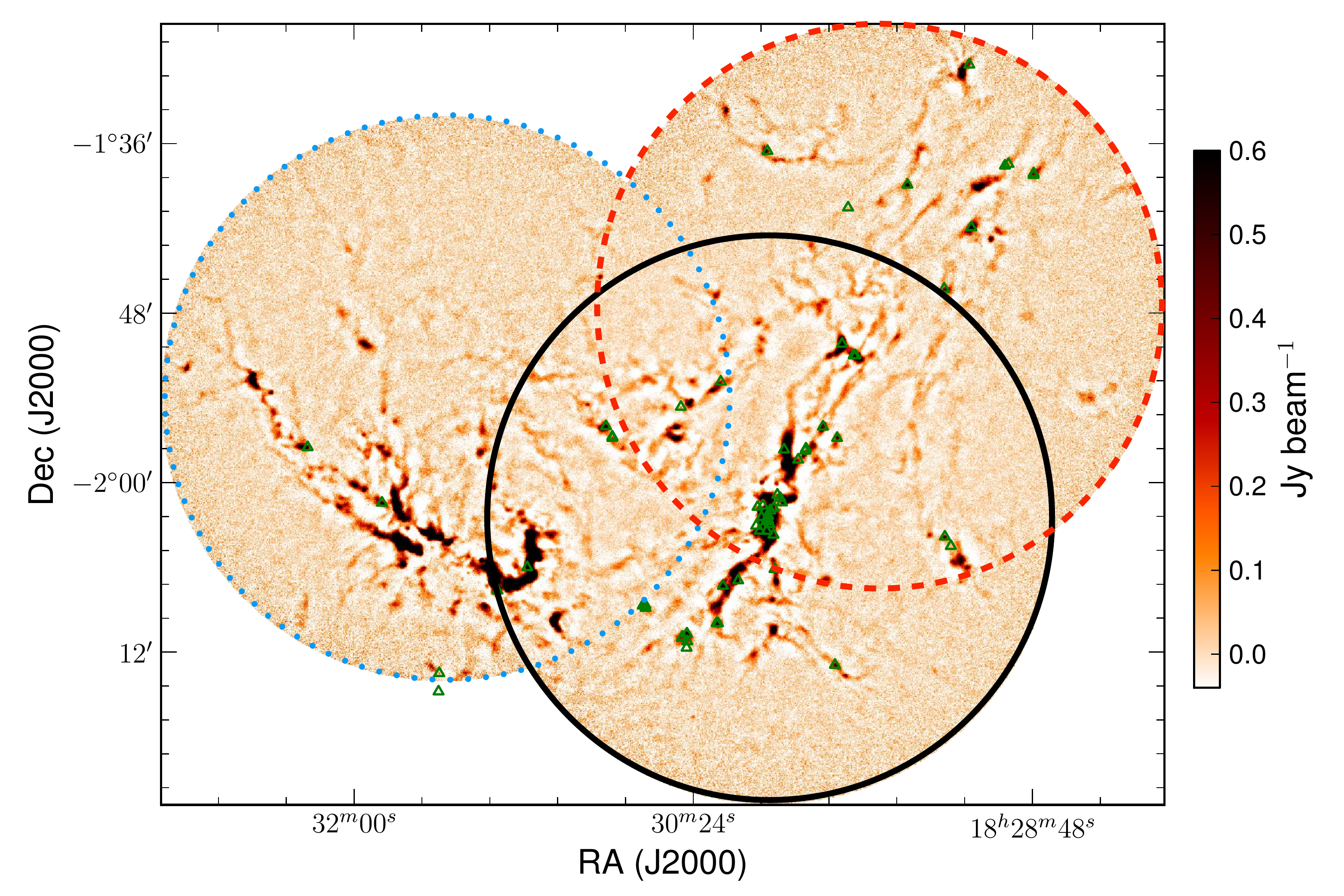}
\caption{Same as Figure \ref{IC348mosfig}, but showing the Serpens South field with its corresponding archival GBS fields. The red (dashed) circle shows the SerpensS-NW GBS field while the blue (dotted) circle shows the SerpensSouthS-NE GBS field.}
\label{SERPENSSOUTHmosfig}
\end{figure*}

For maximum consistency in relative flux calibration across all the regions, we select individual GBS fields and compare each one to its respective Transient Survey field independently. For a given GBS field to be useful for this work, it must: 1) have enough bright, compact sources to undergo self calibration and 2) have significant overlap with a Transient field such that they share at least three sources of interest and a flux calibration between the two datasets can be performed. The first criterion generally causes GBS fields with some Transient field overlap to fail. In order for a specific GBS field to undergo relative flux calibration, we require at least two compact sources brighter than \mbox{0.5 Jy beam$^{-1}$} that are observed to have consistent peak brightnesses with respect to one another across all observations. If two potential calibrator sources are noted, but display discordant flux calibration factors, it is unclear which, if either, of the sources represents the more correct value without further investigation. Where we lack robust sources, we discard the field. Similarly, if only one potential calibrator source is noted, it is unclear whether or not that source is intrinsically variable and, again, we discard the field. A self-calibration failure is generally encountered when the field is very sparse, or when a field is inundated with complex, extended emission without isolated, well-defined point-like sources. In the latter case, the {\sc{Gaussclumps}} algorithm has difficulties in defining source boundaries and properly subtracting background emission (see, for example, the ``Integral Shaped Filament'' in the OMC 2-3 region in Figure \ref{OMC23mosfig}; two GBS fields with the southern extension of the filament present along the noisy edge of their field of view were discarded). 

Note that a clear example of {\sc{Gaussclumps}} having difficulty subtracting the background is in the ORIONBS\_450\_E field. The error bars for many sources in Figure \ref{secondweightedmeanfig} (bottom left) are large due to the inconsistency of the identified source boundaries. To appear in Figures \ref{firstweightedmeanfig} through \ref{lastweightedmeanfig}, a source must be consistently observed in at least 2 GBS observations and at least 2 Transient Survey observations. Recall that we apply minimum peak brightness and maximum radius thresholds. Sources that have sizes that are near this threshold may be culled from many observations, yet detected in two or three as the extraction algorithm fits a slightly different Gaussian model to the same region on different dates, attempting to separate clustered structure. The lack of robust observations increases the uncertainty of the peak measurement but this is mitigated by well-defined, well recovered sources in conjunction with the calculation of a weighted mean to derive the relative FCF.

Source recovery in particularly confused and blended emission regions is also complicated by the fact that the recovery of extended background emission can vary from observation to observation, especially near the edges of the map. Reducing SCUBA-2 data is a complex process. Best practices for the GBS data reduction were developed by \cite{mairs2015} and extended to the Transient Survey by \cite{mairs2017}. Adopting the methodology of \cite{mairs2017} in this paper, we robustly recover compact structure in exchange for less information at extended scales. As discussed in Section \ref{GBStransientdatareducsec}, we perform a two-stage reduction where we employ an external mask in the final stage to help constrain {\sc{makemap}}'s solution. These external masks are created individually for each field in the GBS and Transient surveys, due to the different central coordinates of the overlapping fields. Where the fields from the two surveys overlap (see Figures \ref{IC348mosfig} through \ref{SERPENSSOUTHmosfig}), the masked structures are nearly identical. Outside of these regions, however, the fields have their own structure that can slightly affect how astronomical signal across the rest of the recovered image grows in the final iteration of {\sc{makemap}}. This effect is generally insignificant across the majority of the image but can cause slight differences near the edges of the fields where the noise levels are higher. 



\begin{figure*} 	
\centering
\subfloat{\label{}\includegraphics[width=9.3cm,height=7.2cm]{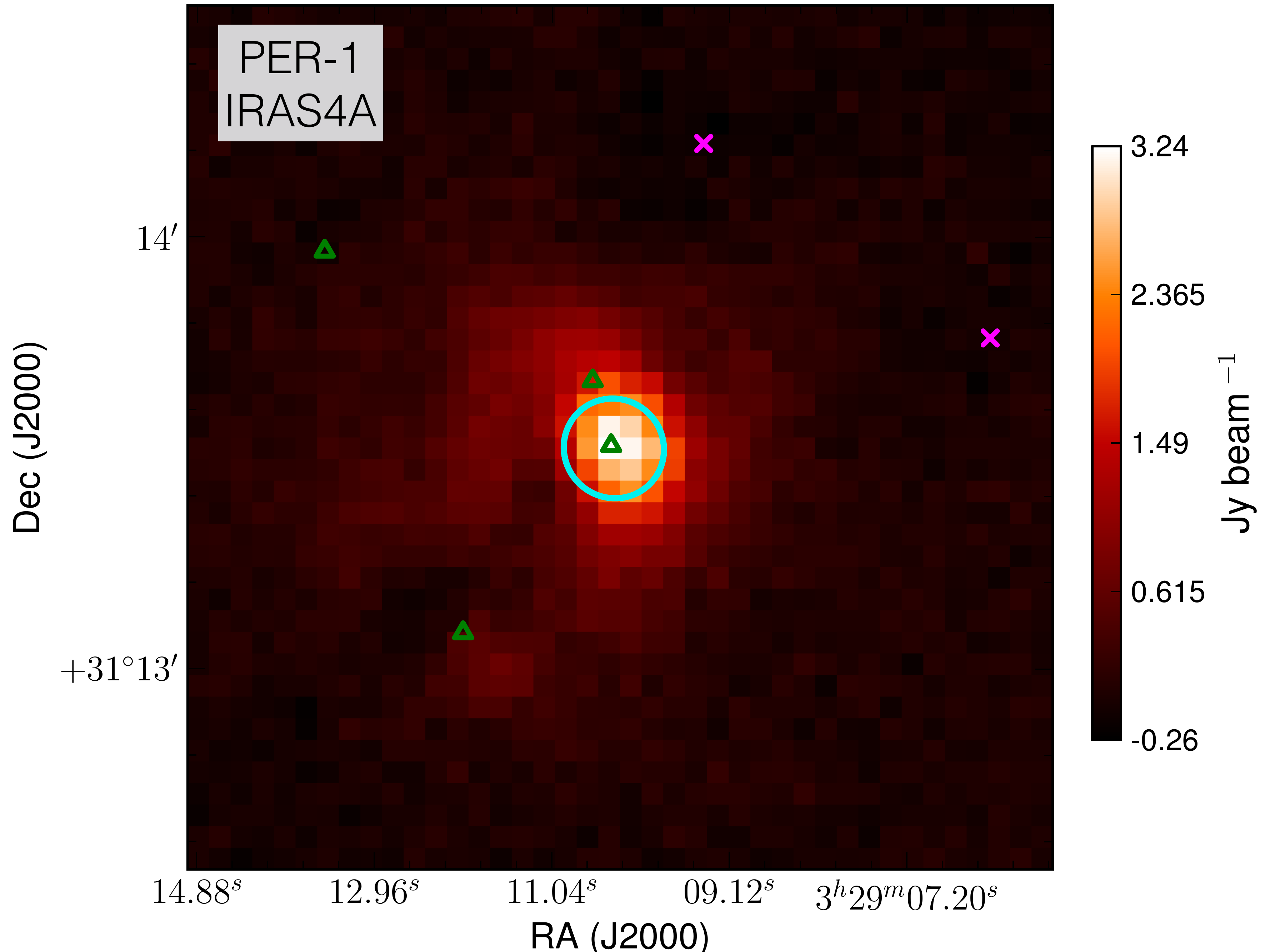}}
\subfloat{\label{}\includegraphics[width=9.3cm,height=7.2cm]{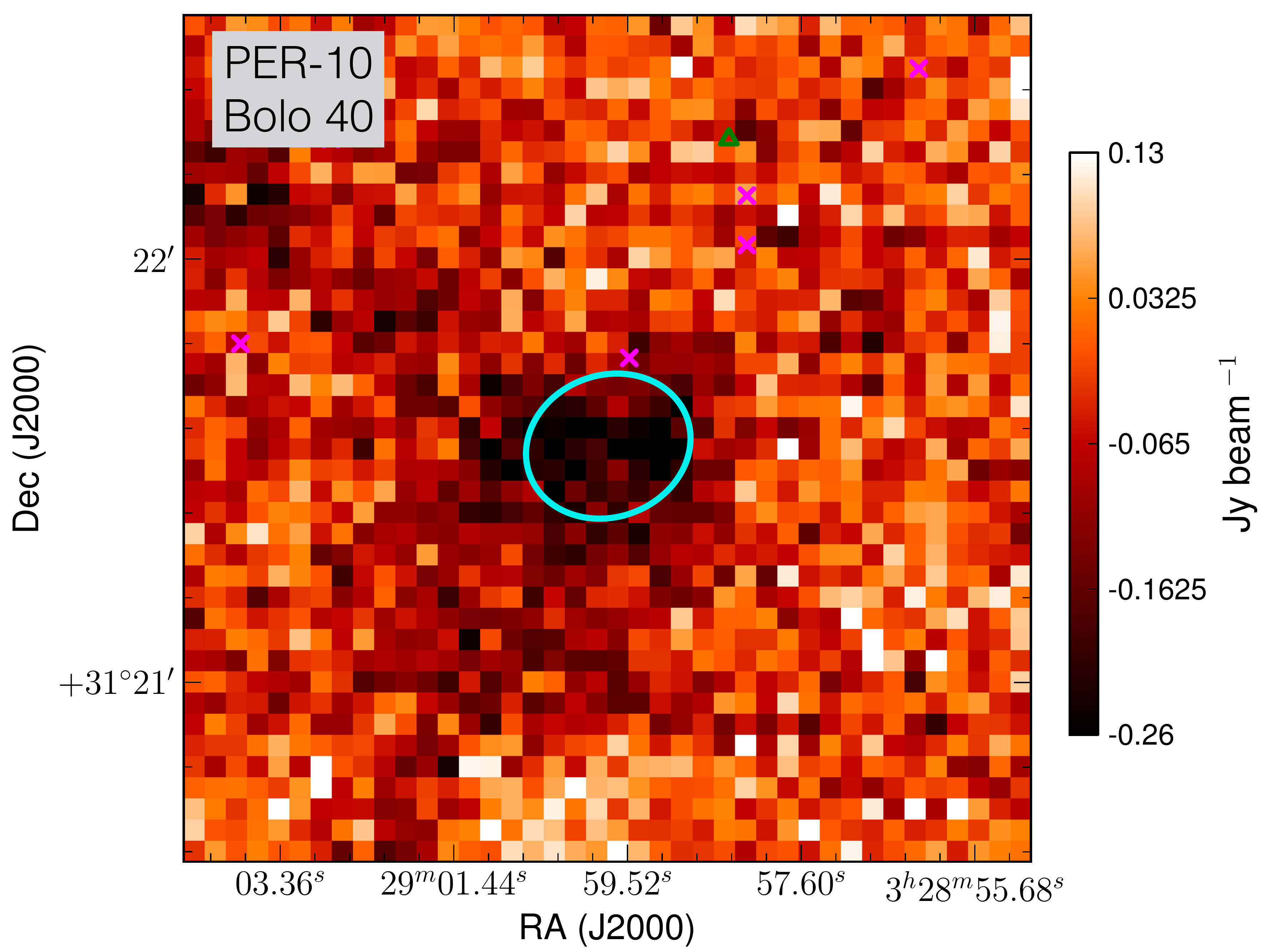}}\\
\subfloat{\label{}\includegraphics[width=9.3cm,height=7.2cm]{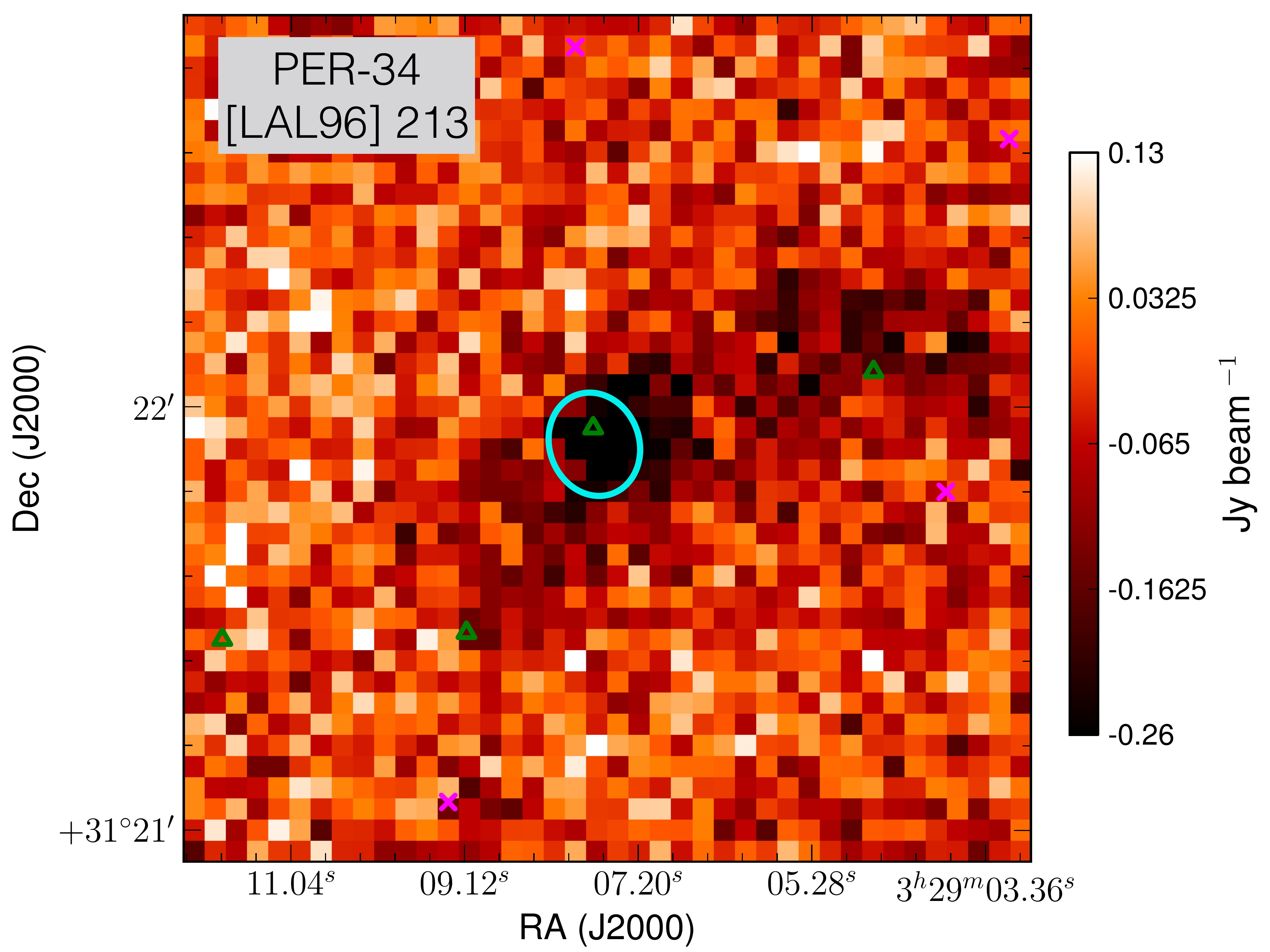}}
\subfloat{\label{}\includegraphics[width=9.3cm,height=7.2cm]{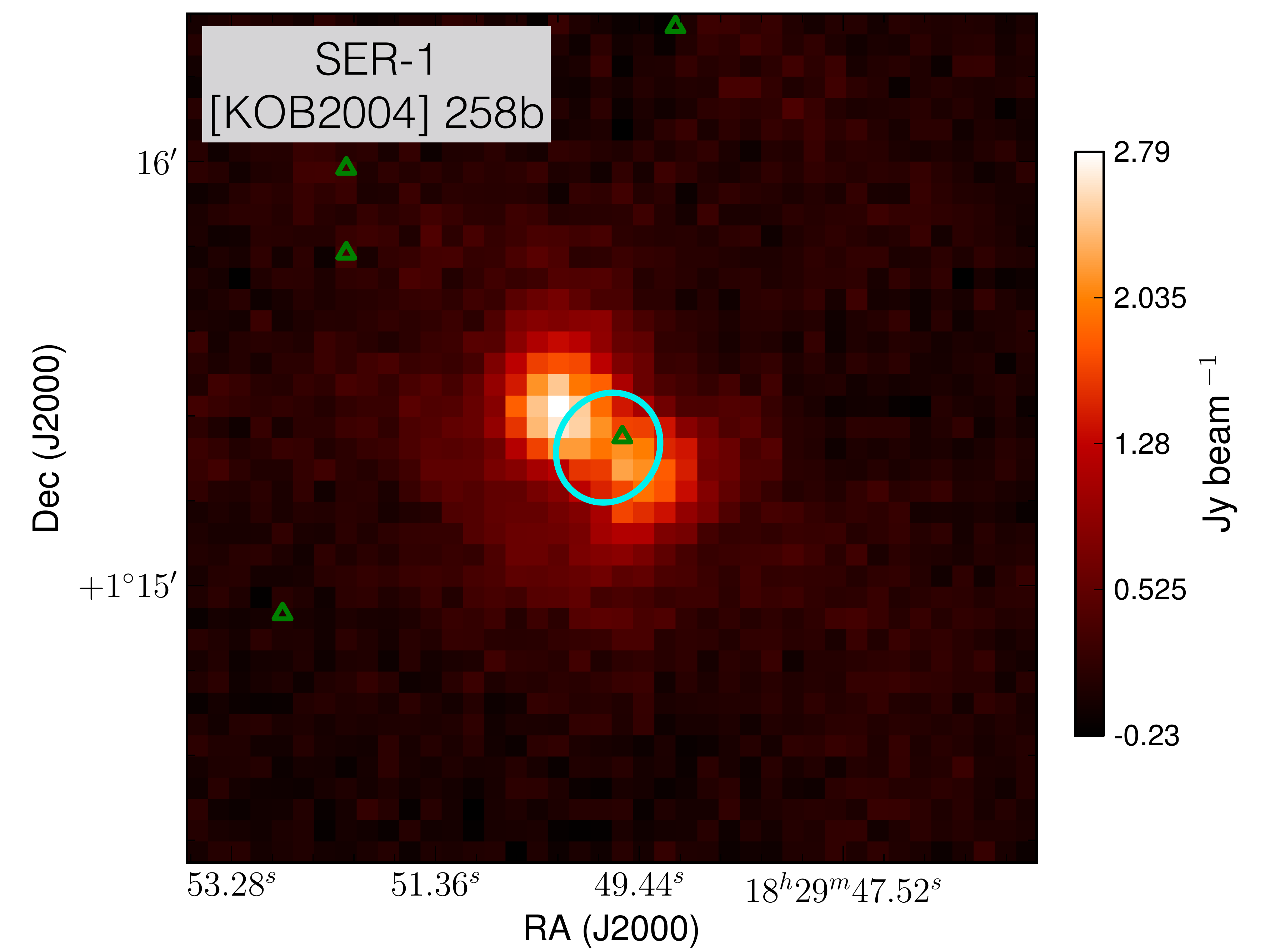}}\\
\subfloat{\label{}\includegraphics[width=9.3cm,height=7.2cm]{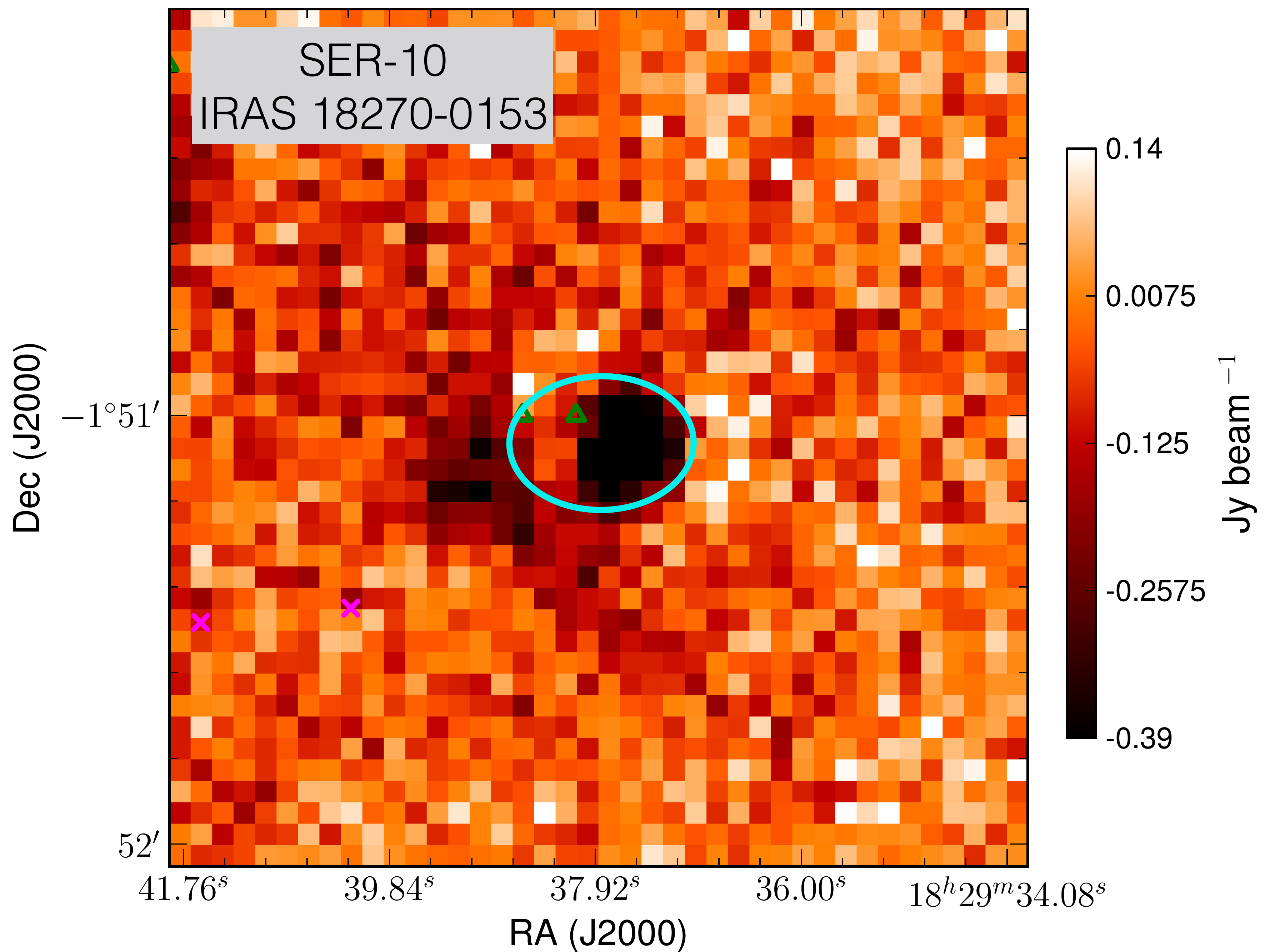}}
\subfloat{\label{}\includegraphics[width=9.3cm,height=7.2cm]{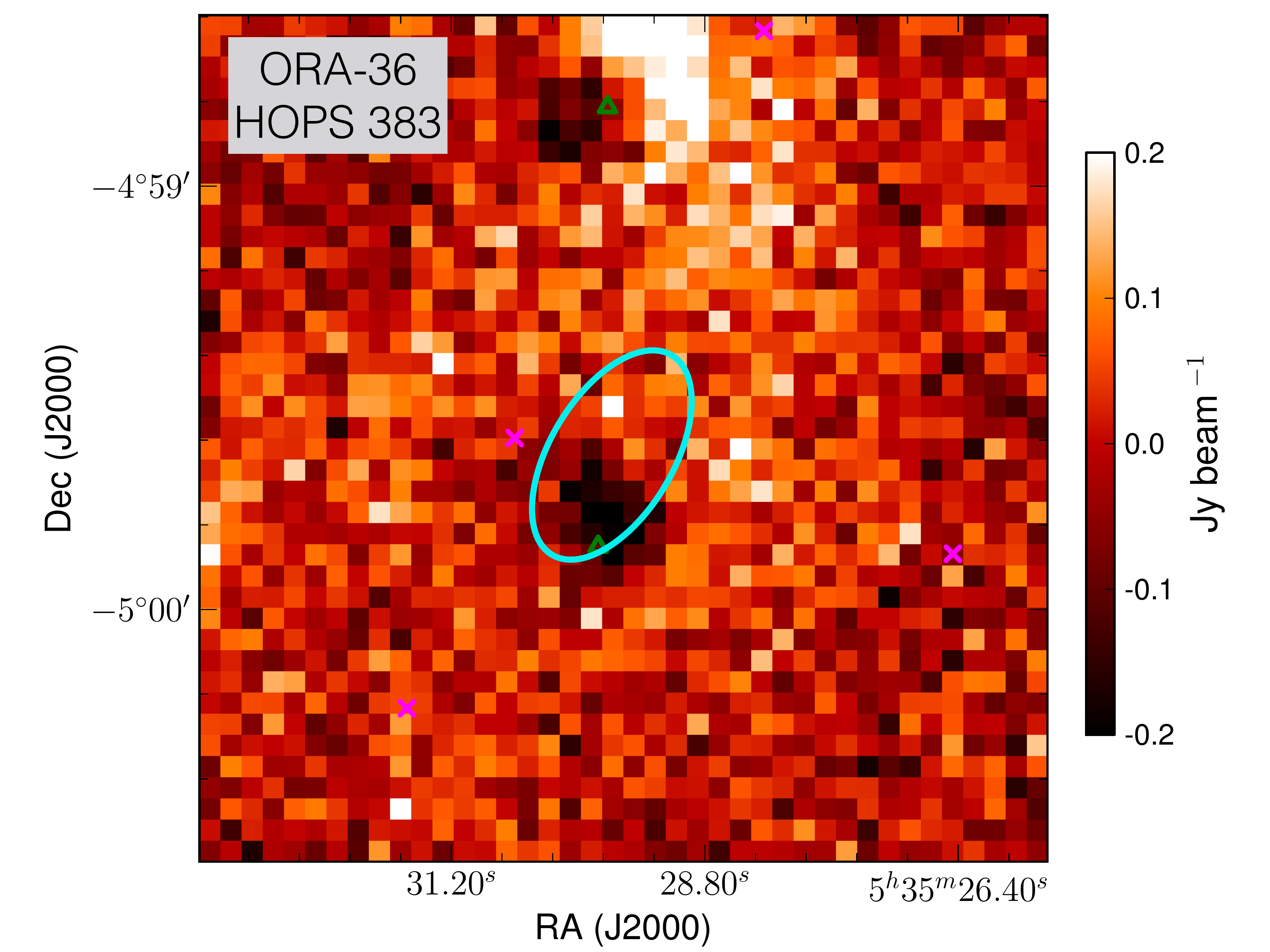}}
\caption{The 5 \textit{Strong} variable candidates and the 1 \textit{Possible} variable candidate (ORA-36/HOPS 383) extracted from difference maps where the GBS co-add has been subtracted from the Transient Survey co-add. Green triangles represent known protostars and magenta crosses represent known disk sources taken from the catalogues of \protect\cite{megeath2012}, \protect\cite{stutz2013}, and \protect\cite{dunham2015}. Cyan boundaries show the fitted 2D Gaussian truncated at the level of 0.5$\sigma_{rms}$.}
\label{difffig1}
\end{figure*}

\begin{figure*} 	
\centering
\subfloat{\label{}\includegraphics[width=9.3cm,height=7.2cm]{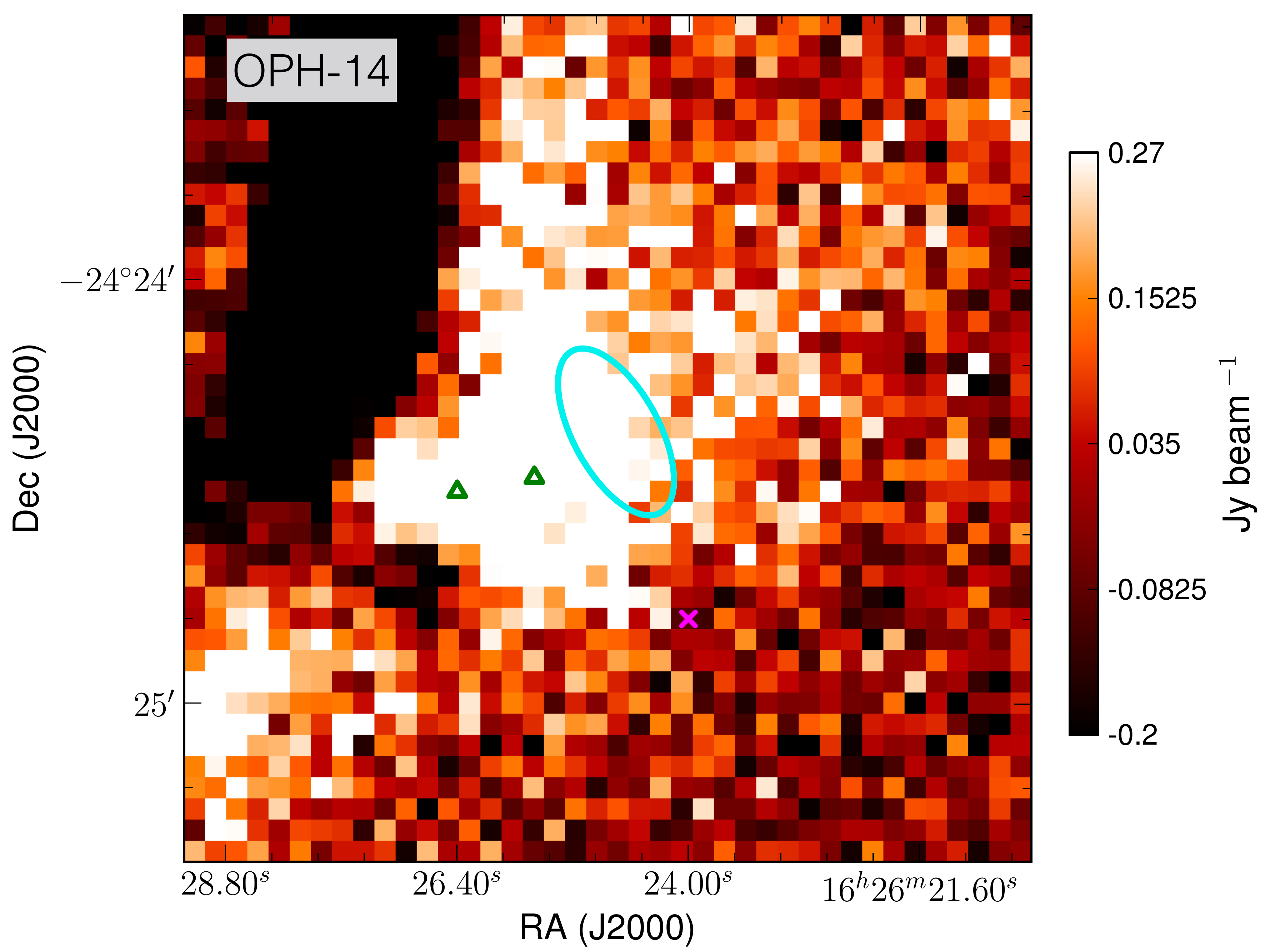}}
\subfloat{\label{}\includegraphics[width=9.3cm,height=7.2cm]{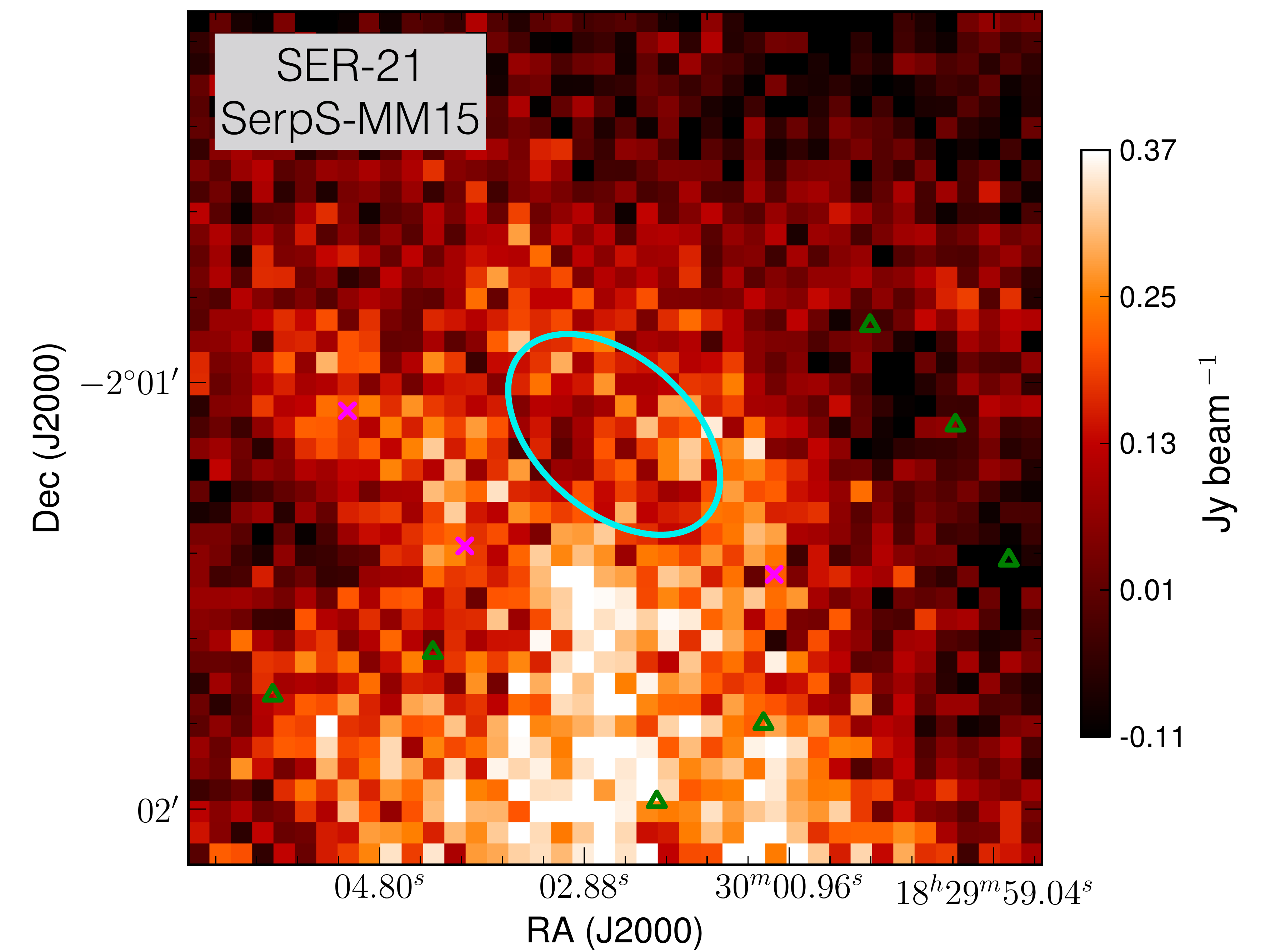}}
\caption{Same as Figure \protect\ref{difffig1}, but for \textit{Extended} variable candidates.}
\label{difffig2}
\end{figure*}

In many instances, the edges of GBS fields overlap with the centre of their associated Transient Survey field and the small differences in extended structure recovery create flux pedestals and negative bowling, which add to the uncertainty of a measured source in those regions \citep[see][for more information]{mairs2015}. Since the telescope has to slow down and speed up near the edges, the time domain filtering has a different effect on structure near the edge of the map with respect to that at the centre of the map. We have addressed this by inspecting each of the variable candidate sources in Table \ref{potvartable} in difference maps we constructed (see Figures \ref{difffig1} and \ref{difffig2}), looking for indications of larger-scale residual structures. We expect that compact, truly varying sources would show significant, point-like structure even in the midst of these extended regions. 


In Figures \ref{difffig1} and \ref{difffig2}, we present thumbnail images extracted from the constructed difference maps (the GBS co-added data subtracted from the Transient Survey co-added data) for all variable candidates in Table \ref{potvartable}. All of the \textit{Strong} variable candidates show significant, compact structure in their respective difference maps. 

\bibliography{GBStransientbib}

\end{document}